\documentclass[jmp,amsmath,amssymb,preprint]{revtex4-1}

\usepackage{graphicx}  
\usepackage{dcolumn}   
\usepackage{bm}        
\usepackage{ulem}      

\usepackage{dsfont}
\newcommand{\mb}[1]{\mathbf{#1}}
\newcommand{\lra}[1]{\langle #1 \rangle }
\newcommand{\bds}[1]{\boldsymbol{#1}}
\newcommand{\mc}[1]{\mathcal{#1}}

\begin{document}

\normalem

\title{Guidelines for the formulation of Lagrangian stochastic models for particle simulations of single-phase and dispersed two-phase turbulent flows}

\author{Jean-Pierre Minier}%
 \email{Jean-Pierre.Minier@edf.fr}
 \affiliation{EDF R$\&$D, M\'{e}canique des Fluides, Energie et Environnement,\\
              6 quai Watier, 78400 Chatou, France}
\author{Sergio Chibbaro}%
 \affiliation{Sorbonne Universit\'es, UPMC Univ Paris 06, CNRS, UMR7190, Institut Jean 
Le Rond d'Alembert, F-75005 Paris, France}%
 \altaffiliation[Also at ]{CNRS UMR7190, 4, place jussieu 75252 Paris Cedex 05, France}

\author{Stephen B. Pope}
\affiliation{Sibley School of Mechanical and Aerospace Engineering, Cornell University, 
254 Upson Hall, New-York 14853, USA}

\begin{abstract}
In this paper, we establish a set of criteria which are applied to discuss various formulations
under which Lagrangian stochastic models can be found. These models are used for the simulation of fluid particles in single-phase turbulence as well as for the fluid seen by discrete particles in dispersed 
turbulent two-phase flows. The purpose of the present work is to provide guidelines, useful for 
experts and non-experts alike, which are shown to be helpful to clarify issues related to the 
form of Lagrangian stochastic models. A central issue is to put forward reliable requirements which must be 
met by Lagrangian stochastic models and a new element brought by the present analysis is to address the
single- and two-phase flow situations from a unified point of view. For that purpose, we consider
first the single-phase flow case and check whether models are fully consistent with the structure of the 
Reynolds-stress models. In the two-phase flow situation, coming up with clear-cut criteria is more difficult 
and the present choice is to require that the single-phase situation be well-retrieved in the fluid-limit 
case, elementary predictive abilities be respected and that some simple statistical features of homogeneous fluid turbulence be correctly reproduced. This analysis does not address the question of the relative 
predictive capacities of different models but concentrates on their formulation since advantages and disadvantages of different formulations are not always clear. Indeed, hidden in the changes from one 
structure to another are some possible pitfalls which can lead to flaws in the construction of practical 
models and to physically-unsound numerical calculations. A first interest of the present approach is illustrated by considering some models proposed in the literature and by showing that these criteria help to assess whether these Lagrangian stochastic models can be regarded as acceptable descriptions. A second interest is to indicate how future developments can be safely built, which is also relevant for stochastic subgrid models for particle-laden flows in the context of Large Eddy Simulations.
\end{abstract}

\pacs{Valid PACS appear here}

\maketitle

\section{Introduction}

Over the last decades, Lagrangian stochastic models have become increasingly used for both single-phase reactive flows and dispersed two-phase turbulent flows (with one phase being present as discrete elements such as solid particles, droplets or bubbles). These approaches are referred to as PDF (Probability Density Function) 
methods~\cite{Pope_1985,Pope_1994,Pope_2000,Minier_2001,Fox_2003,Peirano_2006,Haworth_2010,Jenny_2012} which 
indicates that they are simulating the pdf of the relevant variables which have been retained for the statistical description of either single- or two-phase flows. PDF methods have strong advantages as they treat important phenomena without approximation:in single-phase reactive flows, this corresponds to convective and reactive source terms~\cite{Pope_1985,Pope_1990} while, for dispersed two-phase flows (even for inert particles), 
this corresponds to transport and polydispersity effects (related to the existence of a range of 
particle diameters)~\cite{Pozorski_1999,Minier_2001}. This interplay of modeling issues explains the common interest of PDF descriptions in both situations. 

In the single-phase flow situation, the governing equations are the transport equations for the fluid velocity field $\mb{U}(t,\mb{x})$ and for a set of scalars which gathers the relevant species mass fractions 
$\bds{\phi}(t,\mb{x})=(\phi_{\beta})_{\beta=1,\ldots, N_s}$ to which an equation for the fluid enthalphy is added, along with an equation of state, for compressible flows. For constant-property flows, these equations are
\begin{subequations}
\label{fluid: exact field eqs.}
\begin{align}
&\frac{\partial U_k}{\partial x_k}=0~, \label{fluid: exact field eqs. rho} \\
&\frac{\partial U_i}{\partial t} + U_k\, \frac{\partial U_i}{\partial x_k}
= - \frac{1}{\rho}\frac{\partial P}{\partial x_i} + \nu\, \frac{\partial^2 U_i}{\partial x_k\partial x_k}~, 
\label{fluid: exact field eqs. U} \\
&\frac{\partial \phi_{\beta}}{\partial t} + U_k\, \frac{\partial \phi_{\beta}}{\partial x_k}=
\Gamma \, \frac{\partial^2 \phi_{\beta}}{\partial x_k\partial x_k} + S_{\beta} \label{fluid: exact field eqs. phi}
\end{align}
\end{subequations}
where $\nu$ is the fluid dynamical viscosity, $\Gamma$ the scalar diffusivity and $P(t,\mb{x})$ the
fluid pressure. In Eq.~\eqref{fluid: exact field eqs. phi}, the last term on the rhs (right-hand side)
is the reactive source term $S_{\beta}=\hat{S}_{\beta}(\bds{\phi}(t,\mb{x}))$ which, along with convection, appears in a closed form in a one-point PDF approach.
 
In the disperse two-phase flow case, the basic physical situation is made up by a continuous fluid phase (a gas or a liquid) in which a set of discrete `particles' (solid particles, droplets, bubbles, etc.), having a range of diameters, are embedded. The fluid phase is described by the continuity and Navier-Stokes equations, Eq.~\eqref{fluid: exact field eqs. rho}-Eq.~\eqref{fluid: exact field eqs. U}, to which source terms can be added to account for momentum exchange between the fluid and the discrete particles when two-way coupling (whereby particles influence the fluid phase) is deemed important. For the discrete particles, we limit ourselves to the case of point-like particles or droplets. This approximation is not severe for most industrial applications but usually leaves out bubbles~\cite{Prosperetti_2004}.
For particle diameters of the same order of magnitude as the Kolmogorov length scale, the particle momentum equation involves the well-known pressure-gradient, drag, added-mass and Basset forces~\cite{Gatignol_1983,Maxey_1983}. In the case of particles heavier than the fluid (droplets in a gas, solid particles in a gas or liquid), the particle momentum equation can be simplified to the following form to describe the evolution of particle location $\mb{x}_p(t)$ and velocity $\mb{U}_p(t)$: 
\begin{subequations} 
\label{exact particle eqns}
\begin{align} 
& \frac{d\mb{x}_p}{dt} = \mb{U}_p, \\
& \frac{d\mb{U}_p}{dt} = \frac{1}{\tau_{p}}(\mb{U}_{s}-\mb{U}_{p}) + \mb{g},
\label{exact particle eqns_Up}
\end{align}
\end{subequations}
where, apart from gravity, only the drag force has been retained (other forces can be added but the drag force is sufficient for the present discussion). In Eq.~\eqref{exact particle eqns_Up}, $\mb{g}$ is the gravity acceleration and $\tau_p$ the particle relaxation time defined as
\begin{equation} \label{definition taup}
\tau_{p}=\frac{\rho_p}{\rho_f}\frac{4d_p}{3 C_D |\mb{U_r}|},
\end{equation}
where the local instantaneous relative velocity between the fluid and the particle velocity is
$\mb{U_r}=\mb{U_s}-\mb{U_p}$. The drag coefficient $C_D$ is usually expressed as a non-linear
function of the particle-based Reynolds number, $Re_p=d_p |\mb{U_r}|/\nu$ (where $d_p$ is the particle diameter) based on empirical formulas~\cite{Clift_1978} (apart from the Stokes regime). 
In Eq.~\eqref{exact particle eqns_Up} and in the expression of the particle relaxation time scale, the important variable is $\mb{U}_{s}(t)=\mb{U}(t,\mb{x}_{p}(t))$ which is the `fluid velocity seen', \textit{i.e.} the fluid velocity sampled along the particle trajectory $\mb{x}_{p}(t)$ as it moves across a turbulent flow.
In the limit of the assumptions made above, it is possible to solve the governing equations in the spirit of DNS (Direct Numerical Simulation) for disperse flows with accurate Lagrangian tracking methods~\cite{Toschi_2009,Marchioli_2002,Gualtieri_2009,balachandar_2010}, as the continuation of what is done for the DNS of single-phase flows. This approach is possible in simple geometries and at moderate Reynolds numbers. When such DNS are not available and for practical purposes where only one-point statistics are known, a stochastic model for the velocity of the fluid seen is needed. In the general context of stochastic models and PDF descriptions, we are thus concerned with a one-particle PDF model for $\mb{U}_s(t)$ from which one-point statistics can be derived~\cite{Minier_2001}. Note that when the velocity of the fluid seen is included in the state-vector along with the particle velocity and diameter, the drag force appears in a closed form in the PDF description~\cite{Minier_2001} (it can be seen as the counterpart of the reactive source term in the single-phase flow case).

In the present analysis, we leave out the specific issues related to reactive aspects (gas chemical 
reactions in single-phase flows; burning particles, evaporating droplets, etc. in two-phase flows) 
and concentrate on dynamical aspects in the PDF approach to both situations. Thus, at the core of the PDF 
method lies the specific stochastic model which is used to simulate fluid particle velocities in 
single-phase flows and the fluid velocity `seen' by particles in two-phase flows. In the latter situation,
this notion was introduced as a Lagrangian property attached to each discrete particle~\cite{Simonin_1993},
then included in the particle state-vector~\cite{Pozorski_1999} which led to the standard PDF 
description for two-phase flows~\cite{Minier_2001}. Furthermore, the PDF approach was given a 
complete framework, first in single-phase flows~\cite{Pope_1985} and later in 
two-phase flows~\cite{Minier_2001}, ensuring a continuous link between the choice of the particle state-vector, the formulation of stochastic models as proper stochastic differential equations, the corresponding PDF equation in sample space and the resulting mean-field
equations~\cite{Pope_1985,Pope_1994a,Peirano_2002}.  

At this stage, a first modeling question arises: how should these stochastic models be formulated? 
In single-phase flows, most of the stochastic models have been developed in terms of instantaneous fluid particle velocities~\cite{Haworth_1986,Pope_1994a,Pope_1994}. In two-phase flows, some proposals followed the same road and were made in terms of the instantaneous fluid velocity seen to build so-called Langevin models~\cite{Minier_2001,Minier_2004}. However, the modeling situation is unclear since existing models 
can be expressed with different formulations, for instance in terms of either instantaneous, fluctuating or normalised fluctuating fluid-particle velocities. When particle inertia becomes negligible, the stochastic model for the velocity of the fluid seen becomes a stochastic model for fluid particles and, consequently, this issue overlaps with similar concerns in single-phase flows. This means that both situations are 
impacted by the issue of the formulation of stochastic models. A second modeling question is: what are
the basic properties that such models must respect?

These questions are particularly relevant when the stochastic models used to simulate dynamical
variables are complemented with additional models to address complex-physics problems. For 
instance, in single-phase flows, one could be interested in applying a velocity-composition 
PDF method but having only specifically developed the modeling parts concerned with scalars and 
reactive terms. In two-phase flows, Lagrangian stochastic models can be used to analyze additional 
effects such as thermophoresis, electrophoresis or chemical 
forces~\cite{Henry_2012b,Newman_2005,LoIacono_2005} (due to interface chemistry in liquid medium). 
Other typical examples include droplets or coal/fuel particles where models are added to simulate 
complex combustion or evaporation processes~\cite{Irannejad_2013,Jenny_2012,Mueller_2012,Haworth_2010}. 
Yet, these practical developments can be ruined by a poor formulation of the 
model retained for the velocity of fluid particles or the velocity of the fluid seen. 
It appears therefore important to assess whether stochastic models used for dynamical 
variables have a sound basis. By this assessment step, we do not mean here the (necessary) 
task of outlining the predictive abilities of modeling proposals but the `upstream step' that consists 
in assessing whether the structure of these models respect key properties, regardless of the details 
of specific closures. Going directly to a comparison of computational outcomes between
different model formulations~\cite{Taniere_2014} is interesting but may confuse the status 
of different classes of Lagrangian stochastic models whereas a theoretical analysis can already reveal 
flaws or bring out differences that make these models difficult to compare even-handedly.

Given the subtleties of stochastic calculus~\cite{Gardiner_1990,Ottinger_1996} and the 
particle-based nature of Lagrangian PDF approaches, consistency issues have accompanied the construction
of stochastic models. For example, the issue of so-called spurious drifts and its relation to fluid
mass conservation was addressed a few years ago for models developed for single-phase turbulent 
flows~\cite{Sawford_1986,Pope_1987,Thomson_1987,McInnes_1992}. Relations with Reynolds-stress modeling
were also put forward~\cite{Pope_1994a} with a view towards the interest of stochastic models for
realizable closures. Similar efforts have been made in the two-phase flow situation~\cite{Minier_2001}, though more sparingly. However, there has been no previous attempt at gathering knowledge and addressing 
the validity and the structure of stochastic models by resorting to a systematic list of requirements,
especially when the single- and two-phase situations need to be jointly considered.

With respect to this context, the first purpose of this article is to propose a clear set of requirements, for single-phase as well as for two-phase flows, which must be met by Lagrangian stochastic models in order to be regarded as acceptable descriptions. The second purpose is to discuss the relations between different formulations of a stochastic model for the velocity of fluid particles and to reveal the interests as well as the limitations of some of these formulations. The third purpose is to establish guidelines for future developments, valid for stochastic models developed in classical Reynolds-averaging approaches but also of interest for models considered for subgrid-effects in particle-laden flows where the fluid phase is simulated with a LES (Large Eddy Simulation) approach. In that sense, the present considerations represent an effort to address issues related to single-phase and two-phase flows from a unified perspective.

The paper is organized as follows. The PDF theoretical framework is first recalled in section~\ref{The PDF theoretical framework}, where the simplifying assumptions which define the precise context of this study are stated at the end of section~\ref{From Lagrangian stochastic models to mean-field equations}. Then, the criteria selected in the present analysis are detailed in section~\ref{Criteria for the analysis of stochastic models}, first for the single-phase flow case in section~\ref{Choice of criteria for single-phase flow models} 
and, second, for the two-phase flow case in section~\ref{Choice of criteria for two-phase flow models}. The analysis is first developed for single-phase turbulent flows: different formulations for fluid particle velocities are addressed in section~\ref{Stochastic models for single-phase flows} and analyzed in detail in 
section~\ref{Analysis of different formulations}, while classical scalar modeling (which plays an important role in one criterion for two-phase flows) is recalled in section~\ref{scalar flux}. Drawing on the analysis carried out for the single-phase flow case, an analysis of different modeling proposals for two-phase flows is developed in section~\ref{Stochastic models for two-phase flows}. In particular, new relations for models expressed in terms of fluctuating components, as well as discussions on two-way coupling, are developed in section~\ref{Fluctuating or instantaneous fluid velocity seen}. Present findings are summarized in Table~\ref{table_single-phase-models} for single-phase models and in Table~\ref{table_two-phase-models} for two-phase models. Finally, guidelines for future developments are proposed in the Conclusion.

\section{The PDF theoretical framework}
\label{The PDF theoretical framework}

This framework was first established for single-phase turbulent 
flows~\cite{Pope_1985, Pope_1994, Pope_1994a, Pope_2000} and was later used as a foundation 
for the extension to dispersed two-phase turbulent
flows~\cite{Minier_2001,Peirano_2002,Chibbaro-Minier_2014}. However, for the sake of
a simpler and more compact presentation, we introduce the key aspects of the theoretical 
framework directly from the standpoint of the two-phase flow situation since the fluid-particle 
case can be retrieved as an asymptotic limit.

\subsection{Probabilistic descriptions and stochastic equations}
\label{Probabilistic descriptions and stochastic equations}

The PDF machinery for fluid mechanics starts by the choice of the PDF description (in terms of either one-particle pdf, or two-particle pdf, etc.). In the present context, we consider only one-particle PDF approaches and, thus, the starting point is the selection of the mechanical description retained for each particle or, in other words, the choice of the relevant particle state-vector which gathers the variables of interest attached to each particle. Following the presentation of standard Lagrangian models for dispersed two-phase flows in the Introduction, the particle state-vector is made up by the particle location and velocity as well as the velocity of the fluid seen by the particle, $\mb{Z}=(\mb{x}_{p},\mb{U}_{p},\mb{U}_{s})$, with evolution equations written as
\begin{subequations}
\label{general_stochastic_2phi}
\begin{align}
d\mb{x}_{p} &= \mb{U}_{p}\, dt , \\
d\mb{U}_{p} &= \mb{D}_{p}(t,\mb{Z})\, dt , \\
d\mb{U}_{s} &= \mb{D}_{s}(t,\mb{Z},\mc{F}[\lra{\mb{Z}}],\lra{\Phi})\, dt 
+  \mb{B}_{s}(t,\mb{Z},\mc{F}[\lra{\mb{Z}}],\lra{\Phi})\; d\mb{W} .
\label{general_stochastic_2phi_c}
\end{align}
\end{subequations}
In these equations, $\mb{D}_p$ typically represents the drag and gravity forces, $\mb{D}_p=\left( \mb{U}_s -\mb{U}_p\right)/\tau_p + \mb{g}$ (other forces can also be considered), while the vector $\mb{D}_s$ and the matrix $\mb{B}_s$ are the drift and diffusion coefficients of a stochastic diffusion process which is a typical model for the velocity of the fluid seen. In the drift and diffusion coefficients
in Eq.~(\ref{general_stochastic_2phi_c}), a general notation has been used to indicate that these coefficients can depend on the value of the state-vector $\mb{Z}$ but also on functionals of the mean fields which are calculated from the simulation of that state-vector, written as $\mc{F}[\lra{\mb{Z}}]$, as well as on some external fields represented by $\lra{\Phi}$. A typical example of $\mc{F}[\lra{\mb{Z}}]$ is the particle mean-velocity field while the fluid mean-pressure is another example of what $\lra{\Phi}$ can stand for. For detailed discussions of the modeling issues from a physical point of view, we refer to the relevant literature\cite{Pope_1994, Minier_2001}.
 
The single-phase flow framework is retrieved by considering the particle-tracer limit when particle inertia goes to zero ($\tau_p \to 0$). In that case, the particle velocity tends towards the fluid velocity and, for example, the model system of equations, Eqs.~(\ref{general_stochastic_2phi}), becomes  
\begin{subequations}
\label{general_stochastic_1phi}
\begin{align}
d\mb{x} &= \mb{U}\; dt, \\
d\mb{U} &= \mb{D}(t,\mb{Z},\mc{F}[\lra{\mb{Z}}],\lra{\Phi})\, dt 
+  \mb{B}(t,\mb{Z},\mc{F}[\lra{\mb{Z}}],\lra{\Phi})\; d\mb{W} .
\end{align}
\end{subequations}
where the same notation has been retained to indicate a possible dependence on mean-fields calculated from the solution (such as the fluid mean velocity field) or on external mean fields. A typical example of such an external mean field is the fluid mean dissipation field, $\lra{\epsilon}$. It must be noted that the notion of external fields is, of course, directly dependent upon the choice of the variables entering the state-vector. For instance, for $\mb{Z}=(\mb{x},\mb{U})$ which corresponds to a velocity-PDF description~\cite{Haworth_1986}, 
the mean dissipation is an external field. Yet, if the state-vector is extended to include the instantaneous dissipation, whereby $\mb{Z}$ becomes $\mb{Z}=(\mb{x},\mb{U},\epsilon)$ and corresponds to a velocity-dissipation PDF description~\cite{Pope_1990,Pope_1991}, then the dependence of the drift and diffusion coefficients on the mean dissipation field would appear through the functional form $\mc{F}[\lra{\mb{Z}}]$.

For both the single- and two-phase situations, using a general equation written
\begin{equation}
\label{general_stochastic_diffusion}
d\mb{Z}= \mb{D}_Z\, dt + \mb{B}_Z\, d\mb{W}
\end{equation} 
as a reference model, the key points of the PDF theoretical framework can be unraveled.

The governing equation, Eq.~(\ref{general_stochastic_diffusion}), is to be understood as the evolution equation for a large number of `stochastic particles' (whose behavior mimic the evolution of real particles in a statistical sense) and, in a weak sense~\cite{Ottinger_1996,Talay_1995,Oksendal_2003,Pope_1985,Minier_2001}, a stochastic particle-tracking model is equivalent to a PDF approach. However, it is worth emphasizing that the present case is an extension of the well-established connection between classical Langevin equations, where the drift and diffusion coefficients depend only the chosen state-vector $\mb{Z}$, and the corresponding Fokker-Planck equation in sample space. In that sense, there is an essential difference between the Langevin equations used, for example, in PDF methods and in dispersion studies~\cite{Pope_2000}. The difference is that, in PDF 
methods, the coefficients involve statistics obtained from the particles as indicated by the general notation in Eqs.~(\ref{general_stochastic_2phi}) and~(\ref{general_stochastic_1phi}). In the mathematical literature, these processes are referred to as `McKean diffusion processes'~\cite{Mckean_1969} and in detailed physical presentations~\cite{Ottinger_1996} they are defined as `processes with mean-field interactions'. Compared to classical Monte Carlo methods, numerical implementations lead to handling so-called `weakly interacting processes' but, basically, the classical connection between these generalized Langevin and Fokker-Planck equations remains valid~\cite{Ottinger_1996,Chibbaro-Minier_2014}. In other words, a particle-tracking method amounts to a dynamical Monte Carlo simulation of the corresponding pdf $p_L(t; \mb{z})$, where the index $L$ indicates that we are dealing with a Lagrangian pdf and where $\mb{z}$ stands for the state-vector variables in the corresponding sample space. 
 
\subsection{From Lagrangian stochastic models to mean-field equations}
\label{From Lagrangian stochastic models to mean-field equations}

In order to go from the (Lagrangian) stochastic equations written for each notional particle to the resulting (Eulerian) mean-field equations, a straightforward approach is obtained by introducing MDFs (Mass Density Function)~\cite{Pope_1985} and by defining the Eulerian MDF from the Lagrangian one. This is translated in the following definition where the Eulerian MDF is basically identified with the Lagrangian one at the same location (the identity is for MDFs and not necessarily in terms of pdfs) as detailed elsewhere~\cite{Minier_2001} and where $M_p$ is the total mass of the physical particles in the domain:
\begin{align}
F_{p}^L(t;\mb{y}_p, \mb{z}_c) &= M_p\, p_L(t; \mb{y}_p, \mb{z}_c) ,\\
F_{p}^E(t,\mb{x}; \mb{z}_c) &= F_{p}^L(t;\mb{y}_p=\mb{x}, \mb{z}_c)
                             = \int F_{p}^L(t;\mb{y}_p, \mb{z}_c)
                                               \delta (\mb{y}_p-\mb{x})\, d\mb{y} .
\end{align} 
Since it is useful to distinguish between the particle location and other variables, we have introduced the notation $\mb{Z}=(\mb{x}_p, \mb{Z}_c)$ for the particle state-vector where the particle location is always present and where $\mb{Z}_c$ stands for the complementary part of the chosen state-vector. For example, for the Lagrangian description mentioned above for dispersed two-phase flows, we have $\mb{Z}_c=(\mb{U}_p,\mb{U}_s)$, while for the single-phase flows we would have $\mb{Z}_c=(\mb{U})$. In sample-space, the corresponding variables of the state-vector are noted $\mb{z}=(\mb{y}_p, \mb{z}_c)$. 

For a particle variable written as $H_p(t;\mb{Z}_c)$, its average $\lra{H_p}_m$ (which is a field variable), is defined as
\begin{equation}
\alpha_p(t,\mb{x}) \, \rho_p \lra{H_p}_m(t,\mb{x})= \int H_p(t; \mb{z}_c) 
                           F_p^E(t,\mb{x}; \mb{z}_c)\,d\mb{z}_c.
\end{equation}
where $\rho_p$ is the particle density and $\alpha_p(t,\mb{x})$ is the mean particle volumetric fraction. The fluctuating component is then expressed as $h_p=H_p - \lra{H_p}_m$. The definition of these quantities for the single-phase case are given below, after Eq.~(\ref{eq11}). It is worth emphasizing that $\alpha_p$ is a rigorously-defined probabilistic quantity that represents the average presence of one phase at a given location and should not be confused, in the present framework, with volumetric averages. In a discrete sense, when we handle $N$ stochastic particles, the definitions of Lagrangian and Eulerian MDFs are directly carried out to yield 
\begin{align}
F_{p,N}^L(t\,;\,\mb{y}_p, \mb{z}_c) &=\sum_{i=1}^N m_p^{(i)}\delta (\mb{y}_p-\mb{x}_p^{(i)})
               \,\delta (\mb{z}_c-\mb{Z}_c^{(i)}) \\
F_{p,N}^E(t,\mb{x}\,; \,\mb{z}_c) &= F_{p,N}^L(t\,;\, \mb{y}_p=\mb{x}, \mb{z}_c)
\end{align} 
where $m_p^{(i)}$ is the mass of the particle labeled $(i)$. This shows that in a small volume around location $\mb{x}$ where averages are estimated as the ensemble averages over the $N^p_{\mb{x}}$ particles present in that volume, we get the equivalent of Favre, or mass-weighted, averages
\begin{equation} 
  \lra{H_p}_m(t, \mb{x}) \simeq \lra{H_p}_{m,N} = \frac{\sum_{i=1}^{N^p_{\mb{x}}} m_p^{(i)} 
  H_p(t; \mb{z}_c^{(i)}(t))}{\sum_{i=1}^{N^p_{\mb{x}}} m_p^{(i)}}~,
\end{equation}
which explains the notation $\lra{H_p}_m$ used to indicate mass-weighted averaging for particle variables.

One of the key points is that the Eulerian MDF satisfies the same evolution equation as the Lagrangian MDF. For the general stochastic diffusion model considered in Eq.~(\ref{general_stochastic_diffusion}), this equation is a Fokker-Planck equation
\begin{equation}
\frac{\partial F^E_p}{\partial t} = - \frac{\partial}{\partial z_{k}}\left[ D_{Z,k}\, F^E_p\,\right]  
     + \frac{1}{2}\frac{\partial^2}{\partial z_{k}\partial z_{l}}
           \left[ (B_Z B_Z^T)_{kl}\, F^E_p\,\right].
\end{equation}
For the standard state-vector for dispersed two-phase flows, $\mb{Z}=(\mb{x}_p,\mb{U}_p,\mb{U}_s)$, and for the evolution equations considered in Eqs.~(\ref{general_stochastic_2phi}), the Fokker-Planck equation has the following form where convection appears in closed form:  
\begin{equation}
\label{eq11}
\begin{aligned}
\frac{\partial F^E_p}{\partial t} + V_{p,k}\frac{\partial F^E_p}{\partial y_k} = &
       - \frac{\partial}{\partial V_{p,k}}(D_{p,k}\, F^E_p\,)  \\
     & - \frac{\partial}{\partial V_{s,k}}(D_{s,k}\, F^E_p\,) 
     + \frac{1}{2}\frac{\partial^2}{\partial V_{s,k}\partial V_{s,l}}
           \left((B_s B_s^T)_{kl}\, F^E_p\,\right),
\end{aligned}
\end{equation}
where the variables in sample-space are denoted $(\mb{y},\mb{V}_p,\mb{V}_s)$. It is instructive to consider the first particle-velocity moments obtained from the PDF methodology
\begin{align}
&\frac{\partial \alpha_p \rho_p }{\partial t} 
+ \frac{ \partial \left( \alpha_p\rho_p\lra{U_{p,k}}_m\right)}{\partial x_k} =0~, \\
&\frac{\partial \lra{U_{p,i}}_m}{\partial t}
+ \lra{U_{p,k}}_m \frac{\partial \lra{U_{p,i}}_m}{\partial x_k}
= \frac{1}{\alpha_p\rho_p}
\frac{\partial \left( \alpha_p\rho_p \lra{ u_{p,i}u_{p,k}}_m \right)}{\partial x_k}
+ \left\langle \frac{U_{s,i}-U_{p,i}}{\tau_p} \right\rangle_m + g_i
\label{eq mean particle momentum}
\end{align}
where $\lra{ u_{p,i}u_{p,k}}_m$ is the particle kinetic tensor. In the rhs of the average particle momentum equation, Eq.~\eqref{eq mean particle momentum}, the drag and gravity force terms have been directly introduced instead of the more general notation $\mb{D}_p$. This helps to reveal that in two-phase flows the closure of the drag force term is akin to the closure of the chemical source term in single-phase reactive flows. Indeed, even for inert particles, $\tau_p$ is a complex function of the particle diameter and of the particle and fluid seen velocities (cf. Eq.~\eqref{definition taup}). Thus, for a polydisperse set of particles, the last term on the rhs of Eq.~\eqref{eq mean particle momentum} cannot be closed directly whereas it is handled without approximation with the PDF approach. It is also interesting to consider the second-order equations for the particle kinetic tensor components which correspond to the Reynolds-stress equations in single-phase flows. The exact forms of these equations are detailed elsewhere~\cite{Minier_2001,Peirano_2002} and have an intricate structure which implies further closure issues, such as $\lra{D_{p,i}u_{p,j}}_m$, apart from classical third-order correlation terms. Even when the set of polydisperse particles is separated into classes of monodisperse ones within which $\tau_p$ can be regarded as a (nearly) constant parameter, it is readily seen that, for each class, the equations for $\lra{ u_{p,i}u_{p,k}}_m$ involve the (non-symmetrical) fluid-particle velocities correlation tensor $\lra{u_{s,i}u_{p,j}}_m$ whose equations, in turn, rely on the tensor $\lra{u_{s,i}u_{s,j}}_m$ derived from the velocity of the fluid seen. The important points are that these complete second-order equations (which represents typically 25 coupled partial differential equations (PDE) for each particle class) are actually derived from the stochastic description and involve both the drift and diffusion coefficients of the model used for the velocity of the fluid seen in Eq.~\eqref{general_stochastic_2phi_c}. This indicates that safe guidelines are needed both to assess existing models and to suggest proper formulations.

In the single-phase flow situation, all the relations presented above remain valid. In that case, $\rho_p$ is equal to $\rho$ the fluid density, $M_p$ is replaced by $M_f$ the total fluid mass in the domain, $m_p^{(i)}$ becomes the mass associated to each notional fluid particles $m^{(i)}$, and $\alpha_p(t,\mb{x})=1$ at any point $\mb{x}$. Some comments may be added for the last two points. In dispersed two-phase flows, the particles handled in the numerical simulations represents stochastic particles but since they mimic the `real' particles carried by the fluid turbulent flow, their diameter and their mass can correspond to the real-particle properties (provided no statistical weights are used whereby stochastic particles represent so-called parcels or groups of particles). In that sense, the notion of a stochastic particle may be less surprising at first sight in the two-phase case than in the single-phase one, although the notions are basically the same. In the single-phase case, the mass associated to each stochastic particle appears more as a parameter representing a fraction of the total mass, for example equal mass can be assigned to each particle as often done for incompressible flows. Furthermore,
when the flow occupies the whole domain, the natural limit is to have $\alpha_p(t,\mb{x})=1$. Actually, it will be seen in the next section that this corresponds to the absence of spurious drifts. With these definitions, the above formalism reverts to the developed for single-phase reactive flows~\cite{Pope_1985, Pope_2000}.

In the following, a number of simplifying assumptions are made. First of all, we consider incompressible fluid flows, where the fluid density $\rho$ is constant. In correspondence, particle densities are also taken as constant and, furthermore, we assume that all particles have the same constant density (the case of particle-dependent densities is naturally included since the formalism already accounts for particle-dependent masses). However, it must be noted that, in two-phase flows, variable particle volumetric fractions $\alpha_p(t,\mb{x})$ always induce variable-density effects, which justifies the introduction of MDFs as done in this section. In the present analysis, we limit ourselves to high Reynolds-number flows where viscous effects can be neglected (apart from the non-zero dissipation rate of the fluid turbulent kinetic energy) and, similarly, Brownian effects are not considered for particle motion. In the context of this first analysis, this is done to limit the complexity of the stochastic models and to concentrate on the important convective and dissipative issues. Furthermore, few models have been devised for low-Reynolds particle-laden flows. Note, however, that formulations in the case of low $Re$-number flows were proposed for near-wall fluid flow simulations~\cite{Pop_97,Dreeben_1998,Waclawczyk_2004} and that extensions of the present analysis can be made. For the two-phase flow situations, two-way coupling is not explicitly accounted for, although related aspects are discussed when the form of the equations for the fluctuating velocity components are considered (see section~\ref{Fluctuating or instantaneous fluid velocity seen}) while the effects of particle-particle collisions are left out.

\section{Criteria for the analysis of stochastic models}
\label{Criteria for the analysis of stochastic models}

We consider Lagrangian stochastic models for fluid particles, such as those
characterized by Eqs.~(\ref{general_stochastic_1phi}). The criteria put forward here 
arise from the following question: in what sense can we regard these models as acceptable 
probabilistic descriptions of turbulent flows?

\subsection{Choice of criteria for single-phase flow models}
\label{Choice of criteria for single-phase flow models}

This issue was addressed by Pope~\cite{Pope_1985,Pope_1987} who showed that, for such 
models to be true PDF descriptions (that is for the corresponding pdf to be normalized to one), it is
necessary and sufficient that the mean-pressure gradient be properly introduced and that the mean
continuity condition be respected. This is tantamount to saying that Lagrangian stochastic models
are free from spurious drifts when the mean pressure gradient is properly taken into account\cite{Pope_1987,McInnes_1992,Minier_2001}. The correspondence between Lagrangian 
stochastic models and Reynolds-stress closures was then brought forth~\cite{Haworth_1986,Pope_1994,Pope_1994a}.
Drawing on these relations, we propose to regard a Lagrangian stochastic formulation as an acceptable 
model if three criteria are met:
\begin{enumerate}
\item[(F-1)] the model should be capable of being written in correct Cartesian tensor notation and
should satisfy the relevant invariance principles (such as Galilean and extended Galilean invariance);
\item[(F-2)] the stochastic model should be such that the mean-continuity equation is respected and,
thus, the model must be free of spurious drifts;
\item[(F-3)] all convective terms in the mean Navier-Stokes \textbf{and} in the second-order equations
must be exactly reproduced (as well as dissipative terms).
\end{enumerate}
The first criterion amounts to asking that the basic transformation and invariance properties of the 
Navier-Stokes equations be reflected in the model~\cite{Pope_2000}. The last two criteria mean that 
the first two moments of the velocity fields extracted from the simulated pdf are such that, 
using Reynolds decomposition written as $U_i=\lra{U_i}+u_i$, we have the following structure 
\begin{subequations}
\label{structure_equations_Rij}
\begin{align}
& \frac{\partial \lra{U_k}}{\partial x_k}=0 \\
& \frac{\partial \lra{U_i}}{\partial t} + \lra{U_k}\frac{\partial \lra{U_i}}{\partial x_k}
+ \frac{\partial \lra{u_i\, u_k}}{\partial x_k}=- \frac{1}{\rho}\frac{\partial \lra{P}}{\partial x_i} \\
&\frac{\partial \lra{u_iu_j}}{\partial t} + \lra{U_k}
\frac{\partial \lra{u_iu_j}}{\partial x_k} 
+\frac{\partial \lra{u_iu_ju_k}}{\partial x_k} +
\lra{u_iu_k}\frac{\partial \lra{U_j}}{\partial x_k} 
+ \lra{u_ju_k}\frac{\partial \lra{U_i}}{\partial x_k} = \Psi_{ij}  \label{structure_equations_Rij_c}
\end{align}
\end{subequations}
The first equation is the mass-continuity equation when the fluid density is constant. The second
equation corresponds to the mean Navier-Stokes equation (i.e., the Reynolds equation) and it is seen 
that the mean viscous term, $\nu \Delta U_i$, has been neglected, for the sake of simplicity in 
the present analysis, by assuming sufficiently high Reynolds-number flows. This indicates that we 
are not considering near-wall regions. In Eq.~(\ref{structure_equations_Rij_c}), the rhs (right-hand 
side) $\Psi_{ij}$ is usually decomposed as $\Psi_{ij}=\Phi_{ij} - \epsilon_{ij}$ where $\Phi_{ij}$ 
and $\epsilon_{ij}$ stand for models expressing 
the correlation between the pressure-gradient and fluctuating velocities and the dissipation sink term, 
respectively. These models are subject to classical requirements, for instance that 
$\Phi_{ii}=\partial \lra{p\, u_i}/\partial x_i$ or, when the divergence of the pressure-velocity 
correlation is neglected, to $\Phi_{ii}=0$ and usual model simplifications, such as isotropic 
dissipative terms, which means that we can write $\epsilon_{ij}=-2/3\lra{\epsilon}\delta_{ij}$ with
$\lra{\epsilon}$ the mean turbulent kinetic energy dissipation. In the following, we only require that
half the trace of the rhs has the form $1/2\,\Psi_{ii}=\nabla.I - \epsilon$ with $\epsilon \geq 0$
and where $I$ stands for an expression that represents the pressure-fluctuation correlation. As a 
practical matter, pressure transport is often ignored, and so, for simplicity, we take the requirement 
to be $1/2\,\Psi_{ii}=- \epsilon$. 

It is worth noting the difference with the previous works that established the correspondence between
Lagrangian stochastic models and Reynolds-stress equations~\cite{Haworth_1986,Pope_1994a}. In these
works, the aim was mostly to obtain detailed forms of Reynolds-stress models from the Lagrangian
stochastic approach and, in that sense, the emphasis was put on developing detailed closure proposals 
for $\Phi_{ij}$ and $\epsilon_{ij}$. In the present work, the emphasis is put on a different point.
Indeed, by the criterion (F-3) listed above, it is meant that all the terms appearing on the lhs 
(left-hand side) of Eqs.~(\ref{structure_equations_Rij}) should be exactly reproduced.
This is a rather natural requirement to set forth: indeed, all the terms gathered on the lhs
of Eqs.~(\ref{structure_equations_Rij}) are convective terms which arise from the transport
term in the Navier-Stokes equations. Since Lagrangian approaches are precisely attractive as
they treat convective transport without approximation, it is therefore natural to expect that the
structure of the transport equations for the first two moments of the velocity field will be
exactly reproduced. In other words, failure to do so by a Lagrangian model means that transport 
is poorly described and this is regarded as an unacceptable shortcoming. The criteria 
proposed here are basic physical requirements but represent a step forward with regard to former 
proposals~\cite{Thomson_1987}. 

\subsection{Choice of criteria for two-phase flow models}
\label{Choice of criteria for two-phase flow models}

We now consider Lagrangian stochastic models for the velocity of the fluid seen by discrete particles, as expressed by Eqs.~(\ref{general_stochastic_2phi}), and the issue is to come up with relevant criteria to assess whether these modeling proposals can be retained as acceptable descriptions. However, selecting such a list of criteria for two-phase flows is not straightforward. Indeed, the analysis of single-phase stochastic models can rely on the sound basis provided by given Reynolds-stress models. This is not so in the two-phase situation as it is actually the stochastic model which is used to derive corresponding mean-field or so-called continuum descriptions of two-phase flows~\cite{Simonin_1993,Simonin_2000,Minier_2001}. This lack of a sound reference continuum theory puts a stronger emphasis on the need to have reliable criteria to assess stochastic models for disperse two-phase flows. In spite of this limitation, it is still possible to use the single-phase flow case as a sound basis. More specifically, to analyse stochastic models used for the simulation of the fluid velocity seen $\mb{U}_s$, we propose the following list of criteria:
\begin{enumerate}
\item[(P-1)] the model should be capable of being written in correct Cartesian tensor notation and should satisfy the relevant invariance principles (such as Galilean and extended Galilean invariance); 
\item[(P-2)] the stochastic model must be complete in the sense that the expressions for both the drift vector and the diffusion matrix in Eq.~(\ref{general_stochastic_2phi_c}) must be explicitly given. Furthermore, the stochastic model must have predictive capacities in the sense that it should be applicable to general non-homogeneous situations where fluid or particle statistics are not necessarily known in advance;
\item[(P-3)] the stochastic model used for the fluid velocity seen should revert to an acceptable stochastic model for single-phase flow (thus respecting the criteria (F-1) to (F-3)) when particle inertia goes to zero, that is in the particle-tracer limit;
\item[(P-4)] the complete stochastic model should be such that the predicted mean kinetic energy of the fluid seen respects the same statistical evolution laws as the mean fluid kinetic energy in dilute flows, both in stationary and decaying homogeneous turbulence;
\item[(P-5)] the resulting model for the so-called drift velocity, say $\mb{V}_d$ (which represents the difference between the mean velocity of the fluid seen and the mean fluid velocity at the same particle location, or $V_{d,i}=\lra{U_{s,i}}_m - \lra{U_{i}}$), should be consistent with known scalar dynamics and related modeling;
\item[(P-6)] the model for the velocity of the fluid seen should be consistent with the Equilibrium Eulerian approach formulated in the limit of small particle Stokes numbers~\cite{balachandar_2010,Ferry_2001}. More precisely, the model should be such that the mean conditional increments of the drift vector $\mb{D}_s$ yield the correct behavior for discrete particle velocities for small $\tau_p$. 
\end{enumerate}

The criterion (P-1) is the same criteria as for the fluid case (see (F-1)) carried to the two-phase flow situation. Indeed, since the velocity of the fluid seen is extracted from the velocity field of the fluid carrier phase, the same invariance principles hold. At first sight, the criterion (P-2) seems obvious as it 
simply express what a `model' stands for. Unfortunately, it will be indicated in 
section~\ref{Stochastic models for two-phase flows} that there can be some confusion as to which proposals 
can be truly referred to as a stochastic model or, in other words, as a true PDF description of two-phase 
flows. Thus, this criterion will be shown to be useful to clarify the status of some proposals. To be self-consistent, it is useful to clarify the definition retained here for a predictive model: a model is said to be predictive if it is able to provide information on the future state of a system (which means on the statistics of the variables retained to describe the two-phase flow), given all necessary initial and boundary conditions.

In the list above, the criterion (P-3) is an important one: it translates what is meant when it is said 
that satisfactory models for two-phase flows should be extensions of satisfactory ones for single-phase flows. 
From a physical point of view, this is a sound requirement to make: indeed, small-inertia particles tend to follow closely the surrounding fluid and it appears consistent to expect that statistics derived from the 
fluid and from the discrete particle set become identical and both acceptable descriptions. Furthermore, 
from a numerical point of view when so-called Eulerian/Lagrangian hybrid simulations are used, this will 
help to ensure that consistent predictions are obtained~\cite{Chibbaro_2011}. 

The last criteria involve specific aspects of two-phase flows. The criterion (P-4) states for a statistically homogeneous decaying turbulence, we expect that, for particles which are homogeneously distributed (still in a statistical sense), the fluid seen is such that 
\begin{equation}
\label{eq: criterion (P-4)}
\frac{1}{2}\frac{d\lra{\mb{u}_s \cdot \mb{u}_s}_m}{dt}=-\lra{\epsilon}~.
\end{equation}
This expresses that the mean fluid kinetic energy seen by particles is decaying at the same rate as for the fluid. In the context of the present study, this criterion is meant as a functional requirement and it will be seen in section~\ref{Stochastic models for two-phase flows} that it is particularly useful to lead to correct closure expressions for the diffusion term in Eqs.~(\ref{general_stochastic_2phi}). From a more physical point of view, it can be argued that, even in a homogeneous turbulent fluid flow, particles are not necessarily homogeneously distributed in the whole flow domain but tend to concentrate in some flow regions (this is the particle preferential-concentration effect~\cite{Marchioli_2002, Gualtieri_2009, Toschi_2009, balachandar_2010}). As we are addressing model expressed in the framework of Reynolds-averaging approaches, the issue of whether this effect is specifically well-reproduced in present formulations is not directly addressed. However, it is worth pointing out that particle preferential concentration effects can still be accounted for in the criterion (P-4). This can be achieved by considering that the same requirement be valid, provided that the mean value of the fluid dissipation rate is replaced by the local value, say $\lra{\epsilon_s}$, `seen' by particles in the specific flow regions where particle tend to be located. In that sense, the form of the criterion (P-4) is general enough and is relevant for the statistical models considered in the present analysis. 

The criterion (P-5) refers to the situation when particle inertia can be neglected (thus when the particle relaxation time scale becomes negligible with respect to the fluid turbulence characteristic timescales introduced in the next section) but keeping a non-zero volumetric fraction $\alpha_p$. In that case, it is expected that particles behave as a passive (but non-vanishing) scalar and the meaning of the criterion (P-5) is to require that the resulting model be consistent with classical scalar modeling which will be recalled in section~\ref{scalar flux}. This criterion complements the issue of no-spurious drift effects, or well-mixed condition, (which will be shown to form one of the basis of the criteria (P-3)) and corresponds to the physical situation where particle-tracers are injected so as to create a non-homogeneous distribution. It will be demonstrated in section~\ref{Stochastic models for two-phase flows} that this criterion is helpful to point to acceptable forms of the return-to-equilibrium term which typically enters the closure of the drift vector in Eqs.~(\ref{general_stochastic_2phi}).

The criterion (P-6) is directly related to the Equilibrium Eulerian Model (EEM) proposed for small Stokes numbers~\cite{balachandar_2010,Ferry_2001} (to be defined below), which is also relevant in the present Lagrangian point of view. The formulation expressed by (P-6) can be worked out as follows. We consider the case of a constant particle relaxation timescale $\tau_p$. Then, starting from an initial condition at $t=0$, the particle momentum equation, Eq.~\eqref{exact particle eqns_Up}, can be integrated to give
\begin{align}
\mb{U}_p(t)&=\mb{U}_p(0)e^{-t/\tau_p} + \mb{g}\tau_p\left( 1 - e^{-t/\tau_p} \right)
+\frac{1}{\tau_p}e^{-t/\tau_p}\int_0^t e^{t'/\tau_p}\mb{U}_s(t')\, dt' \\
 &= \mb{U}_s(t) + \left( \mb{U}_p(0) - \mb{U}_s(0) \right)e^{-t/\tau_p}
 + \mb{g}\tau_p\left( 1 - e^{-t/\tau_p} \right) 
 - e^{-t/\tau_p}\int_0^t e^{t'/\tau_p}\frac{d\mb{U}_s(t')}{dt'}\, dt'
\end{align}
where the second line follows from a simple integration by parts. The EEM is obtained by considering small $\tau_p$ so that the memory of initial conditions can be neglected and the derivative in the integral can be approximated by its value at time $t$ (in other words, $d\mb{U}_s(t')/dt' \simeq d\mb{U}_s(t)/dt$ since the integrand is a highly-peaked function when $\tau_p \ll 1$). This yields that
\begin{equation}
\label{eq EEM for Up}
\mb{U}_p(t) \simeq \mb{U}_s(t) + \mb{W}_g - \tau_p \left( \frac{d\mb{U}_s(t)}{dt}\right)
\end{equation}
where $\mb{W}_g=\tau_p\mb{g}$ is the particle settling (or terminal) velocity. It is important to realize that the derivative $d\mb{U}_s(t)/dt$ is the time derivative of the fluid seen along discrete particle trajectories and not the fluid derivative along the fluid particle trajectory located at the same position as the discrete one at time $t$ written as $D\mb{U}/Dt$ (which is the Eulerian notation for the fluid particle acceleration $D\mb{U}/Dt=\partial \mb{U}/\partial t + \mb{U}\cdot \nabla\mb{U}$). By comparing with the classical formulation of the EEM (see Eq.~(1) in~\citet{balachandar_2010}), it is seen that this is equivalent to stating that, for small particle relaxation timescale $\tau_p$, we have 
\begin{equation}
\label{EEM for Us}
\frac{d\mb{U}_s(t)}{dt} \simeq \frac{D\mb{U}}{Dt} + \mb{W}_g\cdot \nabla \mb{U}~.
\end{equation}
While the EEM is mostly expressed as a relation for discrete particle velocities, it is actually a model for the underlying velocity of the fluid seen in the limit of small $\tau_p$. At this stage, three remarks can be made to connect this relation to a workable criterion in our context. First, as we do not consider the added-mass force in the particle momentum equation, the parameter $\beta$ appearing in the EEM~\cite{balachandar_2010,Ferry_2001} is here zero. Second, the expression of the EEM in Eq.~\eqref{eq EEM for Up} is usually written in terms of non-dimensional quantities based on Kolmogorov scales (i.e. the Kolmogorov velocity and time scales, $u_{\eta}$ and $\tau_{\eta}$ respectively), which gives 
\begin{equation}
\label{eq EEM for Up normalized}
\mb{U}^{(n)}_p(t) \simeq \mb{U}^{(n)}_s(t) + \mb{W}^{(n)}_g 
- St_{\eta} \left( \frac{d\mb{U}^{(n)}_s(t)}{dt^{(n)}}\right)
\end{equation}
with $\mb{U}^{(n)}_p=\mb{U}_p/u_{\eta}$ (the same scaling is used for $\mb{U}^{(n)}_s$ and $\mb{W}^{(n)}_g$), $t^{(n)}=t/\tau_{\eta}$, and where $St_{\eta}=\tau_p/\tau_{\eta}$ is the Kolmogorov-based Stokes number, which is a measure of particle inertia. The choice of the Kolmogorov scales is indeed relevant in DNS studies~\cite{balachandar_2010,Ferry_2001}. However, this scaling is not appropriate in the present context where we are considering stochastic models for high-Reynolds number turbulent flows in which a part of fluid particle acceleration is replaced by a white-noise term. For our purpose, it is best to introduce the timescale $T_L$ of fluid velocities (the integral timescale), which will be defined in section~\ref{Analysis of different formulations}, and to refer to the Stokes number defined by $St=\tau_p/T_L$ instead of $St_{\eta}$. Then, the loose statement of `small particle relaxation timescale' can now be properly expressed as meaning that $St \ll 1$ or $\tau_p \ll T_L$. Third, the above formulas have been obtained by considering that the derivative of the velocity of the fluid seen is a sufficiently smooth function. This is not so when stochastic diffusion processes are used and when $d\mb{U}_s(t)/dt$ is white-noise. Using the general form of the stochastic model given in Eq.~\eqref{general_stochastic_2phi_c}, the correct expression of the discrete particle velocity in the limit of small $St$ number can be properly expressed as
\begin{equation}
\mb{U}_p(t) \simeq \mb{U}_s(t) + \mb{W}_g  - e^{-t/\tau_p}\int_0^t e^{t'/\tau_p}\mb{D}_s(t')\, dt' 
 - e^{-t/\tau_p}\int_0^t e^{t'/\tau_p}\mb{B}_s(t')\, d\mb{W}(t')
\end{equation}
from which, using the same approximation, we have 
\begin{equation}
\mb{U}_p(t) \simeq \mb{U}_s(t) + \mb{W}_g  - \tau_p\mb{D}_s(t)
 - e^{-t/\tau_p}\int_0^t e^{t'/\tau_p}\mb{B}_s(t')\, d\mb{W}(t')~.
\end{equation}
If the coefficients $B_{s,kl}$ of the diffusion matrix can be frozen at their value at time $t$, then the random term in this equation can be simulated (in a weak sense) as a sum of Gaussian random variables of zero mean and variance equal to $B_{s,kl}^2 \tau_p/2 \left[ 1 - e^{-2t/\tau_p} \right]$. Yet, in order to complement the criterion (P-4) which corresponds to the diffusion coefficient, we concentrate here on the drift vector. By comparing with Eq.~\eqref{EEM for Us}, we obtain the resulting form of the criterion by taking the conditional average (conditioned on a given value of $\mb{Z}=(\mb{x}_p,\mb{U}_p,\mb{U}_s)$), which gives
\begin{equation}
\lra{ \mb{D}_s\,|\,\mb{Z}} \simeq \lra{ \frac{D\mb{U}}{Dt} \,| \,\mb{Z}} + \mb{W}_g\cdot \nabla \lra{\mb{U}}
\end{equation}
where we have assumed that, for high Reynolds-number turbulent flows, fluid velocity gradients are not strongly correlated with the (large-scale) fluid velocities so that their mean conditional values can be taken as being equal to the average ones. By using the Lagrangian formulation of $D\mb{U}/Dt$ expressed by Eqs.~\eqref{general_stochastic_1phi}, we get therefore that, for $St \ll 1$, the drift vector for the velocity of the fluid seen should be such that we have
\begin{equation}
\label{from of criterion P-6}
\lra{ \mb{D}_s\,|\,\mb{Z}} \simeq \lra{ \mb{D} \,| \,\mb{Z}} + \mb{W}_g\cdot \nabla \lra{\mb{U}}
\end{equation}
where $\mb{D}$ is the drift vector used for the increments of fluid particle velocities. This is the basic form retained for the criterion (P-6). A weaker form can also be considered by taking the unconditional average, in which case we have that, for small $St$ number,
\begin{equation}
\lra{ \mb{D}_s} \simeq \lra{\mb{D}} + \mb{W}_g\cdot \nabla \lra{\mb{U}}~.
\end{equation}
The interest of this weaker form, which is referred to as (P-6bis) from now on, is that, for high Reynolds-number turbulent flows, the mean value of the drift for fluid particle velocities is an exact result and is equal to the fluid mean pressure-gradient (as recalled in section~\ref{Analysis of different formulations}). Thus, the criterion (P-6bis) implies that, for small $St$ number, we should have that
\begin{equation}
\lra{ \mb{D}_s} \simeq -\frac{1}{\rho}\nabla \lra{P} + \mb{W}_g\cdot \nabla \lra{\mb{U}}~.
\end{equation}
Although, Eq.~\eqref{from of criterion P-6} truly embodies the criterion (P-6) in the list of requirements, both forms will be considered and discussed in section~\ref{Stochastic models for two-phase flows}.

\section{Analysis of stochastic models for single-phase flows}
\label{Stochastic models for single-phase flows}

In this section, different formulations in terms of instantaneous, fluctuating or normalized fluctuating velocities are considered to demonstrate the interest of the criteria selected in section~\ref{Choice of criteria for single-phase flow models}.

\subsection{Different formulations of single-phase flow stochastic models}
\label{Analysis of different formulations}

As Lagrangian stochastic models for fluid particle velocities attempt at reproducing some key (one-point) statistical properties of turbulent flows, the starting point is naturally provided by the Navier-Stokes equations. Following the classical notations that use a superscript $^{+}$ to indicate that we are dealing with the exact equation for a fluid particle~\cite{Pope_1985} and using Reynolds decomposition into mean and fluctuating  parts (leaving out the mean viscous term $\nu \Delta \lra{U_i}$ at high Reynolds-number numbers), the Navier-Stokes equations write
\begin{subequations}
\label{Navier-Stokes_decomp}
\begin{align}
\frac{dx_i^+}{dt} &=U_i^+ \\
\frac{dU_i^+}{dt} &=  - \left( \frac{1}{\rho}\frac{\partial \lra{P}}{\partial x_i} \right)^{+}
       - \underbrace{\left(\frac{1}{\rho}\frac{\partial p}{\partial x_i} \right)^{+}
                     + \left( \nu \Delta u_i \right)^+}_{\text{to model}} 
\end{align}
\end{subequations}
where the fluctuating pressure-gradient and viscous terms are to be modeled.

A reference stochastic model is the GLM (Generalised Langevin Model)~\cite{Haworth_1986, Pope_1994, Pope_2000} which represents fluid particle velocities by a stochastic diffusion process whose general form is
\begin{subequations}
\label{model_U_GLM}
\begin{align}
& dx_i = U_i\, dt \\
& dU_i = -\frac{1}{\rho}\frac{\partial \langle P \rangle}{\partial x_i}\, dt +
D_i\, dt + \sqrt{C_0\langle \epsilon \rangle}dW_i .
\end{align}
\end{subequations}
The drift coefficient $D_i$ is usually a function of the difference between instantaneous
and mean velocities at the particle location and is modeled as  
\begin{equation}
\label{drift_model_U_1}
D_i = G_{ij} \left( U_j- \lra{U_j} \right)
    = - \left( \frac{1}{2} + \frac{3}{4}C_0 \right)
\frac{\lra{\epsilon}}{k} (U_i- \langle U_i \rangle) + 
G_{ij}^a \left( U_j- \lra{U_j} \right)
\end{equation}
where the matrix $G_{ij}^a$ represents anisotropic effects and is subject to the condition
that $Tr(G^a R)=0$, with $R_{ij}=\lra{u_i\, u_j}$ the Reynolds-stress tensor. The drift vector
can also be expressed so as to bring forward the timescale $T_L$, which is a measure of the 
integral timescale of large-scale velocity fluctuations, 
\begin{equation}
\label{drift_model_U_2}
D_i= - \frac{ U_i- \lra{U_i}}{T_L} + G_{ij}^a \left( U_j- \lra{U_j} \right)\quad
\text{with} \quad T_L=\frac{1}{ \left(\dfrac{1}{2} + \dfrac{3}{4}C_0 \right)}\frac{k}{\lra{\epsilon}} .
\end{equation}
In these equations, the mean terms are to be understood as being the values of the corresponding 
mean fields at the particle location, for instance $\lra{U_i}=\lra{U_i}(t; \mb{x})$ and, in 
the context of the present study an important point is that the GLM is formulated in terms 
of instantaneous fluid velocities.

So far, the GLM has been used as a reference but the present discussion is not limited to this 
model. Indeed, the GLM is one example of models written in terms of particle instantaneous 
velocities and which can be formulated as
\begin{subequations}
\label{general_form_inst}
\begin{align}
& dx_i = U_i\, dt \\
& dU_i = -\frac{1}{\rho}\frac{\partial \langle P \rangle}{\partial x_i}\, dt +
dM_i (t;\mb{x},\mb{U}) \label{general_form_inst_U}
\end{align}
\end{subequations}
where $\mb{M}$ stands for a stochastic model expressed in terms of the state-vector
$\mb{Z}=(\mb{x},\mb{U})$ based on instantaneous velocities and $dM_i$ its increment over 
a small time interval $dt$. 
The general model $\mb{M}$, whose precise form is irrelevant for the present concern, 
is a model accounting for the fluctuating pressure-gradient and viscous term, as expressed in
Eq.~(\ref{Navier-Stokes_decomp}), and is simply assumed to satisfy the requirements that 
$\lra{dM_i}(t,\mb{x})=0$ and $\lra{u_i\circ dM_i}/dt=-2\,\epsilon$. These two constraints 
translate the fact that the mean `force' $dM_i$ is zero while the mean value of the work 
performed by this force is the mean dissipation sink term. 
In the typical case where $dM_i$ involves a stochastic process, the last expression has been 
written using Statonovich stochastic calculus for the sake of simplicity. In most modeling 
proposals, $dM_i$ is represented by one diffusion stochastic process which can also depend on 
statistics derived from the set of particles as well as on external fields. However, other
models with different structures can be considered~\cite{Pope_2011,Meneveau_2011}: for instance, 
$dM_i$ can represent stochastic models which rely on a two-level stochastic description where 
a random succession of elementary diffusion processes are governed by a parent Poisson 
process~\cite{Guingo_2008}. From the standard methodology recalled 
in section~\ref{From Lagrangian stochastic models to mean-field equations}, it is clear that 
all models belonging to the class represented by Eqs.~(\ref{general_form_inst}) meet the 
criteria listed in section~\ref{Choice of criteria for single-phase flow models} and, therefore,
can be regarded as satisfactory stochastic models for single-phase flows.

In the passage from homogeneous turbulence to general non-homogeneous flows, other modeling roads 
have been followed. In particular, several attempts~\cite{Berlemont_1990,Gouesbet_1999,Matida_2004} 
have been made at formulating a general model by stating that the model developed in homogeneous 
situations corresponds, in the inhomogeneous case, to the model for the fluctuating velocity, $u_i$, 
and that the full model for the instantaneous velocities is simply obtained by writing 
$U_i=\lra{U_i}+u_i$. Other 
attempts~\cite{Dehbi_2008,Dehbi_2010,Mito_2002,Iliopoulos_2003,Iliopoulos_2004}
have been made by considering that the model developed in homogeneous flows retains a similar form 
in inhomogeneous cases when it is expressed for the normalized fluctuating velocities, $u_i/\sigma_{(i)}$ 
where $\sigma_{(i)}$ is the standard deviation for the corresponding velocity component, and saying 
again that instantaneous velocities are retrieved by adding the mean value at particle location
(suffixes in brackets are excluded from the summation convention).  
Therefore, these formulations differ by the choice of the variable
on which they act (instantaneous or fluctuating velocities) and by the structure of the model.
The differences induced by these choices are now discussed. 

\subsubsection{Fluctuating velocity}
\label{Fluctuating velocity versus instantaneous velocity}

To work out the relations between formulations in terms of the instantaneous
velocities and fluctuating parts, it is convenient to start from the general form introduced
in Eqs.~(\ref{general_form_inst}). The Lagrangian derivative is equivalent to the material 
derivative and for a quantity $\phi(t)$ sampled from a field $\Psi(t,\mb{x})$ along a particle 
trajectory, $\phi(t)=\Psi(t,\mb{x}(t))$, we have
\begin{equation}
\frac{d\phi(t)}{dt}= \frac{\partial \Psi}{\partial t} + U_k\frac{\partial \Psi}{\partial x_k} .
\end{equation}
Then, by writing 
\begin{equation}
u_i=U_i - \lra{U_i}~,
\label{eq:fluct}
\end{equation}
 the evolution equation for the fluctuating velocity along 
the same particle trajectory can be obtained from Eq.~(\ref{general_form_inst_U})
\begin{subequations}
\begin{align}
\frac{du_i}{dt}&= \frac{dU_i}{dt} - \frac{d\lra{U_i}}{dt} \\
\frac{du_i}{dt}&= \frac{dU_i}{dt} - \left( \frac{\partial \lra{U_i}}{\partial t} 
                                  + \lra{U_k}\frac{\partial \lra{U_i}}{\partial x_k} 
                                  + u_k\frac{\partial \lra{U_i}}{\partial x_k} \right)\\
\frac{du_i}{dt}&= \frac{dU_i}{dt} +  \frac{1}{\rho}\frac{\partial \lra{P}}{\partial x_i} 
+ \frac{\partial \lra{u_i\, u_k}}{\partial x_k} - u_k\frac{\partial \lra{U_i}}{\partial x_k}
\end{align}
\end{subequations}
The final form is obtained by using the increments of the instantaneous fluid velocities in Eqs.~(\ref{general_form_inst}) and this shows that any stochastic model formulated 
in terms of $\mb{U}$ as 
\begin{subequations}
\label{model_U}
\begin{align}
dx_{i}&= U_i\, dt \\
dU_{i}&= \underbrace{-\frac{1}{\rho}\frac{\partial \lra{P}}{\partial x_i}\, dt}_{
no \, spurious \, drift} 
+ \underbrace{dM_i(t;\mb{x},\mb{U})}_{model} \label{model_U_2}
\end{align}
\end{subequations}
is equivalent to a stochastic model in terms of $\mb{u}$ formulated as~\cite{McInnes_1992,Minier_1999}:
\begin{subequations}
\label{model_u}
\begin{align}
dx_{i}&= \left( \lra{U_i}+ u_{i}\right)\, dt \\
du_{i}&= \underbrace{\frac{\partial \lra{u_iu_k}}{\partial x_k}\, dt}_{
(a)\,no \, spurious \, drift} - \underbrace{u_k\frac{\partial \lra{U_{i}}}{\partial x_k}\, dt}_{
(b)\, production \, term} + \underbrace{dM_i(t;\mb{x},\mb{u})}_{model} . \label{model_u_2}
\end{align}
\end{subequations}

In the formulation in Eq.~(\ref{model_U_2}), the first term on the rhs (right-hand side) is the mean 
pressure-gradient and, when this mean pressure is such that the mean velocity field satisfies the 
divergence-free condition for incompressible flows, this ensures that an initially uniform (fluid) particle 
concentration remains uniform~\cite{Pope_1985,Pope_1987}. In the literature devoted to Lagrangian 
models, failure to maintain such a uniform concentration has been referred to as the ``spurious 
drift effect" or even ``the well-mixed condition problem" in some works~\cite{Thomson_1987}. 
In models formulated in terms of instantaneous velocities, the presence of the mean pressure-gradient 
term is evident and the issue of spurious drifts is thus trivially avoided. However, the situation is 
somewhat more involved in the formulation in terms of fluctuating velocities, Eqs.~(\ref{model_u}). 
The first term on the rhs of Eq.~(\ref{model_u_2}), which involves the spatial derivatives of 
the Reynolds-stress tensor, must be present if spurious drifts are to be avoided. This was not 
the case in the first modeling attempts~\cite{Berlemont_1990,Gouesbet_1999} and, although the 
situation was analyzed from a theoretical point of view~\cite{Pope_1987} and demonstrated from 
a numerical point of view~\cite{McInnes_1992}, this fact remains sometimes missed in dispersed two-phase
flow applications~\cite{Tian_2007,Parker_2008,Sasic_2010}. This has led to a blurred vision of the 
rather simple issue of the spurious drift effect and to the notion that flawed formulations should 
be saved by the addition of `corrected mean terms', yielding so-called drift-corrected 
models~\cite{Taniere_2014}. These analyses are limited to checking that the correct 
form of the Reynolds equation is satisfied whereas, in the context of the present study, the 
consistency of model proposals with the full structure of Reynolds-stress equations has been 
explicitly raised in the criterion (F-3). This implies that all the terms appearing on the rhs 
of Eq.~(\ref{model_u_2}) must be addressed.
 
Indeed, even if the gradient of the Reynolds stress tensor is properly introduced, the second term 
on the rhs of Eq.~(\ref{model_u_2}) is just as compulsory. This is seen by deriving 
the corresponding transport equations for the Reynolds stress components 
$\lra{u_i \, u_j}$ which reads:
\begin{multline} \label{Rij-model_u}
\underbrace{\frac{\partial \langle u_iu_j \rangle}{\partial t}
+ \langle U_k \rangle
\frac{\partial \langle u_iu_j \rangle}{\partial x_k} 
+\frac{\partial \langle u_iu_ju_k \rangle}{\partial x_k}}_{
\lra{d(u_iu_j)}} = \underbrace{
- \langle u_iu_k \rangle\frac{\partial \langle U_j \rangle}
{\partial x_k} - \langle u_ju_k \rangle\frac{\partial \langle U_i\rangle}
{\partial x_k}}_{correct\, production \, term}  \\
+ \frac{1}{dt}\lra{u_i \circ dM_j} + \frac{1}{dt}\lra{u_j \circ dM_i} 
\end{multline} 
With the two constraints satisfied by $dM_i$ (see section~\ref{Analysis of different formulations}),
it is seen that the second term, labelled (b) in Eq.~(\ref{model_u_2}), is essential to obtain 
the correct production term in the Reynolds-stress equations. Failure to account for this 
fluctuating term means that the important production term in the Reynolds-stress equations is 
either missing or badly calculated. Such a formulation would then be inconsistent with the correct 
form of the $R_{ij}-\epsilon$ equations and, following the set of requirements given in 
section~\ref{Choice of criteria for single-phase flow models}, would not be acceptable.  

Finally, once all terms are correctly handled, it is useful to compare the corresponding effort 
which is required in Eqs.~(\ref{model_U}) and Eqs.~(\ref{model_u}). A stochastic model for 
$\mb{U}$ requires 3 gradients (for $\nabla \lra{P}$) while the same stochastic model, written 
for $\mb{u}$, requires 27 gradients (for $\nabla \lra{\mb{U}}$ and $\nabla \lra{u_iu_j}$). 
This represents a considerable amount of additional complexity and computational effort in 
practical calculations. For all the reasons put forward above, it appears therefore that 
formulations made in terms of the instantaneous velocity are both the easiest and the safest 
road for the construction of practical models.

\subsubsection{Normalized velocity}
\label{Normalised velocity}

Other attempts~\cite{Wilson_1981} at going from the homogeneous situations to non-homogeneous 
ones were made through so-called normalized Langevin models. Recent versions of the normalised 
Langevin approach can be written as~\cite{Bocksell_2006,Dehbi_2008,Dehbi_2010}:
\begin{equation}
\label{normalised_Langevin}
d\left( \frac{u_i}{\sigma_{(i)}}\right) = - \frac{u_i}{\tau_L\, \sigma_{(i)}}\, dt +
\sqrt{\frac{2}{\tau_L}}\, dW_i 
+ \frac{\partial}{\partial x_k}\left( \frac{\lra{u_iu_k}}{\sigma_{(i)}}\right) \, dt
\end{equation}
In the following, bracketed indexes are used for the standard deviation 
$\sigma_{(i)}$ of the fluctuating velocity component $u_i$ to indicate that such indexes are 
excluded from the summation convention. 

Before going into the analysis of this formulation in general inhomogeneous situations, we first 
consider homogeneous turbulence where the model becomes
\begin{equation}
\label{normalised_Langevin_origin}
d\left( \frac{u_i}{\sigma_{(i)}}\right) = - \frac{u_i}{\tau_L\, \sigma_{(i)}}\, dt +
\sqrt{\frac{2}{\tau_L}}\, dW_i 
\end{equation}
We can derive the equations for the 
fluctuating velocities $u_i=\sigma_{(i)} \left( \dfrac{u_i}{\sigma_{(i)}}\right)$ by writing
\begin{subequations}
\label{normalised_variable_change}
\begin{align}
du_i &=  \frac{u_i}{\sigma_{(i)}}(d\sigma_{(i)}) + \sigma_{(i)}d\left( \frac{u_i}{\sigma_{(i)}}\right) \\
     &=  \frac{u_i}{\sigma_{(i)}}\left( \frac{\partial \sigma_{(i)}}{\partial t} + 
                     U_k\frac{\partial \sigma_{(i)}}{\partial x_k} \right)\, dt
                     + \sigma_{(i)} d\left( \frac{u_i}{\sigma_{(i)}}\right)
\end{align}
\end{subequations}
Since $\partial \sigma_{(i)}/\partial x_k=0$ for homogeneous flows, this first proposal of the normalized 
Langevin model corresponds to the following equation for the fluctuating velocities:
\begin{equation}
\label{normalised_Langevin_1}
du_i= \left( \frac{1}{\sigma_{(i)}}\frac{\partial \sigma_{(i)}}{\partial t} 
           - \frac{1}{\tau_L} \right)\, u_i\, dt + \sigma_{(i)}\sqrt{\frac{2}{\tau_L}}\, dW .
\end{equation}
For stationary isotropic turbulence (where $\sigma_i^2=2/3\, k$ is a constant), the normalized 
Langevin model yields that
\begin{equation}
du_i = - \frac{u_i}{\tau_L}\, dt + \sqrt{\frac{2\sigma_i^2}{\tau_L}}\, dW_i .
\end{equation}
The timescale $\tau_L$ represents the timescale of velocity fluctuations in the stationary case
and can be expressed as a function of $k$ and $\epsilon$ by introducing a constant $C_0$
\begin{equation}
\tau_L= \frac{4}{3\, C_0}\frac{k}{\lra{\epsilon}}
\end{equation}
from which it results that the equation can also be written as
\begin{equation}
du_i= - \frac{u_i}{\tau_L}\, dt + \sqrt{C_0\, \lra{\epsilon}}\, dW_i
\end{equation}
and since $\tau_L=T_L^{st}$ in the stationary case~\cite{Pope_1994,Minier_2001}, it is seen
that the present normalized Langevin is identical to the SLM. Second, for the case of homogeneous
isotropic but decaying turbulence (such as grid turbulence), the normalized Langevin model
can be written
\begin{equation}
du_i= -\left( \frac{1}{\tau_L} - \frac{1}{2\sigma_{(i)}^2}\frac{d\sigma_{(i)}^2}{dt}\right)\,
u_i\, dt + \sqrt{C_0\, \lra{\epsilon}}\, dW_i 
\end{equation}
where the diffusion coefficient has been re-expressed with the constant $C_0$, as just shown.
For decaying isotropic turbulence, we have that
\begin{equation}
\frac{1}{2\sigma_{(i)}^2}\frac{d\sigma_{(i)}^2}{dt}=\frac{1}{2\, k}\frac{dk}{dt}
                                           =- \frac{\lra{\epsilon}}{2\, k}
\end{equation}
and the equation for the instantaneous velocities (which, in that case, are identical to the
fluctuating ones) becomes
\begin{equation}
dU_i= -\left( \frac{1}{2}+ \frac{3\, C_0}{4} \right) \frac{\lra{\epsilon}}{k}\, dt 
      + \sqrt{C_0\, \lra{\epsilon}}\, dW_i
\end{equation}
By adding a possible term involving a matrix $G_{ij}^a$ as in the previous section, it is seen
that, for the two special cases considered here of stationary and decaying isotropic turbulence,
the normalized Langevin model retrieves the form of the GLM given in 
Eqs.~(\ref{model_U_GLM})-(\ref{drift_model_U_2}).

However, the situation is quite different as soon as we move out of these two simple cases. First
of all, the formulation in Eq.~(\ref{normalised_Langevin_origin}) is inconsistent when it is 
applied to homogeneous anisotropic turbulence since results depend on whether the axis of the 
reference system are aligned with the principal axis of $\lra{u_i\,u_j}$ or not. This can be 
traced to the fact that the model in Eq.~(\ref{normalised_Langevin_origin}) does not respect
the criterion (F-1) (actually, the correct way to define normalized velocities should have been $\widehat{u_i}=(R^{-1/2})_{ij}\,u_j$ where $R^{-1/2}$ is the matrix that stands for the Reynolds 
stress inverse square root $(R^{-1/2})^2R=\mathds{1}$ in matrix notation). Second, for 
general inhomogeneous flows, it appears from the form expressed in Eq.~(\ref{normalised_Langevin_1}), 
that $\lra{du_i}\neq \dfrac{\partial \lra{u_iu_k}}{\partial x_k}\, dt$, which shows that such a 
formulation suffers from spurious drifts. This was recognized in earlier versions~\cite{Wilson_1981}
but attempts to correct the model formulation have been made mainly through the addition of ad-hoc
mean terms~\cite{Dehbi_2008,Dehbi_2010,Mito_2002,Iliopoulos_2003,Iliopoulos_2004}, such as the last 
term on the rhs of Eq.~(\ref{normalised_Langevin}).

We can now consider the complete version given in Eq.~(\ref{normalised_Langevin}) in non-homogeneous
turbulent flows. By following the same derivations as in Eqs.~(\ref{normalised_variable_change}), this 
normalized Langevin model is equivalent to the following equation for the fluctuating velocities
\begin{equation} \label{normalised_Langevin_2}
du_i = \frac{\partial \lra{u_iu_k}}{\partial x_k}\, dt  \underbrace{
- \lra{u_iu_k} \frac{1}{\sigma_{(i)}}\frac{\partial \sigma_{(i)}}{\partial x_k}\, dt
+  u_iU_k \frac{1}{\sigma_{(i)}}\frac{\partial \sigma_{(i)}}{\partial x_k}\, dt}
_{``production \, term"}
\underbrace{ + \frac{1}{\sigma_{(i)}}\frac{\partial \sigma_{(i)}}
{\partial t}\, u_i\, dt -\frac{u_i}{\tau_L}\, dt + \sigma_i\sqrt{\frac{2}{\tau_L}}\, dW}_{Langevin \, term} 
\end{equation}
Since the constraint that $\lra{du_i}= \dfrac{\partial \lra{u_iu_k}}{\partial x_k}\, dt$
is now enforced, there is no spurious drifts and the model is consistent with the Reynolds
equation, indicating that the first criterion put forward in 
section~\ref{Choice of criteria for single-phase flow models} is met. However, as seen by 
comparing Eq.~(\ref{normalised_Langevin_2}) with the exact equation for fluctuating velocities 
in Eq.~(\ref{model_u_2}), it is evident that the exact term 
$- u_k\dfrac{\partial \lra{U_i}}{\partial x_k}$ is not retrieved. This means that the essential 
production terms in the corresponding Reynolds-stress equations are mishandled and, therefore, 
that this model is inconsistent with the correct structure of $R_{ij}-\epsilon$ equations, 
as presented in Eqs.(\ref{structure_equations_Rij}). In other words, present normalized Langevin 
models do not satisfy the criteria (F-3) and the failure to reproduce the correct structure of 
Reynolds-stress equations is a severe limitation.

It is also worth noting that, if the previous equations have been developed using the simplest 
form of the Langevin model, these results does not depend on the special form of the normalized 
model. Indeed, if we consider a similar model written as 
\begin{equation}
d\left( \frac{u_i}{\sigma_{(i)}}\right) =
\frac{\partial}{\partial x_k}\left( \frac{\lra{u_iu_k}}{\sigma_{(i)}}\right) \, dt
+ d\widehat{M}_i(t,\mb{x},\widehat{\mb{u}})
\end{equation}
where $d\widehat{M}_i$ is expressed as a function of the normalized fluctuating velocities 
$\widehat{\mb{u}}$ which, in present formulations are (wrongly) defined as 
$\widehat{u}_i=u_i/\sigma_{(i)}$, and is such that $\lra{dM_i}=0$ and $\lra{u_i\,dM_i}=0$ 
to respect stationary isotropic turbulence conditions, we have for the fluctuating velocity
an equation similar to Eq.~(\ref{normalised_Langevin_2}), namely
\begin{equation} 
du_i = \frac{\partial \lra{u_iu_k}}{\partial x_k}\, dt  
- \lra{u_iu_k} \frac{1}{\sigma_{(i)}}\frac{\partial \sigma_{(i)}}{\partial x_k}\, dt
+  u_iU_k \frac{1}{\sigma_{(i)}}\frac{\partial \sigma_{(i)}}{\partial x_k}\, dt
+ \frac{1}{\sigma_{(i)}}\frac{\partial \sigma_{(i)}}{\partial t} u_i\, dt
+ \sigma_{(i)}\,d\widehat{M}_i \, .
\end{equation}
Thus, whatever the chosen form of the stochastic model $d\widehat{M}_i$ and regardless of variants
in the expression of the diffusion matrix~\cite{Mito_2002,Iliopoulos_2003,Iliopoulos_2004}, the 
same conclusions about the lack of consistency and the failure to respect the criteria (F-3) 
still hold, making these models unacceptable descriptions of turbulent flows.

\subsection{Consistency with scalar modeling}
\label{scalar flux}

Similar issues exist for the formulation of scalar models and, furthermore, consistency with classical scalar modeling is explicitly used in the criteria (P-5) retained in the list of requirements set forth in section~\ref{Choice of criteria for two-phase flow models} for two-phase flow models. It is thus worth discussing the relevant points of classical scalar modeling. 

For this purpose, we consider a passive scalar $\phi(t,\mb{x})$ transported by the flow and which is the solution of the exact advection-diffusion equation
\begin{equation}
\frac{\partial \phi}{\partial t}+ U_k\frac{\partial \phi}{\partial x_k}= 
\Gamma\, \frac{\partial^2 \phi}{\partial x_k^2}
\end{equation}
where $\Gamma$ is the scalar diffusivity. Assuming, still for the sake of simplicity, that we are dealing with high Peclet-number flows, the mean diffusion term can be neglected and the mean scalar equation (using $\phi=\lra{\phi}+\phi'$) has the form
\begin{equation} 
\label{mean-scalar}
\frac{\partial \lra{\phi}}{\partial t} + \lra{U_k}\frac{\partial \lra{\phi}}
{\partial x_k} + \frac{\partial \lra{u_k \, \phi^{'}}}{\partial x_k} =0
\end{equation}
where $\lra{u_k \, \phi'}$ the scalar flux which must be modeled. In second-order turbulence modeling, non-local closures for the scalar flux are obtained by considering the transport equation for the scalar fluxes which write
\begin{equation}
\frac{\partial \lra{u_i\phi^{'}}}{\partial t} 
+ \lra{U_k}\frac{\partial \lra{u_i\,\phi^{'}}}{\partial x_k} +
\frac{\partial \lra{u_iu_k\phi^{'}}}{\partial x_k} = -\lra{u_iu_k}\frac{\partial \lra{\phi}}
{\partial x_k} - \lra{u_k\phi^{'}}\frac{\partial \lra{U_i}}{\partial x_k} + \Psi_i
\end{equation} 
where $\Psi_i$ represents the correlation $\lra{\phi'\left( -\frac{1}{\rho}\frac{\partial p'}{\partial x_i}+\nu\Delta u_i\right)}+\lra{u_i\,\Delta \phi'}$ which is usually modeled as a function of the scalar fluxes, for example $\Psi_i=- B_{ik}\lra{u_k\phi'}$.

A classical approach is to consider the algebraic relations which result from the second-order transport equations when convective terms are neglected~\cite{Rogers_1989}. This approach was developed to show that, in that case, one should obtain expressions consistent with the general diffusivity concept. Indeed, neglecting transport terms in the above equation for the scalar fluxes gives 
\begin{equation}
\left( B_{ik} + \frac{\partial \lra{U_i}}{\partial x_k} \right) \lra{u_k\, \phi'}=
-\lra{u_i\,u_k}\frac{\partial \lra{\phi}}{\partial x_k}
\end{equation}
which yields that
\begin{equation}
\label{algebric_scalar-flux}
\lra{u_i\, \phi'}= - \left( O^{-1}_{ik}R_{kj} \right) \frac{\partial \lra{\phi}}{\partial x_j}
\end{equation}
with $R_{ij}=\lra{u_i\,u_j}$ and $O_{ij}=B_{ij}+\frac{\partial \lra{U_i}}{\partial x_j}$.

This behavior has been found to be realistic both in numerical and experimental investigations\cite{Rogers_1989,Warhaft_2000}. It is thus important to check that various formulations can yield similar expressions when the same hypothesis are made. For example, starting with the formulation in terms of instantaneous variables that was already used at the beginning of section~\ref{Analysis of different formulations}, this consists in adding the instantaneous scalar value attached to a particle trajectory in the state-vector $\mb{Z}=(\mb{x},\mb{U},\phi)$ and in replacing the exact equations by modeled ones. Using for instance the GLM already introduced in section~\ref{Analysis of different formulations} for the dynamical variables and adding a model for $\phi$, we obtain a typical model structure as 
\begin{align}
& dx_i = U_i\, dt \\
& dU_i = -\frac{1}{\rho}\frac{\partial \langle P \rangle}{\partial x_i}\, dt +
D_i\, dt + \sqrt{C_0\langle \epsilon \rangle}dW_i \\
& d\phi= A_{\phi}\, dt .      
\end{align}
In the last equation, $A_{\phi}$ stands for a model for $\Gamma\, \Delta \phi'$ and is subject to the
constraint that $\lra{ A_{\phi}\, |\, \mb{U}}=0$. This is referred to as the micro-mixing modeling issue~\cite{Pope_1994,Pope_2000,Fox_2003} which has been the subject of
ongoing research efforts. For the sake of our present discussion, it is sufficient to use the simple 
IECM (Interaction by Exchange with the Conditional Mean) model~\cite{Fox_2003,Pope_1998} which reads
\begin{equation}
A_{\phi} = - \frac{ \phi - \lra{\phi\, |\, \mb{U}}}{\tau_{\phi}}
\end{equation}
where $\tau_{\phi}$ is the scalar mixing time scale.
Then, using either a proper PDF derivation~\cite{Pope_2000} or a short-cut method as
in section~\ref{Analysis of different formulations}, it is straightforward to show that
the scalar fluxes are the solutions of the transport equations
\begin{multline}
\label{modeled_scalar-fluxes_eq}
\frac{\partial \lra{u_i\, \phi}}{\partial t} 
+ \lra{U_k}\frac{\partial \lra{u_i\,\phi}}{\partial x_k} +
\frac{\partial \lra{u_i\, u_k\, \phi}}{\partial x_k} = -\lra{u_iu_k}\frac{\partial \lra{\phi}}
{\partial x_k} - \lra{u_k\,\phi}\frac{\partial \lra{U_i}}{\partial x_k} \\
- \left( G_{ik} - \frac{1}{2}C_{\phi} \frac{\lra{\epsilon}}
{k} \delta_{ik} \right) \lra{u_k\, \phi} 
\end{multline} 
In this equation, the last term on the rhs represents a model for $\Psi_i$ resulting from
the specific choice of the GLM for particle velocities and the IECM model for $\phi$. Other
closures would result in different expressions but with a similar form. The important point
in the present context is that the formulation in terms of instantaneous variables
gives the correct structure and is, thus, quite consistent with the asymptotic analysis
developed above to derive algebric relations when transport terms are neglected. 

Should alternative formulations be considered, for example in terms of fluctuating scalar, 
the developments presented in section~\ref{Analysis of different formulations} indicate that 
the evolution equation for $\phi'$ is
\begin{equation}
\label{fluctuating-scalar}
d\phi'= \underbrace{\frac{\partial \lra{u_k\,\phi'}}{\partial x_k}\, dt}_{(a)\, no \, spurious \, fluxes} 
        \underbrace{-u_k\frac{\partial \lra{\phi}}{\partial x_k}\, dt}_{(b)\,production\, term} 
        + A_{\phi'}\, dt
\end{equation}
where $A_{\phi'}$ corresponds to the micro-mixing model retained and expressed now in terms
of the fluctuating components. For instance, for the IECM this fluctuating term is given by 
$A_{\phi'}=-\phi'/\tau_{\phi} + (\lra{\phi\, |\, \mb{U}} - \lra{\phi})/\tau_{\phi}$. The first 
term, labeled $(a)$ in Eq.~(\ref{fluctuating-scalar}), means that the correct form of the mean 
scalar equation, Eq.~(\ref{mean-scalar}), is retrieved and is necessary to avoid spurious scalar 
fluxes. The second term, labeled $(b)$ in Eq.~(\ref{fluctuating-scalar}) leads to the correct 
production term, $-\lra{u_iu_k}\dfrac{\partial \lra{\phi}}{\partial x_k}$, in the transport 
equation for the scalar fluxes in Eq.~(\ref{modeled_scalar-fluxes_eq}) and, without this term, 
the asymptotic analysis that retrieves the scalar turbulent diffusivity, as 
in Eq.~(\ref{algebric_scalar-flux}), breaks down.

\section{Analysis of stochastic models for two-phase flows}
\label{Stochastic models for two-phase flows}

In this section, we turn to stochastic models used for the simulation of fluid velocities seen by discrete particles. The discussion is developed by considering some specific models which have been proposed in the literature. However, it is worth repeating that the aim is to help clarifying how formulations can be assessed, what issues are misleadingly taken as important and which ones should be carefully addressed when building new ideas. 

\subsection{Different formulations of two-phase flow stochastic models}
\label{Analysis of different formulations in two-phase flows}

\subsubsection{First Langevin proposals}
\label{First Langevin proposals}

One of the first proposals to introduce a Langevin model for the velocity of the fluid seen relied on a formulation based on the instantaneous velocity along particle trajectory~\cite{Simonin_1993}. We discuss this first proposition since it is interesting to illustrate the stochastic modeling procedure, in spite of some limitations outlined below. The equation was written in a discrete form but, using present notations for the sake of consistency with the other formulations, this model for $\mb{U}_s$ can easily be expressed as
\begin{equation}
\label{Simonin et al 1993}
dU_{s,i} = -\frac{1}{\rho}\frac{\partial \lra{P} }{\partial x_i}\, dt + \nu\, \Delta \lra{U_i}\, dt
+ \left( U_{p,k} - U_{s,k} \right)\frac{\partial \lra{U_{i}}}{\partial x_k}\, dt 
+ G^{*}_{ik} \left( U_{s,k}-\lra{U_{k}} \right) dt + B \, dW_i
\end{equation}
In this equation, the matrix $G^{*}_{ik}$ is defined by the two timescales $T_{L,||}^{*}$ and $T_{L,\bot}^{*}$, which correspond to Csanady's expressions (though the precise form of these Csanady's formulas were not given in the original work) for the timescales of the fluid seen in the directions parallel and perpendicular to the direction of the mean slip velocity $\lra{\mb{U}_r}_m=\lra{\mb{U}_p}_m - \lra{\mb{U}}$ respectively, and is given by~\cite{Simonin_1993}
\begin{equation}
\label{general expression Gij}
G^{*}_{ik}= - \frac{1}{T_{L,\bot}^{*}}\, \delta_{ik} - \left[ \frac{1}{T_{L,||}^{*}} - 
\frac{1}{T_{L,\bot}^{*}} \right]\, r_i\,r_k
\end{equation}
with $r_i= U_{r,i}/|\, \lra{\mb{U}_r}_m \,|$ the normalized vector aligned with the mean slip velocity. The Csanady's formulas for the timescales $T_{L,||}^{*}$ and $T_{L,\bot}^{*}$ are~\cite{Minier_2001}
\begin{equation}
\label{Csanady timescales}
T_{L,||}^{*}= \frac{T_L}{\sqrt{ 1 + \beta^2 \displaystyle \frac{| \lra{\mb{U}_r}_m|^2}{2k/3}}} \, ,
\qquad
T_{L,\bot}^{*}=\frac{T_L}{\sqrt{ 1 + 4\beta^2 \displaystyle \frac{| \lra{\mb{U}_r}_m|^2}{2k/3}}}\, .
\end{equation}
where the expression of $T_L$ is given in Eq.~\eqref{drift_model_U_2} and where $\beta=T_L/T_E$ is the ratio of the Lagrangian integral time scale $T_L$ to the Eulerian one $T_E$.

For two-phase flow modeling, this article~\cite{Simonin_1993} was interesting in that it was one of the first attempts to build bridges between the single- and two-phase flow situations. Another interest is that it introduced the notion of the consistency between the resulting model for the drift velocity $\mb{V}_d$ and classical scalar modeling as detailed in section~\ref{scalar flux}. This consistency issue has been retained in the present list as criteria (P-5). It is seen that in Eq.~(\ref{Simonin et al 1993}) the second term on the rhs of the equation stands for the mean viscous term. In the limit of vanishing particle inertia, this model gives an expression similar to the SLM to which the mean viscous term has been added. Such a formulation is indeed consistent with the low Reynolds-number form of the Reynolds equation but it must be noted that it would not give the correct low Reynolds-number form of the second-order equations~\cite{Dreeben_1998,Waclawczyk_2004}. However, in the present context, we have limited ourselves to high Reynolds-number flows and, therefore, 
the mean viscous term can be neglected. 

This first proposal has often been cited as a `stochastic model'~\cite{Pialat_2007,Arcen_2009,Taniere_2014} to 
be compared on an equal footing with the more complete formulations that will be discussed below. However, for all the interests of the form expressed by Eq.~(\ref{Simonin et al 1993}), this is a misleading presentation and an incorrect statement with regard to its actual status. Indeed, the diffusion coefficient, written as $B$ in Eq.~(\ref{Simonin et al 1993}), is never specified~\cite{Simonin_1993}! In the context of the original study, this is not surprising since, as it transpires from the very title of the article~\cite{Simonin_1993}, its purpose was to derive a set of mean-field equations from this stochastic description. As the set of mean-field equations was chosen to be limited to the particle mean velocity and the fluid-particle correlations, the expression of the diffusion coefficient is not required. If such a proposal leads to as a realizable Eulerian model, it must be stressed that, in no way, can it be called a Lagrangian stochastic model. From the list of criteria in in section~\ref{Choice of criteria for two-phase flow models}, it is indeed obvious that the criterion (P-2) is not met. In the particle-tracer limit, the form given in Eq.~(\ref{Simonin et al 1993}) reverts to the form of the SLM and, in that sense, all the terms appearing on the lhs of Eqs.~(\ref{structure_equations_Rij}) are obtained. However, since the diffusion coefficient is unknown, it cannot be assessed whether the corresponding terms $\Psi_{ij}$ on the rhs of Eq.~(\ref{structure_equations_Rij_c}) is such that $1/2\Psi_{ii}=-\epsilon$ with the constraint that $\epsilon \geq 0$. Therefore, the criterion (P-3) cannot be checked. This is also true for the criterion (P-4) which is directly related to the diffusion coefficient and is therefore not met, whereas the criterion (P-5) is satisfied. 
In order to assess criterion (P-6), we consider the limit of non-vanishing but small Stokes numbers. In that case, we can assume that the mean relative velocity $\lra{\mb{U}_r}_m$ is equal to the particle settling velocity $\mb{W}_g$ introduced in section~\ref{Choice of criteria for two-phase flow models}, that is $\lra{\mb{U}_r}_m\simeq \mb{W}_g$. Then, given the expression of the Csanady's timescales in Eqs.~\eqref{Csanady timescales}, which can be re-written under the form $T_{L,||}^{*}= T_L/\sqrt{1 + x^2}$ with $x=St \times (g\,T_L)/\sqrt{2/3k}$, we have to first order in $St$ that $T_{L,||}^{*}\simeq T_L$ as well as $T_{L,\bot}^{*}\simeq T_L$. This means that the mean conditional average of the drift vector $\lra{ \mb{D}_s\,|\,\mb{Z}}$ which appears in the criterion (P-6) is such that
\begin{equation}
\lra{ D_{s,i}\,|\,\mb{Z}}= -\frac{1}{\rho}\frac{\partial \lra{P}}{\partial x_i} 
- \frac{ U_{s,i} - \lra{U_i}}{T_L} + U_{r,k}\frac{\partial \lra{U_i}}{\partial x_k}~.
\end{equation}
The first two terms on the rhs correspond to the mean conditional increments for fluid particle velocities modeled with the SLM. We can thus re-express this equation as
\begin{equation}
\lra{ D_{s,i}\,|\,\mb{Z}}= \lra{ D_i\,|\,\mb{Z}} + U_{r,k}\frac{\partial \lra{U_i}}{\partial x_k}~.
\end{equation}
where $\mb{D}$ is an acceptable model for the drift vector for fluid velocities, as demonstrated by the analysis carried out in section~\ref{Analysis of different formulations}. By comparing with the formulation of the criterion (P-6) in Eq.~\eqref{from of criterion P-6}, we can see that the second term is not exactly retrieved since the instantaneous relative velocity appears instead of the mean one. Thus, strictly speaking, the criterion (P-6) is not satisfied, though it is also readily seen that the weak form (P-6bis) is respected.

In summary, the proposal in Eq.~\eqref{Simonin et al 1993} must be regarded as an incomplete Langevin model and cannot be retained as a proper PDF description. At this stage, it is an often-seen temptation to close the diffusion coefficient by retaining the closure used in the GLM approach for single-phase flows, that is $B=\sqrt{C_0\, \lra{\epsilon}}$. 
Yet, it can be seen from Eq.~(\ref{Simonin et al 1993}) that the return-to-equilibrium term is not based on a scalar timescale (as in the SLM with $T_L$) but has a non-isotropic form. In other words, using the fluid-limit value of the diffusion coefficient would clearly lead to a violation of the criterion (P-4). This central point in the formulation of stochastic models for the two-phase flow situation was only recognized afterward~\cite{Minier_2001}. 

\subsubsection{Complete Langevin models}
\label{Complete Langevin models}

The velocity along particle trajectory was later called the velocity of the fluid seen~\cite{Pozorski_1999} and models were written as proper stochastic differential equations. The general PDF framework was set up~\cite{Minier_2001} and a new Langevin model was developed as one example of the general methodology. For the sake of a simpler presentation, we assume here that the mean slip velocity $\lra{\mb{U}_r}_m$ is aligned with the first coordinate axis and, based on this assumption, the first complete Langevin model is given by
\begin{multline}
\label{Langevin Minier Peirano 2001}
dU_{s,i} = -\frac{1}{\rho}\frac{\partial \lra{P} }{\partial x_i}\, dt
+ \left( \lra{U_{p,k}}_m - \lra{U_{s,k}}_m \right)
\frac{\partial \lra{U_{i}}}{\partial x_k}\, dt \\
- \frac{ U_{s,i}-\lra{U_{s,i}}_m }{T_{L,(i)}^{*}}\, dt 
+ \sqrt{ \lra{\epsilon}\left( C_0b_{(i)} \tilde{k}/k + \frac{2}{3}( b_{(i)} \tilde{k}/k -1) \right) }\, dW_i
\end{multline}
In this equation, $\tilde{k}$ is a new kinetic energy which stands for the fluid normal kinetic energies weighted by the Csanady's factors and is expressed by:
\begin{equation}
\tilde{k}=\frac{3}{2}\frac{\sum_{i=1}^3 b_i \lra{u_{i}^2}}
{\sum_{i=1}^3 b_i}~, \qquad b_i=\frac{T_L}{T_{L,i}^{*}}~. 
\end{equation}
With the assumption on the coordinate system, the time scales $T_{L,(i)}^{*}$ are equal to
\begin{equation}
T_{L,1}^{*}=T_{L,||}^{*}~, \qquad T_{L,2}^{*}= T_{L,3}^{*}=T_{L,\bot}^{*}
\end{equation}
where $T_{L,||}^{*}$ and $T_{L,\bot}^{*}$ are given in Eqs.~\eqref{Csanady timescales}. The form given
in Eq.~\eqref{Langevin Minier Peirano 2001} in the reference system aligned with the mean relative velocity is best to bring out the characteristic features of the model and the forms of the drift and diffusion coefficients.
Yet, in a general coordinate system, this form of the complete Langevin model is expressed by
\begin{equation}
\label{Langevin Minier Peirano 2001 general}
dU_{s,i} = -\frac{1}{\rho}\frac{\partial \lra{P} }{\partial x_i}\, dt
+ \left( \lra{U_{p,k}}_m - \lra{U_{s,k}}_m \right)
\frac{\partial \lra{U_{i}}}{\partial x_k}\, dt
+ G^{*}_{ik} \left( U_{s,k}-\lra{U_{s,k}}_m \right) dt + B_{ij} \, dW_j
\end{equation}
where $G^{*}_{ik}$ is the same as in Eq.~\eqref{general expression Gij} and where the explicit expression of the diffusion matrix $B_{ij}$ is detailed elsewhere~\cite{Minier_2001}.

It is seen that the closure of the drift vector in Eq.~\eqref{Langevin Minier Peirano 2001} differs slightly from the one chosen in Eq.~(\ref{Simonin et al 1993}). However, the key point is that the diffusion coefficient (actually, the diffusion matrix) is explicitly formulated. Therefore, the first part of the criterion (P-2) is now met, which explains why this model has been referred to as a `complete Langevin model'. It is also clear that such a model allows practical numerical predictions to be carried out~\cite{Minier_2001} and, consequently, the criterion (P-2) is fully satisfied. In that sense, this Langevin model can be truly referred to as a PDF description for two-phase flows. It is also straightforward to show that, when particle inertia becomes negligible, the model given in Eq.~(\ref{Langevin Minier Peirano 2001}) reduces exactly to the SLM for fluid particles which is one of the models that meet the lists of requirements set for single-phase flows. Thus, the criterion (P-3) is satisfied. The specific closure of the diffusion matrix is such that, for a homogeneous turbulent fluid flow laden with discrete particles, the kinetic energy of the fluid seen follows the same statistical law as for the fluid kinetic energy, $d(1/2\,\lra{\mb{u}_s^2}_m)/dt=- \lra{\epsilon}$. As discussed in section~\ref{Choice of criteria for two-phase flow models} when the criterion (P-4) was introduced, this is indeed what can be expected from a physical point of view. This apparently-trivial constraint is actually an important issue in the construction of complete stochastic models for two-phase flows and was stressed accordingly~\cite{Minier_2001}. For our present concern, this means that the criterion (P-4) is also met. However, in the form chosen for the drift vector in Eq.~(\ref{Langevin Minier Peirano 2001}), it is seen that the return-to-equilibrium term was written as a return of the instantaneous fluid velocity seen to its mean value at the same location. This results in a zero-contribution term in the corresponding mean equation for the drift velocity and, therefore, the criterion (P-5) is not respected by the model given in 
Eq.~(\ref{Langevin Minier Peirano 2001})~\cite{Simonin_2001}. 

This limitation was later overcome by expressing the return-to-equilibrium term as a return of the instantaneous value of the fluid velocity seen to the local mean fluid velocity. This led to the final formulation of this complete Langevin model~\cite{Minier_2004} which, using the same simplifying assumption as in Eq.~(\ref{Langevin Minier Peirano 2001}) that $\lra{\mb{U}_r}$ is aligned with the first coordinate axis, is expressed by
\begin{multline}
\label{Minier et al 2004}
dU_{s,i} = -\frac{1}{\rho}\frac{\partial \lra{P} }{\partial x_i}\, dt
+ \left( \lra{U_{p,k}}_m - \lra{U_{k}} \right)\frac{\partial \lra{U_{i}}}{\partial x_k}\, dt \\
- \frac{ U_{s,i}-\lra{U_{i}} }{T_{L,(i)}^{*}}\, dt 
+ \sqrt{ \lra{\epsilon}\left( C_0b_{(i)} \tilde{k}/k + \frac{2}{3}( b_{(i)} \tilde{k}/k -1) \right) }\, dW_i \, .
\end{multline}
whereas the complete formula in a general coordinate system is
\begin{equation}
\label{Langevin Minier et al 2004 general}
dU_{s,i} = -\frac{1}{\rho}\frac{\partial \lra{P} }{\partial x_i}\, dt
+ \left( \lra{U_{p,k}}_m - \lra{U_{k}} \right)\frac{\partial \lra{U_{i}}}{\partial x_k}\, dt
+ G^{*}_{ik} \left( U_{s,k}-\lra{U_{k}} \right)\,dt+ B_{ij} \, dW_j
\end{equation}
with the explicit formulation of the diffusion matrix detailed elsewhere~\cite{Minier_2004}.

Like its preceding version, this formulation fulfills the criteria (P-1) to (P-4). However, with the new form for the return-to-equilibrium term in the drift vector, the corresponding equation for the drift velocity is now
\begin{equation}
\label{drift velocity equation}
\frac{\partial V_{d,i}}{\partial t} + \lra{U_{p,k}}_m\frac{\partial V_{d,i}}{\partial x_k}=
\frac{1}{\alpha_f}\frac{\partial}{\partial x_k}\left[ \alpha_f \lra{u_i\,u_k} \right]
- \frac{1}{\alpha_p}\frac{\partial}{\partial x_k}\left[ \alpha_p \lra{u_{s,i}'\,u_{p,k}}_m \right]
+ G^{*}_{ik}\, V_{d,k}
\end{equation}
where $\alpha_f$ and $\alpha_p$ are the fluid and particle volumetric fractions (with $\alpha_f+\alpha_p=1$). In Eq.~\eqref{drift velocity equation}, the fluctuations of $U_{s,i}$ are defined as $u_{s,i}'=U_{s,i} - \lra{U_i}$ (Note that a specific notation is used for the `fluctuation' of the velocity of the fluid seen while usual notations are kept for particle velocities and for the fluid velocities within the fluid phase, see the discussion on the definition of fluctuating velocities in section~\ref{Fluctuating or instantaneous fluid velocity seen}). It is important to stress that, in the scalar limit case (when the particle relaxation time $\tau_p$ is negligible with respect to the fluid turbulence time scale $T_L$ but with a non-vanishing volumetric fraction), $V_{d,i}$ reduces to the ordinary turbulent correlation between the particle concentration and the fluid velocity, which is the case discussed in section \ref{scalar flux}. By making the same assumptions as in section~\ref{scalar flux}, this equation yields that
\begin{equation}
\label{drift velocity identity}
V_{d,i}= (G^{*})^{-1}_{ik}\left\{ 
\frac{1}{\alpha_p}\frac{\partial}{\partial x_l}\left[ \alpha_p \lra{u_{s,k}'\,u_{p,l}}_m \right]
- \frac{1}{\alpha_f}\frac{\partial}{\partial x_l}\left[ \alpha_f \lra{u_k\,u_l} \right] \right\}
\end{equation}
which is in line with classical scalar modeling as recalled in section~\ref{scalar flux}.
Indeed, for the case of constant correlations, we have that
\begin{equation}
\label{drift velocity identity simplified}
V_{d,i}= (G^{*})^{-1}_{ik}\left\{ 
\frac{1}{\alpha_p}\lra{u_{s,k}'\,u_{p,l}}_m + \frac{1}{\alpha_f}\lra{u_k\,u_l} \right\}
\frac{\partial \alpha_p}{\partial x_l}
\end{equation}
which is similar to the result given in Eq.~(\ref{algebric_scalar-flux}), showing that the criterion 
(P-5) is fulfilled. 
For the analysis of the criterion (P-6), we can draw on the developments obtained in section~\ref{First Langevin proposals} in the limit of small Stokes numbers. Thus, to first order in $St$, we have that $T_{L,||}^{*}\simeq T_L$, $T_{L,\bot}^{*}\simeq T_L$, as well as $b_i \simeq 1$. Using the general expression of the drift vector in the complete model in Eq.~\eqref{Langevin Minier et al 2004 general}, we obtain now that 
the mean conditional average of the drift vector $\lra{ \mb{D}_s\,|\,\mb{Z}}$ is 
\begin{equation}
\lra{ D_{s,i}\,|\,\mb{Z}}= -\frac{1}{\rho}\frac{\partial \lra{P}}{\partial x_i} 
- \frac{ U_{s,i} - \lra{U_i}}{T_L} + \lra{U_{r,k}}_m\frac{\partial \lra{U_i}}{\partial x_k}~.
\end{equation}
Using again the drift vector of the SLM for fluid particle velocities and the fact that, for $St \ll 1$, $\lra{\mb{U}_r}_m\simeq \mb{W}_g$, this is equivalent to
\begin{equation}
\lra{ D_{s,i}\,|\,\mb{Z}}= \lra{ D_i\,|\,\mb{Z}} + W_{g,k}\,\frac{\partial \lra{U_i}}{\partial x_k}~.
\end{equation}
This is indeed the form stated for this criterion, as expressed in Eq.~\eqref{from of criterion P-6}, showing that (P-6) is satisfied, from which (P-6bis) is obviously also met.

Consequently, the complete list of criteria of section~\ref{Choice of criteria for two-phase flow models}
is satisfied by the Langevin model given in Eq.~(\ref{Minier et al 2004}) or in Eq.~\eqref{Langevin Minier et al 2004 general} and, as such, this model can be assessed as being a satisfactory description of two-phase flows.

\subsubsection{Fluctuating or instantaneous fluid velocity seen}
\label{Fluctuating or instantaneous fluid velocity seen}

So far, the models considered in sections~\ref{First Langevin proposals} and \ref{Complete Langevin models} for two-phase flows have been presented in terms of the instantaneous fluid velocity seen. Formulations in terms of the fluctuating component of this fluid velocity seen are, of course, possible but reveal themselves to be not only cumbersome but also trickier than in the single-phase case. One reason is rooted in the fact that `fluctuations' must be carefully defined since averages should be first properly defined as meaningful quantities over the respective fluid and particle phases. For example, two `fluctuations' can be defined from the instantaneous value of the fluid velocity seen: as in section~\ref{Complete Langevin models}, one can introduce fluctuations as the difference with the local mean fluid velocity (properly defined as an average over the fluid phase only), $u_{s,i}'=U_{s,i} - \lra{U_i}$, whose mean value (over the particle phase) is not zero but the drift velocity $V_{d,i}$; or the `real' fluctuation, $u_{s,i}=U_{s,i} - \lra{U_{s,i}}_m$, whose mean value is indeed zero (obviously, these two fluctuations are related through $u_{s,i}=u_{s,i}'-V_{d,i}$. Note that $\lra{u_{s,i}'\,u_{p,k}}_m=\lra{u_{s,i}\,u_{p,k}}_m$ in Eqs.~(\ref{drift velocity equation})-(\ref{drift velocity identity simplified})). These distinctions were made clear in Eulerian descriptions~\cite{Simonin_1993,Simonin_2000} and, in the complete PDF approach to two-phase flows, this means that the complete theoretical framework~\cite{Minier_2001,Peirano_2002} outlined in section~\ref{The PDF theoretical framework} must be carefully followed. 

To illustrate this point, if we consider a stochastic formulation for the instantaneous fluid velocity seen which has the following form:
\begin{equation}
dU_{s,i}= -\frac{1}{\rho}\frac{\partial \lra{P}}{\partial x_i}\, dt + A_i\, dt 
+ G_{ik}^{*}\left( U_{s,k} - \lra{U_k} \right)\, dt + B_{ik}\, dW_k
\end{equation}
then, the corresponding equation for the fluctuating component $u_{s,i}'$ as defined above is
\begin{equation}
\label{eq: fluctuating_velocity_seen_1}
du_{s,i}'= \frac{1}{\alpha_f}\frac{\partial \left[ \, \alpha_f \lra{u_i\,u_k} \,\right]}{\partial x_k}\, dt
- \left[ U_{p,k}-\lra{U_{k}} \right]\frac{\partial \lra{U_i}}{\partial x_k}\, dt 
+ A_i\, dt + G_{ik}^{*} u_{s,k}'\, dt + B_{ik}dW_k
\end{equation}
Handling directly this form of a model raises questions as to the exact formulation of the first term on the rhs of Eq.~\eqref{eq: fluctuating_velocity_seen_1}. Indeed, an alternative formulation can be expressed as 
\begin{equation}
\label{eq: fluctuating_velocity_seen_2}
du_{s,i}'= \frac{\partial \left[ \, \lra{u_i\,u_k} \,\right]}{\partial x_k}\, dt
- \left[ U_{p,k}-\lra{U_{k}} \right]\frac{\partial \lra{U_i}}{\partial x_k}\, dt 
+ A_i\, dt + G_{ik}^{*} u_{s,k}'\, dt + B_{ik}dW_k~.
\end{equation}
This dual form of the equation for the fluctuating part of the velocity of the fluid seen is not related to a particular choice of the drift and diffusion terms. For example, in recent articles~\cite{Pialat_2007, Arcen_2009}, it is said that for the value of the term $A_i$ in the drift vector which corresponds to Eq.~(\ref{Simonin et al 1993}), the equation for this fluctuating component is
\begin{equation}
\label{wrong equation us' Arcen Taniere}
du_{s,i}'= \frac{\partial \left[ \, \lra{u_i\,u_k} \,\right]}{\partial x_k}\, dt 
+ \left( G_{ik}^{*} - \frac{\partial \lra{U_i}}{\partial x_k} \right) u_{s,k}'\, dt + B_{ik}dW_k
\end{equation}
whereas the corresponding first form given above would be
\begin{equation}
\label{right equation us' Arcen Taniere}
du_{s,i}'= \frac{1}{\alpha_f}\frac{\partial \left[ \, \alpha_f \lra{u_i\,u_k} \,\right]}{\partial x_k}\, dt
+ \left( G_{ik}^{*} - \frac{\partial \lra{U_i}}{\partial x_k} \right) u_{s,k}'\, dt + B_{ik}dW_k~.
\end{equation}
The difference between Eq.~\eqref{eq: fluctuating_velocity_seen_1} and Eq.~\eqref{eq: fluctuating_velocity_seen_2} (or between Eq.~\eqref{wrong equation us' Arcen Taniere} and Eq.~\eqref{right equation us' Arcen Taniere}) is related to whether two-way coupling effects are accounted or not. The form in Eq.~\eqref{eq: fluctuating_velocity_seen_1} (and in Eq.~\eqref{right equation us' Arcen Taniere}) is obtained when the mean fluid velocity is given by
\begin{equation}
\label{eq: mean_fluid_velocity_1}
\frac{\partial \lra{U_i}}{\partial t} + \lra{U_k}\frac{\partial \lra{U_i}}{\partial x_k}=
-\frac{1}{\rho}\frac{\partial \lra{P}}{\partial x_i} - 
\frac{1}{\alpha_f}\frac{\partial \left[ \, \alpha_f \lra{u_i\,u_k} \,\right]}{\partial x_k}~,
\end{equation}
which contains the correct expression of the Reynolds stress for the fluid momentum equation in the two-phase flow situation~\cite{Minier_2001,Peirano_2002}. On the other hand, the form in Eq.~\eqref{eq: fluctuating_velocity_seen_2} (and in Eq.~\eqref{wrong equation us' Arcen Taniere}) is obtained by considering that the equation for the mean fluid velocity follows
\begin{equation}
\label{eq: mean_fluid_velocity_2}
\frac{\partial \lra{U_i}}{\partial t} + \lra{U_k}\frac{\partial \lra{U_i}}{\partial x_k}=
-\frac{1}{\rho}\frac{\partial \lra{P}}{\partial x_i} - 
\frac{\partial \left[ \, \lra{u_i\,u_k} \,\right]}{\partial x_k}~.
\end{equation}
Eq.~\eqref{eq: mean_fluid_velocity_2} is, of course, the Reynolds equation for the fluid phase treated as an incompressible single-phase turbulent flow (in the high Reynolds-number limit). On the other hand, in Eq.~\eqref{eq: mean_fluid_velocity_1}, it is seen that a two-way coupling effect is present since the reduced volumetric fraction occupied by the fluid is accounted for through $\alpha_f (t,\mb{x})$. Note, however, that Eq.~\eqref{eq: mean_fluid_velocity_1} represents only a partial account of two-way coupling in the sense that only the volume effect is included. In order to properly account for two-way coupling, momentum exchange terms should be added for the stochastic model for the velocity of the fluid seen and into the Reynolds equation for the fluid phase. To the authors' knowledge, this has been properly proposed only for the complete Langevin model (see detailed presentations~\cite{Minier_2001, Minier_2004,Peirano_2006}) by considering that the term
\begin{equation}
\label{Complete Langevin: added term}
A_{p\rightarrow f,i}=-
\frac{\alpha_p\, \rho_p}{\alpha_f\, \rho}\left( \frac{U_{s,i}-U_{p,i}}{\tau_p} \right)
\end{equation} 
which represents the effect of the particle phase on the fluid velocity seen is added to the rhs of Eq.~\eqref{Minier et al 2004}. In that case, volumetric and momentum two-way coupling are taken into account and the corresponding equation for the fluctuating part of the velocity of the fluid seen is
\begin{multline}
\label{eq: fluctuating_velocity_seen_2way}
du_{s,i}'= \frac{1}{\alpha_f}\frac{\partial \left[ \, \alpha_f \lra{u_i\,u_k} \,\right]}{\partial x_k}\, dt
- \left[ U_{p,k}-\lra{U_{k}} \right]\frac{\partial \lra{U_i}}{\partial x_k}\, dt 
+ A_i\, dt + G_{ik}^{*} u_{s,k}'\, dt + B_{ik}dW_k \\
- \frac{\alpha_p\, \rho_p}{\alpha_f\, \rho}\left[ \left(\frac{U_{s,i}-U_{p,i}}{\tau_p} \right)
- \left\langle \frac{U_{s,i}-U_{p,i}}{\tau_p} \right\rangle_m \,\right]\, dt~.
\end{multline}
Note that, in the usual case of a polydisperse two-phase flow, the last two terms on the rhs of Eq.~\eqref{eq: fluctuating_velocity_seen_2way} cannot be expressed only in terms of the mean and fluctuating fluid and particle velocities (which is a further argument to suggest that formulations in terms of fluctuations are more
cumbersome than the ones in terms of instantaneous values).

Though the purpose of our analysis is not directly about two-way coupling effects, the above discussion is interesting to reveal that, for a given stochastic model, the form of the corresponding equation for the fluctuating velocity components is not immediately determined. Indeed, if we consider for example the complete Langevin model, its formulation depends on whether one wants: (a) to disregard all two-way coupling effects (both volumetric and momentum), which leads to Eq.~\eqref{eq: fluctuating_velocity_seen_2}; (b) to account for the volumetric effect for the fluid phase, which leads to Eq.~\eqref{eq: fluctuating_velocity_seen_1}; (c) to include two-way coupling effects, in which case the model for the instantaneous fluid velocity seen, 
Eq.~\eqref{Minier et al 2004}+Eq.~\eqref{Complete Langevin: added term}, corresponds to Eq.~\eqref{eq: fluctuating_velocity_seen_2way} for the fluctuating component of the velocity of the fluid seen.

Going back to the difference between Eq.~\eqref{eq: fluctuating_velocity_seen_1} and Eq.~\eqref{eq: fluctuating_velocity_seen_2} (or between Eq.~\eqref{wrong equation us' Arcen Taniere} and Eq.~\eqref{right equation us' Arcen Taniere}), this means that an additional term, equal to $\lra{u_i\,u_k} 1/\alpha_f\nabla \alpha_p$ and proportional to the gradients of the volumetric fractions $\nabla \alpha_p$, is introduced in Eq.~\eqref{eq: fluctuating_velocity_seen_2}. Depending on the choice made for two-way coupling effects (see (a) or (b) above),
this (potentially) spurious term may not be negligible as particle distribution is rarely homogeneous. Consequently, numerical simulations which are performed from the stochastic differential equations for the fluctuating parts of the velocity of the fluid seen can be potentially flawed.

Clearly, formulations in terms of so-called `fluctuations' can easily become intricate with many terms, making their manipulation slippery while they do not necessarily clarify the physical picture. Thus, contrary to some statements~\cite{Pialat_2007,Arcen_2009}, a more practical way to account for the velocity of the fluid seen is to retain formulations in terms of the instantaneous velocity.

\subsubsection{Normalized Langevin models}
\label{Normalised Langevin models}

Another category of models consists in proposing Langevin-type of models developed in terms of the normalized fluctuating fluid velocity seen. To give one example, this normalized fluctuating velocity (using the notations introduced in section~\ref{Fluctuating or instantaneous fluid velocity seen}) is sometimes modeled as~\cite{Bocksell_2006,Dehbi_2008,Dehbi_2009}
\begin{equation}
d\left( \frac{u_i'}{\sigma_{(i)}}\right) = - \frac{u_i'}{\tau_L\, \sigma_{(i)}}\, dt +
\sqrt{\frac{2}{\tau_L}}\, dW_i 
+ \frac{\partial}{\partial x_k}\left( \frac{\lra{u_i'\,u_k'}}{\sigma_{(i)}}\right) 
\left( \frac{1}{1 + St} \right)\, dt
\end{equation}
Such formulations have been mostly used for boundary-layer simulations often in relation with particle deposition issues~\cite{Dehbi_2008,Dehbi_2010}. It is seen that these propositions include terms written as functions of the particle Stokes number, which with present notation is $St=\tau_p/\tau_L$. Such models are clearly able to predict particle statistics and, in that sense, they satisfy the criterion (P-2). However, when the Stokes number goes to zero (that is when particle inertia becomes negligible), these models revert to the normalized models which have been discussed in section~\ref{Fluctuating velocity versus instantaneous velocity}. As analyzed there, these normalized Langevin models do not have the correct transformation and invariance properties and are inconsistent with the structure of Reynolds-stress equations: they are, therefore, unsatisfactory descriptions for single-phase turbulent flows. This means that the criteria (P-3) and (P-1), as well as the two forms of the criterion (P-6), are not respected. Consequently, these models cannot be regarded as acceptable descriptions for two-phase flows.

\subsubsection{Hybrid DNS-stochastic approach}
\label{Hybrid DNS-stochastic approach}

As already mentioned above, a recent proposal introduced a new formulation for the velocity of the fluid seen~\cite{Arcen_2009}. Using the present notations, this proposal consists in simulating $\mb{U}_s$ as the solution of the following stochastic differential equation
\begin{multline}
\label{equation Arcen Taniere 2009}
dU_{s,i} = -\frac{1}{\rho}\frac{\partial \lra{P} }{\partial x_i}\, dt + \nu\, \Delta \lra{U_i}\, dt
+ \left( U_{p,k} - U_{s,k} \right)\frac{\partial \lra{U_{i}}}{\partial x_k}\, dt 
+ A'_i \, dt\\
+ G^{*}_{ik} \left( U_{s,k}-\lra{U_{k}} \right)dt + B_{ik} \, dW_k
\end{multline}
where a new term $A'_i$ is added to the drift vector in Eq.~(\ref{Simonin et al 1993}) and is given by 
\begin{equation}
\label{Arcen_added_drift}
A_i'=\frac{\partial \lra{u_{s,i}'u_{p,k}}_m}{\partial x_k}
- \frac{\partial \lra{u_iu_k}}{\partial x_k} \, .
\end{equation}
This model is expressed in terms of the fluctuating components of the velocity of the fluid seen and the discussions in section~\ref{Fluctuating or instantaneous fluid velocity seen} are thus relevant. In particular, the formulation put forward~\cite{Arcen_2009} is along the one expressed in Eq.~\eqref{wrong equation us' Arcen Taniere} to which the extra term $A'_i$ is added, which results in
\begin{equation}
\label{wrong equation us' Arcen 2009}
du_{s,i}'= \frac{\partial \lra{u_{s,i}'u_{p,k}}_m}{\partial x_k}\, dt 
+ \left( G_{ik}^{*} - \frac{\partial \lra{U_i}}{\partial x_k} \right) u_{s,k}'\, dt + B_{ik}dW_k~.
\end{equation}
In this formulation, the matrix $G_{ik}^{*}$ is expressed as the inverse of the matrix $\mc{T}_{ik}$ which is the matrix of the decorrelation time scales of the fluid seen and given by~\cite{Arcen_2009}
\begin{equation}
\label{matrix G Arcen Taniere}
G^{*}_{ik}=\widetilde{G_{ik}^{*}} + \frac{\partial \lra{U_i}}{\partial x_k} \, , \;
\widetilde{G_{ik}^{*}}=- \left( \mb{\mc{T}}\right)^{-1}_{ki} \, , \;
\mc{T}_{ik}=\int_0^{\infty} \lra{u_{s,i}'u_{s,l}'}_m^{-1} \lra{u_{s,l}'(0)u_{s,k}'(t)}_m \,dt
\end{equation}
where the values of the matrix $\mc{T}_{ki}$ are obtained as statistics extracted from a DNS of the same particle-laden flow that is considered. Then, the diffusion matrix is determined by the following equality~\cite{Arcen_2009}:
\begin{equation}
\label{diffusion B Arcen Taniere}
B^2_{ik}=B_{il}B_{kl}=- \widetilde{G^{*}_{il}}\lra{u_{s,l}' u_{s,k}'}_m
- \widetilde{G^{*}_{kl}}\lra{u_{s,l}' u_{s,i}'}_m \, .
\end{equation}
Note that the notation $B^2_{ik}$ which was used~\cite{Arcen_2009} should in fact refer to $(B B^{T})_{ik}$.
This proposal introduces new elements and, since DNS results are built-in, it is referred to as an `hybrid DNS-stochastic' approach. Some more comments are in order about the role of DNS in this hybrid approach. In this formulation, the matrices $\mc{T}_{ik}$ and $G^{*}_{ik}$ are not obtained first on some simple test cases (thus from some sample DNS) and then used as a predictive model in any geometry. Actually, for each geometry and for each flow considered, a DNS must be run beforehand in order to obtain the correlation matrix which is used to extract the matrix $G^{*}_{ik}$ fed into the governing equation for the velocity of the fluid seen, Eq.~\eqref{equation Arcen Taniere 2009}. This means that a concurrent run of a DNS and of the PDF model must be carried out.

Compared to the propositions addressed in the previous subsections, the additional term $A_i'$ in the drift vector is also a new element. This term is motivated by an analysis of the form of the exact equation satisfied by the drift velocity in the limit of high-inertia particles ($St\gg 1$ or $\tau_p \to \infty$), while the two expressions that yield the matrix $G^{*}_{ik}$ as well as the diffusion matrix $B_{ik}$ in Eqs.~(\ref{matrix G Arcen Taniere}) and (\ref{diffusion B Arcen Taniere}) respectively are based on a so-called `local homogeneity assumption'. In Eq.~(\ref{equation Arcen Taniere 2009}), it is seen that the mean viscous term is present as in the proposal in Eq.~(\ref{Simonin et al 1993}). However, as already indicated in section~\ref{First Langevin proposals}, this is inconsistent with the proper low Reynolds-number form of the Reynolds-stress equations and should be avoided, though it seems to be systematically presented~\cite{Taniere_2014}. Since we are only considering high Reynolds-number flows, this term is disregarded in the present analysis.

From a physical point of view, some of these assumptions raise questions. The local homogeneity assumption means that the underlying fluid turbulence is assumed to remain homogeneous and stationary so that the turbulent characteristics seen by particles (which are functions of the elapsed time) can be taken as stationary processes during a `long-enough' period (at least of the order of the particle relaxation timescale $\tau_p$) for these statistics to reach their local `equilibrium values'. For high-inertia particles which can typically cover large distances during a time of the order of their relaxation timescale, this amounts to assuming that the fluid turbulence remains homogeneous and stationary over considerable length and time scales, for the hypothesis to apply. This hybrid-DNS approach bears some similarities with one development proposed by Pope~\cite{Pope_2002}. However, that study~\cite{Pope_2002} was for homogeneous shear fluid flows where, once correctly rescaled, fluid particle velocities constitute truly statistically-stationary processes. Furthermore, its purpose was mainly to demonstrate the potential of properly-defined Langevin models, suggesting that there is still room for considerable improvement in the development of particle stochastic models~\cite{Pope_2002}, rather than as a 
general methodology to apply in any flows. On the other hand, the issue of whether consistency results are to be expected in the high-inertia limit is worth noting and will be taken up in the next subsection. In the context of the present analysis, we now concentrate on how this new proposal stands with respect to the criteria retained in section~\ref{Choice of criteria for two-phase flow models}. 

Since the drift vector and diffusion matrix are explicit in Eqs.~(\ref{equation Arcen Taniere 2009})-(\ref{diffusion B Arcen Taniere}), the stochastic model is complete. From the choice of the drift term, it is also seen that this proposal is such that the criterion (P-5) is satisfied and that the fluid limit yields a model whose form is acceptable. In the limit of small $St$, the additional term $A_i'$ in Eq.~\eqref{Arcen_added_drift} does not contribute to the first-order development in $St$ and, in that limit, the form of the drift vector is identical to the one given in the incomplete Langevin proposal in Eq.~\eqref{Simonin et al 1993}. Consequently, the analysis performed in section~\ref{First Langevin proposals} applies here, showing that (P-6) is not satified but that (P-6bis) is.
However, the main characteristic of this proposal is that the matrix entering the return-to-equilibrium term in the drift vector is provided by the solution of a DNS for the same flow which is modeled by the proposal in Eq~(\ref{equation Arcen Taniere 2009}). As such, it is clear that the criterion (P-2) is violated and that this proposal cannot be referred to as a `model'. Indeed, as indicated above, the formulation requires that a DNS solution be performed beforehand for each flow to feed the drift term of a model which is supposed to provide predictions on this very flow. If such formulations can indeed be considered as making interesting consistency checks for specific closure propositions, it cannot be 
accepted as a proper stochastic model for two-phase flow simulations. Furthermore, it appears that the present formulation suffers from some inconsistencies that limit its applicability. Indeed, if we consider one of the simplest situations, namely of homogeneous isotropic decaying turbulence, the present hybrid DNS-stochastic approach would consist in running first a DNS to obtain the timescales necessary to define the matrix 
$\widetilde{G^{*}_{ik}}$ in  Eq.~(\ref{matrix G Arcen Taniere}), from which the diffusion matrix is derived through Eq.~(\ref{diffusion B Arcen Taniere}). The latter equations is the expression of the classical fluctuation-dissipation theorem for stationary processes~\cite{Gardiner_1990,Ottinger_1996}, which means that we have $d\lra{\mb{u}_s^2}_m/dt=0$ instead of the natural decaying law for the turbulent kinetic energy. Thus, the criterion (P-4) is also violated. It is also evident that the proper fluid limit is not retrieved as the decay law for single-phase turbulent homogeneous flows is not correctly retrieved, showing that (P-3) is also violated. For an approach based on DNS results, this is a severe shortcoming. Furthermore, it can be noted that the procedure of coupling DNS and Langevin models raises consistency questions, especially in the fluid limit\cite{Chibbaro_2011}.

\subsection{Discussion on additional criteria for two-phase flow modeling}
\label{Open issues at stake in two-phase flow modeling}

In the list of criteria set forth in section~\ref{Choice of criteria for two-phase flow models}, it is seen that the fluid, or particle-tracer ($St \ll 1$), limit has been particularly emphasized (in the criterion (P-3)). It can be wondered whether similar consistency limits are to be selected in the other limit, when particle inertia is very high ($St \gg 1$). However, three remarks can be made. First, it must be remembered that the present form of Langevin models (i.e. modeling the velocity of the fluid seen by a diffusion stochastic process) has less justifications than in the fluid case. For high-inertia particles, the well-known frozen-turbulence hypothesis can be applied and it is easy to show that the increments of the velocity of the fluid seen should then be governed by spatial correlations. In that case, $\lra{(d\mb{U}_s)^2}$ scales as $\left(\lra{\epsilon} |\mb{U}_r| \Delta t\right)^{2/3}$ over a time interval  $\Delta t$, provided that the relative mean distance covered by particles $\Delta r=|\mb{U}_r| \Delta t$ remains in the inertial range length $\Delta r \ll L$. It remains to be seen whether such limits are encountered as they imply the existence of a regime where $|\mb{U}_r|$ is high enough so that Lagrangian fluid correlations can be neglected while we still have that 
$\Delta r=|\mb{U}_r| \Delta t \ll L$. If present, such limits are indeed not reproduced by present Langevin formulations as discussed at length in the construction of the complete Langevin model~\cite{Minier_2001}. This suggests that, in the high-inertia limit, the issue would be to devise altogether new stochastic model formulations rather than imposing requirements on the present ones. On the other hand, when dealing with high-inertia particles, a second remark is that the detailed form of a Langevin-type of model is less an issue if macroscopic particle statistical properties, such as diffusion coefficients, kinetic energies, etc., are sought. For these properties, the important element is mainly to retrieve the correct limit of the integral time scale of the velocity of the fluid seen in this limit. This is indeed what expressions such as Csanady's formulas in Eqs.~(\ref{Csanady timescales}) are doing without having to change the form of the Langevin equation itself. 
Finally, with respect to the particle relaxation timescale $\tau_p$ and since $St \gg 1$ means that 
$T_L^{*} \ll \tau_p$, the fluid velocity seen becomes a fast-variable and can be safely taken as acting as a white-noise on discrete particle trajectories. In that case, the form of the diffusion matrix in Langevin formulations is more important than details of the drift vector (apart from the return-to-equilibrium term and the integral time scale). Among other asymptotic cases, a comprehensive discussion of this high-inertia limit was
proposed~\cite{Peirano_2006} along with the corresponding consequences on the development of suitable numerical schemes~\cite{Peirano_2006}.

Thus, the development of new stochastic models that would capture both Lagrangian statistics (time-spectrum) as well as Eulerian ones (space-spectrum) is still an open issue. Yet, this question must be weighted against what would be gained from such developments. 

\subsection{Open issues on stochastic models for the velocity of the fluid seen}
\label{open issues for Us}

At the end of the analysis of stochastic models for polydisperse two-phase flows, it is worth emphasizing that modeling the velocity of the fluid seen remains an open issue. Coming up with tractable models that still represent proper descriptions of the physics of turbulent flows is not an easy task. Indeed, it must be remembered that the velocity of the fluid seen is $\mb{U}_s(t)=\mb{U}(t,\mb{x}_p(t))$ which means that, though $\mb{U}_s(t)$ is a particle-attached variable, it nevertheless involves time and space correlations of the carrier turbulent fluid flow. Thus, the crossing-trajectory effect (CTE)~\cite{Minier_2001}, which is related to particle and fluid velocity slips, induces several challenges for the formulation of a model for $\mb{U}_s(t)$.

With respect to this situation, the standpoint chosen in the present work is to consider \textit{jointly} a set of criteria that points to acceptable forms of the different terms entering a stochastic diffusion model retained for the velocity of the fluid seen. As the complexity of the physics involved can, unfortunately, lead to overlooking the basic properties of a model, the criteria (P-1) and (P-2) have been useful to clarify some situations. Then, based on proper formulations of the fluid limit and on the GLM (which is the essence of (P-3)), (P-4) is important for the closure of the diffusion coefficient, (P-5) helps to bring out a correct form of the return-to-equilibrium term of a Langevin model, while (P-6) indicates possible expressions for the additional term of the drift vector related to the slip velocity between the fluid and the particles. Yet, they remain suggestions and indications. It must not be forgotten that present models still contain assumptions. More precisely, the timescales of the velocity of the fluid seen are inputs (relying on the Csanady's expressions) and, to the best of the authors' knowledge, these expressions have not been worked out from first principles (see detailed discussions in sections 7.4 and 9.3.4 in~\citet{Minier_2001} or first attempts in~\cite{Pozorski_1998}). This has consequences. For example, (P-6) is helpful to reveal the existence of a supplementary drift term, which basically comes from a first-order development of the fluid velocity seen in the limit of small $St$. However, an important point is that it may not be relevant to push the analysis too far concentrating on that sole issue, when other aspects are disregarded. For example, arguing on whether the mean or instantaneous slip velocities should enter the additional drift term becomes secondary if the diffusion coefficient is not correctly closed and while it remains unclear whether the difference is accounted for (or not) by the Csanady's formulas. The main messages are therefore: first, it is essential to address together the different terms entering a stochastic model and, second, that new ideas based on more fundamental derivations would be of value.

\section{Conclusion}
\label{conclusion}

In this paper, a new approach which consists in selecting a set of criteria has been described for Lagrangian stochastic models. Gathering present knowledge into a comprehensive set of guidelines has been shown to be useful both to assess whether existing models satisfy basic properties and also to help future developments.
For that purpose, criteria have been put forward for Lagrangian stochastic models in single-phase (see (F-1) to (F-3) in section~\ref{Choice of criteria for single-phase flow models}) and in two-phase flows (see (P-1) to (P-6) in section~\ref{Choice of criteria for two-phase flow models}). In the single-phase flow situation, the emphasis is basically put on requiring that stochastic models be fully consistent with the high Reynolds-number structure of the Reynolds-stress equations. In the two-phase flow situation, as the available information is 
different, the criteria insist on respecting the particle-tracer limit, as well as basic properties of turbulence and what models should stand for. Although these criteria are somewhat different in the single- and two-phase cases, the approach followed in the present paper represents an attempt at addressing stochastic models devised for each situation from a unified standpoint. 

It is believed that the present list of criteria is made up by simple and physically-meaningful requirements that must be met by Lagrangian stochastic models. Yet, from the analyses which are summarized in Table~\ref{table_single-phase-models} for single-phase models and in Table~\ref{table_two-phase-models} for two-phase models, first conclusions can be drawn. For single-phase flows, it was brought out that present normalized Langevin models are flawed since they do not respect convective transport in general non-homogeneous flows, as shown in section~\ref{Normalised velocity}. For two-phase flows, the situation is more confused and, in that respect, the analysis carried out in section~\ref{Analysis of different formulations in two-phase flows} is helpful to clarify the modeling picture. It appears that, among those considered there, only one formulation is acceptable, in the sense that all the criteria chosen in section~\ref{Choice of criteria for two-phase flow models} are met. Not surprisingly, this model remains the only two-phase PDF model which has been validated in engineering configurations~\cite{Peirano_2006}. This is clearly a very poor and unsatisfactory state and much work remains to be done. In that sense, another conclusion is that a safe approach is to formulate models in terms of instantaneous fluid particle velocities, as for the GLM in single-phase flows. 

It is hoped that the guidelines put forward here will lead to improved Lagrangian stochastic models. For example, it appears that the complete Langevin model in two-phase flows is based on the SLM and, therefore, one option could be to devise extensions to obtain the counterpart for the velocity of the fluid seen of the GLM for the velocity of fluid particles. Furthermore, the present set of criteria is not meant as a definitive choice and could be extended in further works. For instance, low-Reynolds, Brownian effects or specific aspects of particle preferential concentration effects could be considered through the formulation of new criteria to guide model assessment in these limits.

Another important aim of the present approach is to provide guidelines for future model developments and in related subjects. In particular, the present analysis has been carried out in the framework of RANS approaches where a complete formalism is available. Yet, the methodology followed here, as well as the conclusions which have been reached, have direct implications for particle-laden turbulent flows simulated with a LES (Large-Eddy Simulation) method. Indeed, in the LES approach, a specific model is needed to account for the unresolved part of the fluid velocity seen by discrete particles~\cite{Kuerten_2006,Fede_2006} which corresponds to the fluctuating velocity in the RANS context. Among the few models developed so far, most are written in terms of this unresolved part of the fluid velocity seen~\cite{Fede_2006,Michalek_2012} but some formulations~\cite{Fede_2006} include terms similar to the ones appearing on the rhs of Eq.~(\ref{model_u_2}) while others retain only the simplest form of a Langevin equation~\cite{Michalek_2012}. Furthermore, some proposals disregard effects due to particle inertia or the CTE effect and write an equation with a diffusion coefficient as for fluid particles~\cite{Fede_2006} whereas others assume that the unresolved part is at equilibrium to close the diffusion coefficient of the Langevin model~\cite{Michalek_2012}. On the other hand, it is interesting to note that, for engineering applications, another proposal~\cite{Berrouk_2007,Berrouk_2008} was formulated in terms of the instantaneous fluid velocity seen with a model that is an extension to the LES context of the complete Langevin model discussed in section~\ref{Complete Langevin models} with corresponding expressions of the Csanady's time scales and diffusion terms but without the mean slip term in the drift vector that is essential to the criterion (P-6). Clearly, there is some uncertainty and the issues addressed in this paper concerning the formulation of Lagrangian stochastic models (instantaneous versus fluctuating or unresolved fluid velocity components, closures of the drift and diffusion terms in the two-phase flow case, etc.) are relevant for the question on how to express subgrid effects to simulate two-phase flows with the LES method. However, as the statistical operator involved is different (a spatial filtering is applied instead of a mathematically well-defined probabilistic expectation), a rigorous formalism is needed to address these issues. The situation is more advanced for single-phase flows where the FDF (Filtered Density Function) formalism has been been developed~\cite{Gicquel_2002,Sheikhi_2003}. First steps have been proposed in the two-phase case~\cite{Chibbaro_2011b} but additional work is needed to clarify the formulations of Lagrangian stochastic models in the LES approach to particle-laden turbulent flows. 

\begin{acknowledgments}
The contributions of S. B. Pope to this work are supported by the U.S. Department of Energy, Office of Science, Office of Basic Energy Sciences under Award Number DE-FG02-90 ER14128.
\end{acknowledgments}

\newpage

\begin{table}
\begin{ruledtabular}
\begin{tabular}{|c||l|l|l|}
 & (F-1) & (F-2) & (F-3) \\
\hline
\hline
Generalised Langevin model~\footnote{as analysed in section~\ref{Analysis of different formulations}} 
& $\checkmark$  & $\checkmark$ & $\checkmark$  \\
\hline
Incomplete Langevin model for fluctuating velocities~\footnote{as analysed in 
section~\ref{Fluctuating velocity versus instantaneous velocity} where the term
labelled (a) in Eq.~(\ref{model_u_2}) is missing}
& $\checkmark$ & \textsf{X} & $\checkmark$  \\
\hline
Incomplete Langevin model for fluctuating velocities~\footnote{as analysed in 
section~\ref{Fluctuating velocity versus instantaneous velocity} where the term
labelled (b) in Eq.~(\ref{model_u_2}) is missing}
& $\checkmark$ & $\checkmark$ & \textsf{X}  \\
\hline
Complete Langevin model for fluctuating velocities~\footnote{as analysed in 
section~\ref{Fluctuating velocity versus instantaneous velocity} where all the terms on the rhs
of Eq.~(\ref{model_u_2}) are present} 
& $\checkmark$ & $\checkmark$ & $\checkmark$  \\
\hline
Normalised Langevin model~\footnote{as analysed in 
section~\ref{Normalised velocity}}  
& \textsf{X} & $\checkmark$ & \textsf{X}  \\
\hline
\end{tabular}
\end{ruledtabular}
\caption{\label{table_single-phase-models} Summary of the properties of the stochastic models for 
single-phase flows considered in section~\ref{Stochastic models for single-phase flows}
with respect to the criteria listed in section~\ref{Choice of criteria for single-phase flow models}.}
\end{table}

\vspace*{1cm}

\begin{table}
\begin{ruledtabular}
\begin{tabular}{|c||l|l|l|l|l|l|c|}
 &(P-1)&(P-2)&(P-3)&(P-4)&(P- 5)&(P-6)&(P-6bis) \\
\hline
\hline
First Langevin model~\footnote{as analysed in section~\ref{First Langevin proposals}
(where the sign (-) indicates that the criterion cannot be checked)}
& $\checkmark$  & \textsf{X} & - & - & $\checkmark$ & \textsf{X} & $\checkmark$ \\
\hline
Complete Langevin model (2001)~\footnote{as analysed in section~\ref{Complete Langevin models}} 
& $\checkmark$  & $\checkmark$ & $\checkmark$ & $\checkmark$ & \textsf{X} & $\checkmark$ & $\checkmark$ \\
\hline
Complete Langevin model (2004)~\footnote{as analysed in section~\ref{Complete Langevin models}}  
& $\checkmark$ & $\checkmark$ & $\checkmark$ & $\checkmark$ & $\checkmark$ & $\checkmark$ & $\checkmark$\\
\hline
Normalised Langevin model~\footnote{as analysed in section~\ref{Normalised Langevin models}} 
& \textsf{X} & $\checkmark$  & \textsf{X} & $\checkmark$ & $\checkmark$ & \textsf{X} & \textsf{X}\\
\hline
Hybrid-DNS model~\footnote{as analysed in section~\ref{Hybrid DNS-stochastic approach}}
& $\checkmark$ &  \textsf{X} & \textsf{X} & \textsf{X} & $\checkmark$ & \textsf{X} & $\checkmark$\\
\end{tabular}
\end{ruledtabular}
\caption{\label{table_two-phase-models} Summary of the properties of the stochastic models for 
two-phase flows considered in section~\ref{Stochastic models for two-phase flows}
with respect to the criteria listed in section~\ref{Choice of criteria for two-phase flow models}.}
\end{table}

\vspace*{1cm}



\begin{thebibliography}{80}%
\makeatletter
\providecommand \@ifxundefined [1]{%
 \@ifx{#1\undefined}
}%
\providecommand \@ifnum [1]{%
 \ifnum #1\expandafter \@firstoftwo
 \else \expandafter \@secondoftwo
 \fi
}%
\providecommand \@ifx [1]{%
 \ifx #1\expandafter \@firstoftwo
 \else \expandafter \@secondoftwo
 \fi
}%
\providecommand \natexlab [1]{#1}%
\providecommand \enquote  [1]{``#1''}%
\providecommand \bibnamefont  [1]{#1}%
\providecommand \bibfnamefont [1]{#1}%
\providecommand \citenamefont [1]{#1}%
\providecommand \href@noop [0]{\@secondoftwo}%
\providecommand \href [0]{\begingroup \@sanitize@url \@href}%
\providecommand \@href[1]{\@@startlink{#1}\@@href}%
\providecommand \@@href[1]{\endgroup#1\@@endlink}%
\providecommand \@sanitize@url [0]{\catcode `\\12\catcode `\$12\catcode
  `\&12\catcode `\#12\catcode `\^12\catcode `\_12\catcode `\%12\relax}%
\providecommand \@@startlink[1]{}%
\providecommand \@@endlink[0]{}%
\providecommand \url  [0]{\begingroup\@sanitize@url \@url }%
\providecommand \@url [1]{\endgroup\@href {#1}{\urlprefix }}%
\providecommand \urlprefix  [0]{URL }%
\providecommand \Eprint [0]{\href }%
\providecommand \doibase [0]{http://dx.doi.org/}%
\providecommand \selectlanguage [0]{\@gobble}%
\providecommand \bibinfo  [0]{\@secondoftwo}%
\providecommand \bibfield  [0]{\@secondoftwo}%
\providecommand \translation [1]{[#1]}%
\providecommand \BibitemOpen [0]{}%
\providecommand \bibitemStop [0]{}%
\providecommand \bibitemNoStop [0]{.\EOS\space}%
\providecommand \EOS [0]{\spacefactor3000\relax}%
\providecommand \BibitemShut  [1]{\csname bibitem#1\endcsname}%
\let\auto@bib@innerbib\@empty
\bibitem [{\citenamefont {Pope}(1985)}]{Pope_1985}%
  \BibitemOpen
  \bibfield  {author} {\bibinfo {author} {\bibfnamefont {S.~B.}\ \bibnamefont
  {Pope}},\ }\bibfield  {title} {\enquote {\bibinfo {title} {Pdf methods for
  turbulent reactive flows},}\ }\href@noop {} {\bibfield  {journal} {\bibinfo
  {journal} {Prog. Energy Combust. Sci.}\ }\textbf {\bibinfo {volume} {11}},\
  \bibinfo {pages} {119--192} (\bibinfo {year} {1985})}\BibitemShut {NoStop}%
\bibitem [{\citenamefont {Pope}(1994{\natexlab{a}})}]{Pope_1994}%
  \BibitemOpen
  \bibfield  {author} {\bibinfo {author} {\bibfnamefont {S.~B.}\ \bibnamefont
  {Pope}},\ }\bibfield  {title} {\enquote {\bibinfo {title} {Lagrangian pdf
  methods for turbulent reactive flows},}\ }\href@noop {} {\bibfield  {journal}
  {\bibinfo  {journal} {Ann. Rev. Fluid Mech.}\ }\textbf {\bibinfo {volume}
  {26}},\ \bibinfo {pages} {23--63} (\bibinfo {year}
  {1994}{\natexlab{a}})}\BibitemShut {NoStop}%
\bibitem [{\citenamefont {Pope}(2000)}]{Pope_2000}%
  \BibitemOpen
  \bibfield  {author} {\bibinfo {author} {\bibfnamefont {S.~B.}\ \bibnamefont
  {Pope}},\ }\href@noop {} {\emph {\bibinfo {title} {Turbulent Flows}}}\
  (\bibinfo  {publisher} {Cambridge University Press},\ \bibinfo {year}
  {2000})\BibitemShut {NoStop}%
\bibitem [{\citenamefont {Minier}\ and\ \citenamefont
  {Peirano}(2001)}]{Minier_2001}%
  \BibitemOpen
  \bibfield  {author} {\bibinfo {author} {\bibfnamefont {J.-P.}\ \bibnamefont
  {Minier}}\ and\ \bibinfo {author} {\bibfnamefont {E.}~\bibnamefont
  {Peirano}},\ }\bibfield  {title} {\enquote {\bibinfo {title} {The {PDF}
  approach to turbulent and polydispersed two-phase flows},}\ }\href@noop {}
  {\bibfield  {journal} {\bibinfo  {journal} {Phys. Rep.}\ }\textbf {\bibinfo
  {volume} {352}},\ \bibinfo {pages} {1--214} (\bibinfo {year}
  {2001})}\BibitemShut {NoStop}%
\bibitem [{\citenamefont {Fox}(2003)}]{Fox_2003}%
  \BibitemOpen
  \bibfield  {author} {\bibinfo {author} {\bibfnamefont {R.~O.}\ \bibnamefont
  {Fox}},\ }\href@noop {} {\emph {\bibinfo {title} {Computational models for
  turbulent reacting flows}}}\ (\bibinfo  {publisher} {Cambridge University
  Press},\ \bibinfo {year} {2003})\BibitemShut {NoStop}%
\bibitem [{\citenamefont {Peirano}\ \emph {et~al.}(2006)\citenamefont
  {Peirano}, \citenamefont {Chibbaro}, \citenamefont {Pozorski},\ and\
  \citenamefont {Minier}}]{Peirano_2006}%
  \BibitemOpen
  \bibfield  {author} {\bibinfo {author} {\bibfnamefont {E.}~\bibnamefont
  {Peirano}}, \bibinfo {author} {\bibfnamefont {S.}~\bibnamefont {Chibbaro}},
  \bibinfo {author} {\bibfnamefont {J.}~\bibnamefont {Pozorski}}, \ and\
  \bibinfo {author} {\bibfnamefont {J.-P.}\ \bibnamefont {Minier}},\ }\bibfield
   {title} {\enquote {\bibinfo {title} {Mean-field/{PDF} numerical approach for
  polydispersed turbulent two-phase flows},}\ }\href@noop {} {\bibfield
  {journal} {\bibinfo  {journal} {Prog. Energy Combust. Sci.}\ }\textbf
  {\bibinfo {volume} {32}},\ \bibinfo {pages} {315--371} (\bibinfo {year}
  {2006})}\BibitemShut {NoStop}%
\bibitem [{\citenamefont {Haworth}(2010)}]{Haworth_2010}%
  \BibitemOpen
  \bibfield  {author} {\bibinfo {author} {\bibfnamefont {D.~C.}\ \bibnamefont
  {Haworth}},\ }\bibfield  {title} {\enquote {\bibinfo {title} {Progress in
  probability density function methods for turbulent reacting flows},}\
  }\href@noop {} {\bibfield  {journal} {\bibinfo  {journal} {Prog. Energy
  Combust. Sci.}\ }\textbf {\bibinfo {volume} {36}},\ \bibinfo {pages}
  {168--259} (\bibinfo {year} {2010})}\BibitemShut {NoStop}%
\bibitem [{\citenamefont {Jenny}, \citenamefont {Roekaerts},\ and\
  \citenamefont {Beishuizen}(2012)}]{Jenny_2012}%
  \BibitemOpen
  \bibfield  {author} {\bibinfo {author} {\bibfnamefont {J.}~\bibnamefont
  {Jenny}}, \bibinfo {author} {\bibfnamefont {D.}~\bibnamefont {Roekaerts}}, \
  and\ \bibinfo {author} {\bibfnamefont {N.}~\bibnamefont {Beishuizen}},\
  }\bibfield  {title} {\enquote {\bibinfo {title} {Modeling of turbulent dilute
  spray combustion},}\ }\href@noop {} {\bibfield  {journal} {\bibinfo
  {journal} {Prog. Energy Combust. Sci.}\ }\textbf {\bibinfo {volume} {38}},\
  \bibinfo {pages} {846--887} (\bibinfo {year} {2012})}\BibitemShut {NoStop}%
\bibitem [{\citenamefont {Pope}\ and\ \citenamefont {Chen}(1990)}]{Pope_1990}%
  \BibitemOpen
  \bibfield  {author} {\bibinfo {author} {\bibfnamefont {S.~B.}\ \bibnamefont
  {Pope}}\ and\ \bibinfo {author} {\bibfnamefont {Y.~L.}\ \bibnamefont
  {Chen}},\ }\bibfield  {title} {\enquote {\bibinfo {title} {The
  velocity-dissipation probability density function model for turbulent
  flows},}\ }\href@noop {} {\bibfield  {journal} {\bibinfo  {journal} {Phys.
  Fluids A}\ }\textbf {\bibinfo {volume} {2}},\ \bibinfo {pages} {1437}
  (\bibinfo {year} {1990})}\BibitemShut {NoStop}%
\bibitem [{\citenamefont {Pozorski}\ and\ \citenamefont
  {Minier}(1998{\natexlab{a}})}]{Pozorski_1999}%
  \BibitemOpen
  \bibfield  {author} {\bibinfo {author} {\bibfnamefont {J.}~\bibnamefont
  {Pozorski}}\ and\ \bibinfo {author} {\bibfnamefont {J.-P.}\ \bibnamefont
  {Minier}},\ }\bibfield  {title} {\enquote {\bibinfo {title} {Probability
  density function modelling of dispersed two-phase turbulent flows},}\
  }\href@noop {} {\bibfield  {journal} {\bibinfo  {journal} {Phys. Rev. E}\
  }\textbf {\bibinfo {volume} {59}},\ \bibinfo {pages} {855--863} (\bibinfo
  {year} {1998}{\natexlab{a}})}\BibitemShut {NoStop}%
\bibitem [{\citenamefont {Prosperetti}(2004)}]{Prosperetti_2004}%
  \BibitemOpen
  \bibfield  {author} {\bibinfo {author} {\bibfnamefont {A.}~\bibnamefont
  {Prosperetti}},\ }\bibfield  {title} {\enquote {\bibinfo {title} {Bubbles},}\
  }\href@noop {} {\bibfield  {journal} {\bibinfo  {journal} {Phys. Fluids}\
  }\textbf {\bibinfo {volume} {16}},\ \bibinfo {pages} {1852--1865} (\bibinfo
  {year} {2004})}\BibitemShut {NoStop}%
\bibitem [{\citenamefont {Gatignol}(1983)}]{Gatignol_1983}%
  \BibitemOpen
  \bibfield  {author} {\bibinfo {author} {\bibfnamefont {R.}~\bibnamefont
  {Gatignol}},\ }\bibfield  {title} {\enquote {\bibinfo {title} {The
  {F}ax{\'e}n formulae for a rigid particle in an unsteady non-uniform {S}tokes
  flow},}\ }\href@noop {} {\bibfield  {journal} {\bibinfo  {journal} {J. Mec.
  Theor. Appl.}\ }\textbf {\bibinfo {volume} {1}},\ \bibinfo {pages} {143--160}
  (\bibinfo {year} {1983})}\BibitemShut {NoStop}%
\bibitem [{\citenamefont {Maxey}\ and\ \citenamefont
  {Riley}(1983)}]{Maxey_1983}%
  \BibitemOpen
  \bibfield  {author} {\bibinfo {author} {\bibfnamefont {M.~R.}\ \bibnamefont
  {Maxey}}\ and\ \bibinfo {author} {\bibfnamefont {J.~J.}\ \bibnamefont
  {Riley}},\ }\bibfield  {title} {\enquote {\bibinfo {title} {Equation of
  motion for a small rigid sphere in a nonuniform flow},}\ }\href@noop {}
  {\bibfield  {journal} {\bibinfo  {journal} {Phys. Fluids}\ }\textbf {\bibinfo
  {volume} {26}},\ \bibinfo {pages} {883--889} (\bibinfo {year}
  {1983})}\BibitemShut {NoStop}%
\bibitem [{\citenamefont {Clift}, \citenamefont {Grace},\ and\ \citenamefont
  {Weber}(1978)}]{Clift_1978}%
  \BibitemOpen
  \bibfield  {author} {\bibinfo {author} {\bibfnamefont {R.}~\bibnamefont
  {Clift}}, \bibinfo {author} {\bibfnamefont {J.~R.}\ \bibnamefont {Grace}}, \
  and\ \bibinfo {author} {\bibfnamefont {M.~E.}\ \bibnamefont {Weber}},\
  }\href@noop {} {\emph {\bibinfo {title} {Bubbles, Drops and Particles}}}\
  (\bibinfo  {publisher} {Academic Press. New York},\ \bibinfo {year}
  {1978})\BibitemShut {NoStop}%
\bibitem [{\citenamefont {Toschi}\ and\ \citenamefont
  {Bodenschatz}(2009)}]{Toschi_2009}%
  \BibitemOpen
  \bibfield  {author} {\bibinfo {author} {\bibfnamefont {F.}~\bibnamefont
  {Toschi}}\ and\ \bibinfo {author} {\bibfnamefont {E.}~\bibnamefont
  {Bodenschatz}},\ }\bibfield  {title} {\enquote {\bibinfo {title} {Lagrangian
  properties of particles in turbulence},}\ }\href@noop {} {\bibfield
  {journal} {\bibinfo  {journal} {Ann. Rev. Fluid Mech.}\ }\textbf {\bibinfo
  {volume} {41}},\ \bibinfo {pages} {375--404} (\bibinfo {year}
  {2009})}\BibitemShut {NoStop}%
\bibitem [{\citenamefont {Marchioli}\ and\ \citenamefont
  {Soldati}(2002)}]{Marchioli_2002}%
  \BibitemOpen
  \bibfield  {author} {\bibinfo {author} {\bibfnamefont {C.}~\bibnamefont
  {Marchioli}}\ and\ \bibinfo {author} {\bibfnamefont {A.}~\bibnamefont
  {Soldati}},\ }\bibfield  {title} {\enquote {\bibinfo {title} {Mechanisms for
  particle transfer and segregation in a turbulent boundary layer},}\
  }\href@noop {} {\bibfield  {journal} {\bibinfo  {journal} {J. Fluid Mech.}\
  }\textbf {\bibinfo {volume} {468}},\ \bibinfo {pages} {283--315} (\bibinfo
  {year} {2002})}\BibitemShut {NoStop}%
\bibitem [{\citenamefont {Gualtieri}, \citenamefont {Picano},\ and\
  \citenamefont {Casciola}(2009)}]{Gualtieri_2009}%
  \BibitemOpen
  \bibfield  {author} {\bibinfo {author} {\bibfnamefont {P.}~\bibnamefont
  {Gualtieri}}, \bibinfo {author} {\bibfnamefont {F.}~\bibnamefont {Picano}}, \
  and\ \bibinfo {author} {\bibfnamefont {C.~M.}\ \bibnamefont {Casciola}},\
  }\bibfield  {title} {\enquote {\bibinfo {title} {Anisotropic clustering of
  inertial particles in homogeneous shear flow},}\ }\href@noop {} {\bibfield
  {journal} {\bibinfo  {journal} {J. Fluid Mech.}\ }\textbf {\bibinfo {volume}
  {629}},\ \bibinfo {pages} {25--39} (\bibinfo {year} {2009})}\BibitemShut
  {NoStop}%
\bibitem [{\citenamefont {Balachandar}\ and\ \citenamefont
  {Eaton}(2010)}]{balachandar_2010}%
  \BibitemOpen
  \bibfield  {author} {\bibinfo {author} {\bibfnamefont {S.}~\bibnamefont
  {Balachandar}}\ and\ \bibinfo {author} {\bibfnamefont {J.~K.}\ \bibnamefont
  {Eaton}},\ }\bibfield  {title} {\enquote {\bibinfo {title} {Turbulent
  dispersed multiphase flow},}\ }\href@noop {} {\bibfield  {journal} {\bibinfo
  {journal} {Ann. Rev. Fluid Mech.}\ }\textbf {\bibinfo {volume} {42}},\
  \bibinfo {pages} {111--133} (\bibinfo {year} {2010})}\BibitemShut {NoStop}%
\bibitem [{\citenamefont {Simonin}, \citenamefont {Deutsch},\ and\
  \citenamefont {Minier}(1993)}]{Simonin_1993}%
  \BibitemOpen
  \bibfield  {author} {\bibinfo {author} {\bibfnamefont {O.}~\bibnamefont
  {Simonin}}, \bibinfo {author} {\bibfnamefont {E.}~\bibnamefont {Deutsch}}, \
  and\ \bibinfo {author} {\bibfnamefont {J.-P.}\ \bibnamefont {Minier}},\
  }\bibfield  {title} {\enquote {\bibinfo {title} {Eulerian prediction of the
  fluid/particle correlated motion in turbulent two-phase flows},}\ }\href@noop
  {} {\bibfield  {journal} {\bibinfo  {journal} {Appl. Sci. Res.}\ }\textbf
  {\bibinfo {volume} {51}},\ \bibinfo {pages} {275--283} (\bibinfo {year}
  {1993})}\BibitemShut {NoStop}%
\bibitem [{\citenamefont {Pope}(1994{\natexlab{b}})}]{Pope_1994a}%
  \BibitemOpen
  \bibfield  {author} {\bibinfo {author} {\bibfnamefont {S.~B.}\ \bibnamefont
  {Pope}},\ }\bibfield  {title} {\enquote {\bibinfo {title} {On the
  relationship between stochastic {L}agrangian models of turbulence and
  second-order closures},}\ }\href@noop {} {\bibfield  {journal} {\bibinfo
  {journal} {Phys. Fluids}\ }\textbf {\bibinfo {volume} {6}},\ \bibinfo {pages}
  {973--985} (\bibinfo {year} {1994}{\natexlab{b}})}\BibitemShut {NoStop}%
\bibitem [{\citenamefont {Peirano}\ and\ \citenamefont
  {Minier}(2002)}]{Peirano_2002}%
  \BibitemOpen
  \bibfield  {author} {\bibinfo {author} {\bibfnamefont {E.}~\bibnamefont
  {Peirano}}\ and\ \bibinfo {author} {\bibfnamefont {J.-P.}\ \bibnamefont
  {Minier}},\ }\bibfield  {title} {\enquote {\bibinfo {title} {A probabilistic
  formalism and hierarchy of models for polydispersed turbulent two-phase
  flows},}\ }\href@noop {} {\bibfield  {journal} {\bibinfo  {journal} {Phys.
  Rev. E}\ }\textbf {\bibinfo {volume} {65}} (\bibinfo {year}
  {2002})}\BibitemShut {NoStop}%
\bibitem [{\citenamefont {Haworth}\ and\ \citenamefont
  {Pope}(1986)}]{Haworth_1986}%
  \BibitemOpen
  \bibfield  {author} {\bibinfo {author} {\bibfnamefont {D.~C.}\ \bibnamefont
  {Haworth}}\ and\ \bibinfo {author} {\bibfnamefont {S.~B.}\ \bibnamefont
  {Pope}},\ }\bibfield  {title} {\enquote {\bibinfo {title} {A generalized
  {L}angevin model for turbulent flows},}\ }\href@noop {} {\bibfield  {journal}
  {\bibinfo  {journal} {Phys. Fluids}\ }\textbf {\bibinfo {volume} {30}},\
  \bibinfo {pages} {387} (\bibinfo {year} {1986})}\BibitemShut {NoStop}%
\bibitem [{\citenamefont {Minier}, \citenamefont {Peirano},\ and\ \citenamefont
  {Chibbaro}(2004)}]{Minier_2004}%
  \BibitemOpen
  \bibfield  {author} {\bibinfo {author} {\bibfnamefont {J.-P.}\ \bibnamefont
  {Minier}}, \bibinfo {author} {\bibfnamefont {E.}~\bibnamefont {Peirano}}, \
  and\ \bibinfo {author} {\bibfnamefont {S.}~\bibnamefont {Chibbaro}},\
  }\bibfield  {title} {\enquote {\bibinfo {title} {Pdf model based on
  {L}angevin equation for polydispersed two-phase flows applied to a bluff-body
  gas-solid flow},}\ }\href@noop {} {\bibfield  {journal} {\bibinfo  {journal}
  {Phys. Fluids}\ }\textbf {\bibinfo {volume} {16}},\ \bibinfo {pages} {2419}
  (\bibinfo {year} {2004})}\BibitemShut {NoStop}%
\bibitem [{\citenamefont {Henry}, \citenamefont {Minier},\ and\ \citenamefont
  {Lef\`{e}vre}(2012)}]{Henry_2012b}%
  \BibitemOpen
  \bibfield  {author} {\bibinfo {author} {\bibfnamefont {C.}~\bibnamefont
  {Henry}}, \bibinfo {author} {\bibfnamefont {J.-P.}\ \bibnamefont {Minier}}, \
  and\ \bibinfo {author} {\bibfnamefont {G.}~\bibnamefont {Lef\`{e}vre}},\
  }\bibfield  {title} {\enquote {\bibinfo {title} {Towards a description of
  particulate fouling: from single-particle deposition to clogging},}\
  }\href@noop {} {\bibfield  {journal} {\bibinfo  {journal} {Adv. Colloid
  Interface Sci.}\ }\textbf {\bibinfo {volume} {185-186}},\ \bibinfo {pages}
  {34--76} (\bibinfo {year} {2012})}\BibitemShut {NoStop}%
\bibitem [{\citenamefont {Newman}\ and\ \citenamefont
  {McGuffin}(2005)}]{Newman_2005}%
  \BibitemOpen
  \bibfield  {author} {\bibinfo {author} {\bibfnamefont {C.~I.~D.}\
  \bibnamefont {Newman}}\ and\ \bibinfo {author} {\bibfnamefont {V.~L.}\
  \bibnamefont {McGuffin}},\ }\bibfield  {title} {\enquote {\bibinfo {title}
  {Stochastic simulation of reactive separations in capillary
  electrophoresis},}\ }\href@noop {} {\bibfield  {journal} {\bibinfo  {journal}
  {Electrophoresis}\ }\textbf {\bibinfo {volume} {26}},\ \bibinfo {pages}
  {537--547} (\bibinfo {year} {2005})}\BibitemShut {NoStop}%
\bibitem [{\citenamefont {Lo~Iacono}\ and\ \citenamefont
  {Reynolds}(2005)}]{LoIacono_2005}%
  \BibitemOpen
  \bibfield  {author} {\bibinfo {author} {\bibfnamefont {G.}~\bibnamefont
  {Lo~Iacono}}\ and\ \bibinfo {author} {\bibfnamefont {A.~M.}\ \bibnamefont
  {Reynolds}},\ }\bibfield  {title} {\enquote {\bibinfo {title} {A {L}agrangian
  stochastic model for the dispersion and deposition of {B}rownian particles in
  the presence of a temperature gradient},}\ }\href@noop {} {\bibfield
  {journal} {\bibinfo  {journal} {J. Aerosol Sci.}\ }\textbf {\bibinfo {volume}
  {36}},\ \bibinfo {pages} {1238--1250} (\bibinfo {year} {2005})}\BibitemShut
  {NoStop}%
\bibitem [{\citenamefont {Irannejad}\ and\ \citenamefont
  {Jaberi}(2013)}]{Irannejad_2013}%
  \BibitemOpen
  \bibfield  {author} {\bibinfo {author} {\bibfnamefont {A.}~\bibnamefont
  {Irannejad}}\ and\ \bibinfo {author} {\bibfnamefont {F.}~\bibnamefont
  {Jaberi}},\ }\bibfield  {title} {\enquote {\bibinfo {title} {Large eddy
  simulation of evaporating spray with a stochastic breakup model},}\
  }\href@noop {} {\bibfield  {journal} {\bibinfo  {journal} {Training}\
  }\textbf {\bibinfo {volume} {2013}},\ \bibinfo {pages} {09--30} (\bibinfo
  {year} {2013})}\BibitemShut {NoStop}%
\bibitem [{\citenamefont {Mueller}, \citenamefont {Iaccarino},\ and\
  \citenamefont {Pitsch}(2013)}]{Mueller_2012}%
  \BibitemOpen
  \bibfield  {author} {\bibinfo {author} {\bibfnamefont {M.~E.}\ \bibnamefont
  {Mueller}}, \bibinfo {author} {\bibfnamefont {G.}~\bibnamefont {Iaccarino}},
  \ and\ \bibinfo {author} {\bibfnamefont {H.}~\bibnamefont {Pitsch}},\
  }\bibfield  {title} {\enquote {\bibinfo {title} {Chemical kinetic uncertainty
  quantification for large eddy simulation of turbulent nonpremixed
  combustion},}\ }\href@noop {} {\bibfield  {journal} {\bibinfo  {journal} {P.
  Combust. Inst.}\ }\textbf {\bibinfo {volume} {34}},\ \bibinfo {pages}
  {1299--1306} (\bibinfo {year} {2013})}\BibitemShut {NoStop}%
\bibitem [{\citenamefont {Tani\`{e}re}\ and\ \citenamefont
  {Arcen}(2014)}]{Taniere_2014}%
  \BibitemOpen
  \bibfield  {author} {\bibinfo {author} {\bibfnamefont {A.}~\bibnamefont
  {Tani\`{e}re}}\ and\ \bibinfo {author} {\bibfnamefont {B.}~\bibnamefont
  {Arcen}},\ }\bibfield  {title} {\enquote {\bibinfo {title} {Prediction of a
  particle-laden turbulent channel flow: examination of two classes of
  stochastic dispersion models},}\ }\href@noop {} {\bibfield  {journal}
  {\bibinfo  {journal} {Int. J. Multiphase Flow}\ }\textbf {\bibinfo {volume}
  {60}},\ \bibinfo {pages} {1--10} (\bibinfo {year} {2014})}\BibitemShut
  {NoStop}%
\bibitem [{\citenamefont {Gardiner}(1990)}]{Gardiner_1990}%
  \BibitemOpen
  \bibfield  {author} {\bibinfo {author} {\bibfnamefont {C.~W.}\ \bibnamefont
  {Gardiner}},\ }\href@noop {} {\emph {\bibinfo {title} {Handbook of Stochastic
  Methods for Physics, Chemistry and the Natural Sciences}}}\ (\bibinfo
  {publisher} {Springer},\ \bibinfo {year} {1990})\BibitemShut {NoStop}%
\bibitem [{\citenamefont {\"{O}ttinger}(1996)}]{Ottinger_1996}%
  \BibitemOpen
  \bibfield  {author} {\bibinfo {author} {\bibfnamefont {H.~C.}\ \bibnamefont
  {\"{O}ttinger}},\ }\href@noop {} {\emph {\bibinfo {title} {Stochastic
  {P}rocesses in {P}olymeric {F}luids. Tools and Examples for Developing
  Simulation Algorithms}}}\ (\bibinfo  {publisher} {Springer, Berlin},\
  \bibinfo {year} {1996})\BibitemShut {NoStop}%
\bibitem [{\citenamefont {Sawford}(1986)}]{Sawford_1986}%
  \BibitemOpen
  \bibfield  {author} {\bibinfo {author} {\bibfnamefont {B.}~\bibnamefont
  {Sawford}},\ }\bibfield  {title} {\enquote {\bibinfo {title} {Generalized
  random forcing in random-walk turbulent dispersion models},}\ }\href@noop {}
  {\bibfield  {journal} {\bibinfo  {journal} {Phys. Fluids}\ }\textbf {\bibinfo
  {volume} {29}},\ \bibinfo {pages} {3582} (\bibinfo {year}
  {1986})}\BibitemShut {NoStop}%
\bibitem [{\citenamefont {Pope}(1987)}]{Pope_1987}%
  \BibitemOpen
  \bibfield  {author} {\bibinfo {author} {\bibfnamefont {S.~B.}\ \bibnamefont
  {Pope}},\ }\bibfield  {title} {\enquote {\bibinfo {title} {Consistency
  conditions for random-walk models of turbulent dispersion},}\ }\href@noop {}
  {\bibfield  {journal} {\bibinfo  {journal} {Phys. Fluids}\ }\textbf {\bibinfo
  {volume} {30}},\ \bibinfo {pages} {2374--2378} (\bibinfo {year}
  {1987})}\BibitemShut {NoStop}%
\bibitem [{\citenamefont {Thomson}(1987)}]{Thomson_1987}%
  \BibitemOpen
  \bibfield  {author} {\bibinfo {author} {\bibfnamefont {D.~J.}\ \bibnamefont
  {Thomson}},\ }\bibfield  {title} {\enquote {\bibinfo {title} {Criteria for
  the selection of stochastic models of particle trajectories in turbulent
  flows},}\ }\href@noop {} {\bibfield  {journal} {\bibinfo  {journal} {J. Fluid
  Mech.}\ }\textbf {\bibinfo {volume} {180}},\ \bibinfo {pages} {529--556}
  (\bibinfo {year} {1987})}\BibitemShut {NoStop}%
\bibitem [{\citenamefont {McInnes}\ and\ \citenamefont
  {Bracco}(1992)}]{McInnes_1992}%
  \BibitemOpen
  \bibfield  {author} {\bibinfo {author} {\bibfnamefont {J.~M.}\ \bibnamefont
  {McInnes}}\ and\ \bibinfo {author} {\bibfnamefont {F.~V.}\ \bibnamefont
  {Bracco}},\ }\bibfield  {title} {\enquote {\bibinfo {title} {Stochastic
  particle dispersion modeling and the tracer-particle limit},}\ }\href@noop {}
  {\bibfield  {journal} {\bibinfo  {journal} {Phys. Fluids A}\ }\textbf
  {\bibinfo {volume} {4}},\ \bibinfo {pages} {2809} (\bibinfo {year}
  {1992})}\BibitemShut {NoStop}%
\bibitem [{\citenamefont {Chibbaro}\ and\ \citenamefont
  {Minier}(2014)}]{Chibbaro-Minier_2014}%
  \BibitemOpen
  \bibfield  {author} {\bibinfo {author} {\bibfnamefont {S.}~\bibnamefont
  {Chibbaro}}\ and\ \bibinfo {author} {\bibfnamefont {J.-P.}\ \bibnamefont
  {Minier}},\ }\href@noop {} {\emph {\bibinfo {title} {Stochastic Methods for
  Fluid Mechanics}}}\ (\bibinfo  {publisher} {CISM, International Centre for
  Mechanical Sciences, Vol. 548, Springer Verlag, Berlin},\ \bibinfo {year}
  {2014})\BibitemShut {NoStop}%
\bibitem [{\citenamefont {Pope}(1991)}]{Pope_1991}%
  \BibitemOpen
  \bibfield  {author} {\bibinfo {author} {\bibfnamefont {S.~B.}\ \bibnamefont
  {Pope}},\ }\bibfield  {title} {\enquote {\bibinfo {title} {Application of the
  velocity-dissipation probability density function model to inhomogeneous
  turbulent flows},}\ }\href@noop {} {\bibfield  {journal} {\bibinfo  {journal}
  {Phys. Fluids A}\ }\textbf {\bibinfo {volume} {3}},\ \bibinfo {pages} {1947}
  (\bibinfo {year} {1991})}\BibitemShut {NoStop}%
\bibitem [{\citenamefont {Talay}(1995)}]{Talay_1995}%
  \BibitemOpen
  \bibfield  {author} {\bibinfo {author} {\bibfnamefont {D.}~\bibnamefont
  {Talay}},\ }\bibfield  {title} {\enquote {\bibinfo {title} {Simulation of
  stochastic differential systems},}\ }in\ \href@noop {} {\emph {\bibinfo
  {booktitle} {Probabilistic methods in applied physics}}}\ (\bibinfo
  {publisher} {Springer},\ \bibinfo {year} {1995})\ pp.\ \bibinfo {pages}
  {54--96}\BibitemShut {NoStop}%
\bibitem [{\citenamefont {{\O}ksendal}(2003)}]{Oksendal_2003}%
  \BibitemOpen
  \bibfield  {author} {\bibinfo {author} {\bibfnamefont {B.}~\bibnamefont
  {{\O}ksendal}},\ }\href@noop {} {\emph {\bibinfo {title} {Stochastic
  differential equations}}}\ (\bibinfo  {publisher} {Springer},\ \bibinfo
  {year} {2003})\BibitemShut {NoStop}%
\bibitem [{\citenamefont {McKean}(1969)}]{Mckean_1969}%
  \BibitemOpen
  \bibfield  {author} {\bibinfo {author} {\bibfnamefont {H.~P.}\ \bibnamefont
  {McKean}},\ }\href@noop {} {\emph {\bibinfo {title} {Stochastic
  integrals}}},\ Vol.\ \bibinfo {volume} {353}\ (\bibinfo  {publisher}
  {American Mathematical Soc.},\ \bibinfo {year} {1969})\BibitemShut {NoStop}%
\bibitem [{\citenamefont {Dreeben}\ and\ \citenamefont {Pope}(1997)}]{Pop_97}%
  \BibitemOpen
  \bibfield  {author} {\bibinfo {author} {\bibfnamefont {T.~D.}\ \bibnamefont
  {Dreeben}}\ and\ \bibinfo {author} {\bibfnamefont {S.~B.}\ \bibnamefont
  {Pope}},\ }\bibfield  {title} {\enquote {\bibinfo {title} {Probability
  density function and reynolds-stress modeling of new near-wall turbulent
  flows},}\ }\href@noop {} {\bibfield  {journal} {\bibinfo  {journal} {Phys.
  Fluids}\ }\textbf {\bibinfo {volume} {9}},\ \bibinfo {pages} {154} (\bibinfo
  {year} {1997})}\BibitemShut {NoStop}%
\bibitem [{\citenamefont {Dreeben}\ and\ \citenamefont
  {Pope}(1998)}]{Dreeben_1998}%
  \BibitemOpen
  \bibfield  {author} {\bibinfo {author} {\bibfnamefont {T.~D.}\ \bibnamefont
  {Dreeben}}\ and\ \bibinfo {author} {\bibfnamefont {S.~B.}\ \bibnamefont
  {Pope}},\ }\bibfield  {title} {\enquote {\bibinfo {title} {Probability
  density function/{M}onte {C}arlo simulation of near-wall turbulent flows},}\
  }\href@noop {} {\bibfield  {journal} {\bibinfo  {journal} {J. Fluid Mech.}\
  }\textbf {\bibinfo {volume} {357}},\ \bibinfo {pages} {141} (\bibinfo {year}
  {1998})}\BibitemShut {NoStop}%
\bibitem [{\citenamefont {Wac\l{}awczyk}, \citenamefont {Pozorski},\ and\
  \citenamefont {Minier}(2004)}]{Waclawczyk_2004}%
  \BibitemOpen
  \bibfield  {author} {\bibinfo {author} {\bibfnamefont {M.}~\bibnamefont
  {Wac\l{}awczyk}}, \bibinfo {author} {\bibfnamefont {J.}~\bibnamefont
  {Pozorski}}, \ and\ \bibinfo {author} {\bibfnamefont {J.-P.}\ \bibnamefont
  {Minier}},\ }\bibfield  {title} {\enquote {\bibinfo {title} {Probability
  density function computation of turbulent flows with a new near-wall
  model},}\ }\href@noop {} {\bibfield  {journal} {\bibinfo  {journal} {Phys.
  Fluids}\ }\textbf {\bibinfo {volume} {16}},\ \bibinfo {pages} {1410--1422}
  (\bibinfo {year} {2004})}\BibitemShut {NoStop}%
\bibitem [{\citenamefont {Simonin}(2000)}]{Simonin_2000}%
  \BibitemOpen
  \bibfield  {author} {\bibinfo {author} {\bibfnamefont {O.}~\bibnamefont
  {Simonin}},\ }\bibfield  {title} {\enquote {\bibinfo {title} {Statistical and
  continuum modelling of turbulent reactive particulate flows},}\ }in\
  \href@noop {} {\emph {\bibinfo {booktitle} {Lecture Series 2000-06}}}\
  (\bibinfo  {publisher} {Von Karman Institute for Fluid Dynamics},\ \bibinfo
  {year} {2000})\BibitemShut {NoStop}%
\bibitem [{\citenamefont {Ferry}\ and\ \citenamefont
  {Balachandar}(2001)}]{Ferry_2001}%
  \BibitemOpen
  \bibfield  {author} {\bibinfo {author} {\bibfnamefont {J.}~\bibnamefont
  {Ferry}}\ and\ \bibinfo {author} {\bibfnamefont {S.}~\bibnamefont
  {Balachandar}},\ }\bibfield  {title} {\enquote {\bibinfo {title} {A fast
  eulerian model ofr two-phase flow},}\ }\href@noop {} {\bibfield  {journal}
  {\bibinfo  {journal} {Int. J. Multiphase Flow}\ }\textbf {\bibinfo {volume}
  {27}},\ \bibinfo {pages} {199--226} (\bibinfo {year} {2001})}\BibitemShut
  {NoStop}%
\bibitem [{\citenamefont {Chibbaro}\ and\ \citenamefont
  {Minier}(2011{\natexlab{a}})}]{Chibbaro_2011}%
  \BibitemOpen
  \bibfield  {author} {\bibinfo {author} {\bibfnamefont {S.}~\bibnamefont
  {Chibbaro}}\ and\ \bibinfo {author} {\bibfnamefont {J.-P.}\ \bibnamefont
  {Minier}},\ }\bibfield  {title} {\enquote {\bibinfo {title} {A note on the
  consistency of hybrid {E}ulerian/{L}agrangian approach to multiphase
  flows},}\ }\href@noop {} {\bibfield  {journal} {\bibinfo  {journal} {Int. J.
  Multiphase Flow}\ }\textbf {\bibinfo {volume} {37}},\ \bibinfo {pages}
  {293--297} (\bibinfo {year} {2011}{\natexlab{a}})}\BibitemShut {NoStop}%
\bibitem [{\citenamefont {Pope}(2011)}]{Pope_2011}%
  \BibitemOpen
  \bibfield  {author} {\bibinfo {author} {\bibfnamefont {S.~B.}\ \bibnamefont
  {Pope}},\ }\bibfield  {title} {\enquote {\bibinfo {title} {Simple models of
  turbulent flowsa)},}\ }\href@noop {} {\bibfield  {journal} {\bibinfo
  {journal} {Phys. Fluids}\ }\textbf {\bibinfo {volume} {23}},\ \bibinfo
  {pages} {011301} (\bibinfo {year} {2011})}\BibitemShut {NoStop}%
\bibitem [{\citenamefont {Meneveau}(2011)}]{Meneveau_2011}%
  \BibitemOpen
  \bibfield  {author} {\bibinfo {author} {\bibfnamefont {C.}~\bibnamefont
  {Meneveau}},\ }\bibfield  {title} {\enquote {\bibinfo {title} {Lagrangian
  dynamics and models of the velocity gradient tensor in turbulent flows},}\
  }\href@noop {} {\bibfield  {journal} {\bibinfo  {journal} {Ann. Rev. Fluid
  Mech.}\ }\textbf {\bibinfo {volume} {43}},\ \bibinfo {pages} {219--245}
  (\bibinfo {year} {2011})}\BibitemShut {NoStop}%
\bibitem [{\citenamefont {Guingo}\ and\ \citenamefont
  {Minier}(2008)}]{Guingo_2008}%
  \BibitemOpen
  \bibfield  {author} {\bibinfo {author} {\bibfnamefont {M.}~\bibnamefont
  {Guingo}}\ and\ \bibinfo {author} {\bibfnamefont {J.-P.}\ \bibnamefont
  {Minier}},\ }\bibfield  {title} {\enquote {\bibinfo {title} {A stochastic
  model of coherent structures for particle deposition in turbulent flows},}\
  }\href@noop {} {\bibfield  {journal} {\bibinfo  {journal} {Phys. Fluids}\
  }\textbf {\bibinfo {volume} {20}},\ \bibinfo {pages} {053303} (\bibinfo
  {year} {2008})}\BibitemShut {NoStop}%
\bibitem [{\citenamefont {Berlemont}, \citenamefont {Desjonqueres},\ and\
  \citenamefont {Gouesbet}(1990)}]{Berlemont_1990}%
  \BibitemOpen
  \bibfield  {author} {\bibinfo {author} {\bibfnamefont {A.}~\bibnamefont
  {Berlemont}}, \bibinfo {author} {\bibfnamefont {P.}~\bibnamefont
  {Desjonqueres}}, \ and\ \bibinfo {author} {\bibfnamefont {G.}~\bibnamefont
  {Gouesbet}},\ }\bibfield  {title} {\enquote {\bibinfo {title} {Particle
  {L}agrangian simulation in turbulent flows},}\ }\href@noop {} {\bibfield
  {journal} {\bibinfo  {journal} {Int. J. Multiphase Flow}\ }\textbf {\bibinfo
  {volume} {16}},\ \bibinfo {pages} {19--34} (\bibinfo {year}
  {1990})}\BibitemShut {NoStop}%
\bibitem [{\citenamefont {Gouesbet}\ and\ \citenamefont
  {Berlemont}(1999)}]{Gouesbet_1999}%
  \BibitemOpen
  \bibfield  {author} {\bibinfo {author} {\bibfnamefont {G.}~\bibnamefont
  {Gouesbet}}\ and\ \bibinfo {author} {\bibfnamefont {A.}~\bibnamefont
  {Berlemont}},\ }\bibfield  {title} {\enquote {\bibinfo {title} {Eulerian and
  {L}agrangian approaches for predicting the behaviour of discrete particles in
  turbulent flows},}\ }\href@noop {} {\bibfield  {journal} {\bibinfo  {journal}
  {Prog. Energy Combust. Sci.}\ }\textbf {\bibinfo {volume} {25}},\ \bibinfo
  {pages} {133--159} (\bibinfo {year} {1999})}\BibitemShut {NoStop}%
\bibitem [{\citenamefont {Matida}\ \emph {et~al.}(2004)\citenamefont {Matida},
  \citenamefont {Finlay}, \citenamefont {Lange},\ and\ \citenamefont
  {Grgic}}]{Matida_2004}%
  \BibitemOpen
  \bibfield  {author} {\bibinfo {author} {\bibfnamefont {E.~A.}\ \bibnamefont
  {Matida}}, \bibinfo {author} {\bibfnamefont {W.~H.}\ \bibnamefont {Finlay}},
  \bibinfo {author} {\bibfnamefont {C.~F.}\ \bibnamefont {Lange}}, \ and\
  \bibinfo {author} {\bibfnamefont {B.}~\bibnamefont {Grgic}},\ }\bibfield
  {title} {\enquote {\bibinfo {title} {Improved numerical simulation of aerosol
  deposition in an idealized mouth-throat},}\ }\href@noop {} {\bibfield
  {journal} {\bibinfo  {journal} {J. Aerosol Sci.}\ }\textbf {\bibinfo {volume}
  {35}},\ \bibinfo {pages} {1--19} (\bibinfo {year} {2004})}\BibitemShut
  {NoStop}%
\bibitem [{\citenamefont {Dehbi}(2008)}]{Dehbi_2008}%
  \BibitemOpen
  \bibfield  {author} {\bibinfo {author} {\bibfnamefont {A.}~\bibnamefont
  {Dehbi}},\ }\bibfield  {title} {\enquote {\bibinfo {title} {Turbulent
  particle dispersion in arbitrary wall-bounded geometries: a coupled cfd
  {L}angevin equation based approach},}\ }\href@noop {} {\bibfield  {journal}
  {\bibinfo  {journal} {Int. J. Multiphase Flow}\ }\textbf {\bibinfo {volume}
  {34}},\ \bibinfo {pages} {819--828} (\bibinfo {year} {2008})}\BibitemShut
  {NoStop}%
\bibitem [{\citenamefont {Dehbi}(2010)}]{Dehbi_2010}%
  \BibitemOpen
  \bibfield  {author} {\bibinfo {author} {\bibfnamefont {A.}~\bibnamefont
  {Dehbi}},\ }\bibfield  {title} {\enquote {\bibinfo {title} {Validation
  against dns statistics of the normalized {L}angevin model for particle
  transport in turbulent channel flows},}\ }\href@noop {} {\bibfield  {journal}
  {\bibinfo  {journal} {Powder Technol.}\ }\textbf {\bibinfo {volume} {200}},\
  \bibinfo {pages} {60--68} (\bibinfo {year} {2010})}\BibitemShut {NoStop}%
\bibitem [{\citenamefont {Mito}\ and\ \citenamefont
  {Hanratty}(2002)}]{Mito_2002}%
  \BibitemOpen
  \bibfield  {author} {\bibinfo {author} {\bibfnamefont {Y.}~\bibnamefont
  {Mito}}\ and\ \bibinfo {author} {\bibfnamefont {T.~J.}\ \bibnamefont
  {Hanratty}},\ }\bibfield  {title} {\enquote {\bibinfo {title} {Use of a
  modified {L}angevin equation to describe turbulent dispersion of fluid
  particles in a channel flow},}\ }\href@noop {} {\bibfield  {journal}
  {\bibinfo  {journal} {Flow Turbul. Combust.}\ }\textbf {\bibinfo {volume}
  {68}},\ \bibinfo {pages} {1--26} (\bibinfo {year} {2002})}\BibitemShut
  {NoStop}%
\bibitem [{\citenamefont {Iliopoulos}, \citenamefont {Mito},\ and\
  \citenamefont {Hanratty}(2003)}]{Iliopoulos_2003}%
  \BibitemOpen
  \bibfield  {author} {\bibinfo {author} {\bibfnamefont {I.}~\bibnamefont
  {Iliopoulos}}, \bibinfo {author} {\bibfnamefont {Y.}~\bibnamefont {Mito}}, \
  and\ \bibinfo {author} {\bibfnamefont {T.~J.}\ \bibnamefont {Hanratty}},\
  }\bibfield  {title} {\enquote {\bibinfo {title} {A stochastic model for solid
  particle dispersion in a nonhomogeneous turbulent field},}\ }\href@noop {}
  {\bibfield  {journal} {\bibinfo  {journal} {Int. J. Multiphase Flow}\
  }\textbf {\bibinfo {volume} {29}},\ \bibinfo {pages} {375--394} (\bibinfo
  {year} {2003})}\BibitemShut {NoStop}%
\bibitem [{\citenamefont {Iliopoulos}, \citenamefont {Mito},\ and\
  \citenamefont {Hanratty}(2004)}]{Iliopoulos_2004}%
  \BibitemOpen
  \bibfield  {author} {\bibinfo {author} {\bibfnamefont {I.}~\bibnamefont
  {Iliopoulos}}, \bibinfo {author} {\bibfnamefont {Y.}~\bibnamefont {Mito}}, \
  and\ \bibinfo {author} {\bibfnamefont {T.~J.}\ \bibnamefont {Hanratty}},\
  }\bibfield  {title} {\enquote {\bibinfo {title} {A non-{G}aussian stochastic
  model to describe passive tracer dispersion and its comparison to a direct
  numerical simulation},}\ }\href@noop {} {\bibfield  {journal} {\bibinfo
  {journal} {Phys. Fluids}\ }\textbf {\bibinfo {volume} {16}},\ \bibinfo
  {pages} {3006--3030} (\bibinfo {year} {2004})}\BibitemShut {NoStop}%
\bibitem [{\citenamefont {Minier}\ and\ \citenamefont
  {Pozorski}(1999)}]{Minier_1999}%
  \BibitemOpen
  \bibfield  {author} {\bibinfo {author} {\bibfnamefont {J.-P.}\ \bibnamefont
  {Minier}}\ and\ \bibinfo {author} {\bibfnamefont {J.}~\bibnamefont
  {Pozorski}},\ }\bibfield  {title} {\enquote {\bibinfo {title} {Wall boundary
  conditions in the pdf method and application to a turbulent channel flow},}\
  }\href@noop {} {\bibfield  {journal} {\bibinfo  {journal} {Phys. Fluids}\
  }\textbf {\bibinfo {volume} {11}},\ \bibinfo {pages} {2632--2644} (\bibinfo
  {year} {1999})}\BibitemShut {NoStop}%
\bibitem [{\citenamefont {Tian}\ and\ \citenamefont
  {Ahmadi}(2007)}]{Tian_2007}%
  \BibitemOpen
  \bibfield  {author} {\bibinfo {author} {\bibfnamefont {L.}~\bibnamefont
  {Tian}}\ and\ \bibinfo {author} {\bibfnamefont {G.}~\bibnamefont {Ahmadi}},\
  }\bibfield  {title} {\enquote {\bibinfo {title} {Particle deposition in
  turbulent duct flows-comparisons of different model predictions},}\
  }\href@noop {} {\bibfield  {journal} {\bibinfo  {journal} {J. Aerosol Sci.}\
  }\textbf {\bibinfo {volume} {38}},\ \bibinfo {pages} {377--397} (\bibinfo
  {year} {2007})}\BibitemShut {NoStop}%
\bibitem [{\citenamefont {Parker}, \citenamefont {Foat},\ and\ \citenamefont
  {Preston}(2008)}]{Parker_2008}%
  \BibitemOpen
  \bibfield  {author} {\bibinfo {author} {\bibfnamefont {S.}~\bibnamefont
  {Parker}}, \bibinfo {author} {\bibfnamefont {T.}~\bibnamefont {Foat}}, \ and\
  \bibinfo {author} {\bibfnamefont {S.}~\bibnamefont {Preston}},\ }\bibfield
  {title} {\enquote {\bibinfo {title} {Towards quantitative prediction of
  aerosol deposition from turbulent flows},}\ }\href@noop {} {\bibfield
  {journal} {\bibinfo  {journal} {J. Aerosol Sci.}\ }\textbf {\bibinfo {volume}
  {39}},\ \bibinfo {pages} {99--112} (\bibinfo {year} {2008})}\BibitemShut
  {NoStop}%
\bibitem [{\citenamefont {Sasic}\ and\ \citenamefont
  {Almstedt}(2010)}]{Sasic_2010}%
  \BibitemOpen
  \bibfield  {author} {\bibinfo {author} {\bibfnamefont {S.}~\bibnamefont
  {Sasic}}\ and\ \bibinfo {author} {\bibfnamefont {A.-E.}\ \bibnamefont
  {Almstedt}},\ }\bibfield  {title} {\enquote {\bibinfo {title} {Dynamics of
  fobres in a turbulent flow field - a particle-level simulation technique},}\
  }\href@noop {} {\bibfield  {journal} {\bibinfo  {journal} {Int. J. Multiphase
  Flow}\ }\textbf {\bibinfo {volume} {31}},\ \bibinfo {pages} {1058--1064}
  (\bibinfo {year} {2010})}\BibitemShut {NoStop}%
\bibitem [{\citenamefont {Wilson}, \citenamefont {Thurtell},\ and\
  \citenamefont {Kidd}(1981)}]{Wilson_1981}%
  \BibitemOpen
  \bibfield  {author} {\bibinfo {author} {\bibfnamefont {J.~D.}\ \bibnamefont
  {Wilson}}, \bibinfo {author} {\bibfnamefont {G.~W.}\ \bibnamefont
  {Thurtell}}, \ and\ \bibinfo {author} {\bibfnamefont {G.~E.}\ \bibnamefont
  {Kidd}},\ }\bibfield  {title} {\enquote {\bibinfo {title} {Numerical
  simulation of particle trajectories in inhomogeneous turbulence: systems with
  variable tubulent velocity scale},}\ }\href@noop {} {\bibfield  {journal}
  {\bibinfo  {journal} {Boundary-Layer Meteorol.}\ }\textbf {\bibinfo {volume}
  {21}},\ \bibinfo {pages} {423--441} (\bibinfo {year} {1981})}\BibitemShut
  {NoStop}%
\bibitem [{\citenamefont {Bocksell}\ and\ \citenamefont
  {Loth}(2006)}]{Bocksell_2006}%
  \BibitemOpen
  \bibfield  {author} {\bibinfo {author} {\bibfnamefont {T.~L.}\ \bibnamefont
  {Bocksell}}\ and\ \bibinfo {author} {\bibfnamefont {E.}~\bibnamefont
  {Loth}},\ }\bibfield  {title} {\enquote {\bibinfo {title} {Stochastic
  modeling of particle diffusion in a turbulent boundary layer},}\ }\href@noop
  {} {\bibfield  {journal} {\bibinfo  {journal} {Int. J. Multiphase Flow}\
  }\textbf {\bibinfo {volume} {32}},\ \bibinfo {pages} {1234--1253} (\bibinfo
  {year} {2006})}\BibitemShut {NoStop}%
\bibitem [{\citenamefont {Rogers}, \citenamefont {Mansour},\ and\ \citenamefont
  {Reynolds}(1989)}]{Rogers_1989}%
  \BibitemOpen
  \bibfield  {author} {\bibinfo {author} {\bibfnamefont {M.~M.}\ \bibnamefont
  {Rogers}}, \bibinfo {author} {\bibfnamefont {N.~N.}\ \bibnamefont {Mansour}},
  \ and\ \bibinfo {author} {\bibfnamefont {W.~C.}\ \bibnamefont {Reynolds}},\
  }\bibfield  {title} {\enquote {\bibinfo {title} {An algebrix model for the
  turbulent flux of a passive scalar},}\ }\href@noop {} {\bibfield  {journal}
  {\bibinfo  {journal} {J. Fluid Mech.}\ }\textbf {\bibinfo {volume} {203}},\
  \bibinfo {pages} {77--101} (\bibinfo {year} {1989})}\BibitemShut {NoStop}%
\bibitem [{\citenamefont {Warhaft}(2000)}]{Warhaft_2000}%
  \BibitemOpen
  \bibfield  {author} {\bibinfo {author} {\bibfnamefont {Z.}~\bibnamefont
  {Warhaft}},\ }\bibfield  {title} {\enquote {\bibinfo {title} {Passive scalars
  in turbulent flows},}\ }\href@noop {} {\bibfield  {journal} {\bibinfo
  {journal} {Ann. Rev. Fluid Mech.}\ }\textbf {\bibinfo {volume} {32}},\
  \bibinfo {pages} {203--240} (\bibinfo {year} {2000})}\BibitemShut {NoStop}%
\bibitem [{\citenamefont {Pope}(1998)}]{Pope_1998}%
  \BibitemOpen
  \bibfield  {author} {\bibinfo {author} {\bibfnamefont {S.~B.}\ \bibnamefont
  {Pope}},\ }\bibfield  {title} {\enquote {\bibinfo {title} {The vanishing
  effect of molecular diffusivity on turbulent dispersion: implications for
  turbulent mixing and the scalar flux},}\ }\href@noop {} {\bibfield  {journal}
  {\bibinfo  {journal} {J. Fluid Mech.}\ }\textbf {\bibinfo {volume} {359}},\
  \bibinfo {pages} {299} (\bibinfo {year} {1998})}\BibitemShut {NoStop}%
\bibitem [{\citenamefont {Pialat}, \citenamefont {Simonin},\ and\ \citenamefont
  {Villedieu}(2007)}]{Pialat_2007}%
  \BibitemOpen
  \bibfield  {author} {\bibinfo {author} {\bibfnamefont {X.}~\bibnamefont
  {Pialat}}, \bibinfo {author} {\bibfnamefont {O.}~\bibnamefont {Simonin}}, \
  and\ \bibinfo {author} {\bibfnamefont {P.}~\bibnamefont {Villedieu}},\
  }\bibfield  {title} {\enquote {\bibinfo {title} {A hybrid
  {E}ulerian-{L}agrangian method to simulate the dispersed phase in turbulent
  gas-particle flows},}\ }\href@noop {} {\bibfield  {journal} {\bibinfo
  {journal} {Int. J. Multiphase Flow}\ }\textbf {\bibinfo {volume} {33}},\
  \bibinfo {pages} {766--788} (\bibinfo {year} {2007})}\BibitemShut {NoStop}%
\bibitem [{\citenamefont {Arcen}\ and\ \citenamefont
  {Tani\`{e}re}(2009)}]{Arcen_2009}%
  \BibitemOpen
  \bibfield  {author} {\bibinfo {author} {\bibfnamefont {B.}~\bibnamefont
  {Arcen}}\ and\ \bibinfo {author} {\bibfnamefont {A.}~\bibnamefont
  {Tani\`{e}re}},\ }\bibfield  {title} {\enquote {\bibinfo {title} {Simulation
  of a particle-laden turbulent channel flow using an improved stochastic
  {L}agrangian model},}\ }\href@noop {} {\bibfield  {journal} {\bibinfo
  {journal} {Phys. Fluids}\ }\textbf {\bibinfo {volume} {21}},\ \bibinfo
  {pages} {043303} (\bibinfo {year} {2009})}\BibitemShut {NoStop}%
\bibitem [{\citenamefont {Simonin}(2001)}]{Simonin_2001}%
  \BibitemOpen
  \bibfield  {author} {\bibinfo {author} {\bibfnamefont {O.}~\bibnamefont
  {Simonin}},\ }\href@noop {} {} (\bibinfo {year} {2001}),\ \bibinfo {note}
  {private communication}\BibitemShut {NoStop}%
\bibitem [{\citenamefont {Dehbi}(2009)}]{Dehbi_2009}%
  \BibitemOpen
  \bibfield  {author} {\bibinfo {author} {\bibfnamefont {A.}~\bibnamefont
  {Dehbi}},\ }\bibfield  {title} {\enquote {\bibinfo {title} {A stochastic
  {L}angevin model of turbulent particle dispersion in the presence of
  thermophoresis},}\ }\href@noop {} {\bibfield  {journal} {\bibinfo  {journal}
  {Int. J. Multiphase Flow}\ }\textbf {\bibinfo {volume} {35}},\ \bibinfo
  {pages} {219--226} (\bibinfo {year} {2009})}\BibitemShut {NoStop}%
\bibitem [{\citenamefont {Pope}(2002)}]{Pope_2002}%
  \BibitemOpen
  \bibfield  {author} {\bibinfo {author} {\bibfnamefont {S.~B.}\ \bibnamefont
  {Pope}},\ }\bibfield  {title} {\enquote {\bibinfo {title} {Stochastic
  {L}agrangian models of velocity in homogeneous turbulent shear flow},}\
  }\href@noop {} {\bibfield  {journal} {\bibinfo  {journal} {Phys. Fluids}\
  }\textbf {\bibinfo {volume} {14}},\ \bibinfo {pages} {1696--1702} (\bibinfo
  {year} {2002})}\BibitemShut {NoStop}%
\bibitem [{\citenamefont {Pozorski}\ and\ \citenamefont
  {Minier}(1998{\natexlab{b}})}]{Pozorski_1998}%
  \BibitemOpen
  \bibfield  {author} {\bibinfo {author} {\bibfnamefont {J.}~\bibnamefont
  {Pozorski}}\ and\ \bibinfo {author} {\bibfnamefont {J.-P.}\ \bibnamefont
  {Minier}},\ }\bibfield  {title} {\enquote {\bibinfo {title} {On the
  {L}agrangian turbulent dispersion models based on the {L}angevin equation},}\
  }\href@noop {} {\bibfield  {journal} {\bibinfo  {journal} {Int. J. Multiphase
  Flow}\ }\textbf {\bibinfo {volume} {24}},\ \bibinfo {pages} {913--945}
  (\bibinfo {year} {1998}{\natexlab{b}})}\BibitemShut {NoStop}%
\bibitem [{\citenamefont {Kuerten}(2006)}]{Kuerten_2006}%
  \BibitemOpen
  \bibfield  {author} {\bibinfo {author} {\bibfnamefont {J.~G.~M.}\
  \bibnamefont {Kuerten}},\ }\bibfield  {title} {\enquote {\bibinfo {title}
  {Subgrid modeling in particle-laden channel flow},}\ }\href@noop {}
  {\bibfield  {journal} {\bibinfo  {journal} {Phys. Fluids}\ }\textbf {\bibinfo
  {volume} {18}},\ \bibinfo {pages} {025108} (\bibinfo {year}
  {2006})}\BibitemShut {NoStop}%
\bibitem [{\citenamefont {Fede}\ \emph {et~al.}(2006)\citenamefont {Fede},
  \citenamefont {Simonin}, \citenamefont {Villedieu},\ and\ \citenamefont
  {Squires}}]{Fede_2006}%
  \BibitemOpen
  \bibfield  {author} {\bibinfo {author} {\bibfnamefont {P.}~\bibnamefont
  {Fede}}, \bibinfo {author} {\bibfnamefont {O.}~\bibnamefont {Simonin}},
  \bibinfo {author} {\bibfnamefont {P.}~\bibnamefont {Villedieu}}, \ and\
  \bibinfo {author} {\bibfnamefont {K.}~\bibnamefont {Squires}},\ }\bibfield
  {title} {\enquote {\bibinfo {title} {Stochastic modeling of the turbulent
  subgrid fluid velocity along inertial particle trajectories},}\ }in\
  \href@noop {} {\emph {\bibinfo {booktitle} {Proceedings of Summer Program
  2006}}}\ (\bibinfo  {publisher} {Center for Turbulence Research},\ \bibinfo
  {year} {2006})\ pp.\ \bibinfo {pages} {247--258}\BibitemShut {NoStop}%
\bibitem [{\citenamefont {Michalek}\ \emph {et~al.}(2012)\citenamefont
  {Michalek}, \citenamefont {Kuerten}, \citenamefont {Zeegers}, \citenamefont
  {Liew}, \citenamefont {Pozorski},\ and\ \citenamefont
  {Geurts}}]{Michalek_2012}%
  \BibitemOpen
  \bibfield  {author} {\bibinfo {author} {\bibfnamefont {W.~R.}\ \bibnamefont
  {Michalek}}, \bibinfo {author} {\bibfnamefont {J.~G.~M.}\ \bibnamefont
  {Kuerten}}, \bibinfo {author} {\bibfnamefont {J.~C.~H.}\ \bibnamefont
  {Zeegers}}, \bibinfo {author} {\bibfnamefont {R.}~\bibnamefont {Liew}},
  \bibinfo {author} {\bibfnamefont {J.}~\bibnamefont {Pozorski}}, \ and\
  \bibinfo {author} {\bibfnamefont {B.~J.}\ \bibnamefont {Geurts}},\ }\bibfield
   {title} {\enquote {\bibinfo {title} {A hybrid stochastic-deconvolution model
  for large-eddy simulation of particle-laden flow},}\ }\href@noop {}
  {\bibfield  {journal} {\bibinfo  {journal} {Phys. Fluids}\ }\textbf {\bibinfo
  {volume} {25}},\ \bibinfo {pages} {123302} (\bibinfo {year}
  {2012})}\BibitemShut {NoStop}%
\bibitem [{\citenamefont {Berrouk}\ \emph {et~al.}(2007)\citenamefont
  {Berrouk}, \citenamefont {Laurence}, \citenamefont {Riley},\ and\
  \citenamefont {Stock}}]{Berrouk_2007}%
  \BibitemOpen
  \bibfield  {author} {\bibinfo {author} {\bibfnamefont {A.~S.}\ \bibnamefont
  {Berrouk}}, \bibinfo {author} {\bibfnamefont {D.}~\bibnamefont {Laurence}},
  \bibinfo {author} {\bibfnamefont {J.~J.}\ \bibnamefont {Riley}}, \ and\
  \bibinfo {author} {\bibfnamefont {D.~E.}\ \bibnamefont {Stock}},\ }\bibfield
  {title} {\enquote {\bibinfo {title} {Stochastic modelling of inertial
  particle dispersion by subgrid motion for les of high {R}eynolds number pipe
  flow},}\ }\href@noop {} {\bibfield  {journal} {\bibinfo  {journal} {J.
  Turbul.}\ }\textbf {\bibinfo {volume} {8}},\ \bibinfo {pages} {916--923}
  (\bibinfo {year} {2007})}\BibitemShut {NoStop}%
\bibitem [{\citenamefont {Berrouk}\ \emph {et~al.}(2008)\citenamefont
  {Berrouk}, \citenamefont {Stock}, \citenamefont {Laurence},\ and\
  \citenamefont {Riley}}]{Berrouk_2008}%
  \BibitemOpen
  \bibfield  {author} {\bibinfo {author} {\bibfnamefont {A.~S.}\ \bibnamefont
  {Berrouk}}, \bibinfo {author} {\bibfnamefont {D.~E.}\ \bibnamefont {Stock}},
  \bibinfo {author} {\bibfnamefont {D.}~\bibnamefont {Laurence}}, \ and\
  \bibinfo {author} {\bibfnamefont {J.~J.}\ \bibnamefont {Riley}},\ }\bibfield
  {title} {\enquote {\bibinfo {title} {Heavy particle dispersion from a point
  source in turbulent pipe flow},}\ }\href@noop {} {\bibfield  {journal}
  {\bibinfo  {journal} {Int. J. Multiphase Flow}\ }\textbf {\bibinfo {volume}
  {34}},\ \bibinfo {pages} {916--923} (\bibinfo {year} {2008})}\BibitemShut
  {NoStop}%
\bibitem [{\citenamefont {Gicquel}\ \emph {et~al.}(2002)\citenamefont
  {Gicquel}, \citenamefont {Givi}, \citenamefont {Jaberi},\ and\ \citenamefont
  {Pope}}]{Gicquel_2002}%
  \BibitemOpen
  \bibfield  {author} {\bibinfo {author} {\bibfnamefont {L.~Y.~M.}\
  \bibnamefont {Gicquel}}, \bibinfo {author} {\bibfnamefont {P.}~\bibnamefont
  {Givi}}, \bibinfo {author} {\bibfnamefont {F.~A.}\ \bibnamefont {Jaberi}}, \
  and\ \bibinfo {author} {\bibfnamefont {S.~B.}\ \bibnamefont {Pope}},\
  }\bibfield  {title} {\enquote {\bibinfo {title} {Velocity filtered density
  function for large eddy simulation of turbulent flows},}\ }\href@noop {}
  {\bibfield  {journal} {\bibinfo  {journal} {Phys. Fluids}\ }\textbf {\bibinfo
  {volume} {14}},\ \bibinfo {pages} {1196--1213} (\bibinfo {year}
  {2002})}\BibitemShut {NoStop}%
\bibitem [{\citenamefont {Sheikhi}, \citenamefont {Drozda},\ and\ \citenamefont
  {Pope}(2003)}]{Sheikhi_2003}%
  \BibitemOpen
  \bibfield  {author} {\bibinfo {author} {\bibfnamefont {M.~R.~H.}\
  \bibnamefont {Sheikhi}}, \bibinfo {author} {\bibfnamefont {P.}~\bibnamefont
  {Drozda}, \bibfnamefont {T.~G.~Givi}}, \ and\ \bibinfo {author}
  {\bibfnamefont {S.~B.}\ \bibnamefont {Pope}},\ }\bibfield  {title} {\enquote
  {\bibinfo {title} {Velocity-scalar filtered density function for large eddy
  simulation of turbulent flows},}\ }\href@noop {} {\bibfield  {journal}
  {\bibinfo  {journal} {Phys. Fluids}\ }\textbf {\bibinfo {volume} {15}},\
  \bibinfo {pages} {2321--2337} (\bibinfo {year} {2003})}\BibitemShut {NoStop}%
\bibitem [{\citenamefont {Chibbaro}\ and\ \citenamefont
  {Minier}(2011{\natexlab{b}})}]{Chibbaro_2011b}%
  \BibitemOpen
  \bibfield  {author} {\bibinfo {author} {\bibfnamefont {S.}~\bibnamefont
  {Chibbaro}}\ and\ \bibinfo {author} {\bibfnamefont {J.-P.}\ \bibnamefont
  {Minier}},\ }\bibfield  {title} {\enquote {\bibinfo {title} {The {FDF} or
  {LES}/{PDF} method for turbulent two-phase flows},}\ }\href@noop {}
  {\bibfield  {journal} {\bibinfo  {journal} {J. Phys.: Conf. Ser.}\ }\textbf
  {\bibinfo {volume} {318}},\ \bibinfo {pages} {042049} (\bibinfo {year}
  {2011}{\natexlab{b}})}\BibitemShut {NoStop}%
\end{thebibliography}

%

\end{document}